\documentclass[usenatbib]{mn2e}
\usepackage{amssymb,amsmath,graphicx,natbib,setspace,pifont}
\newcommand{\HI}{\hbox{{\sc H}{\sc i}} }

\newcommand{\CIV}{\hbox{{\sc C}{\sc iv}} }
\newcommand{\CII}{\hbox{{\sc C}{\sc ii}} }
\newcommand{\MgII}{\hbox{{\sc M}g{\sc ii}} }
\newcommand{\SiIV}{\hbox{{\sc S}i{\sc iv}} }
\newcommand{\OVI}{\hbox{{\sc O}{\sc vi}} }
\newcommand{\NV}{\hbox{{\sc N}{\sc v}} }
\newcommand{\NeVIII}{\hbox{{\sc N}e{\sc viii}} }

\newcommand{\NHI}{{N_{\rm HI}}}
\newcommand{\Nion}{{N_{\rm Ion}}}

\newcommand{\cmsq}{\,{\rm cm^{-2}}}

\newcommand{\cmcb}{\,{\rm cm^{-3}}}

\newcommand{\nH}{n_{\rm _{H}} }

\newcommand{\Ob}{\Omega_{\rm b} }
\newcommand{\Om}{\Omega_{\rm m} }

\newcommand{\Ol}{\Omega_{\Lambda} }
\newcommand{\ns}{n_{\rm s} }
\newcommand{\sigeight}{\sigma_{\rm 8} }

\newcommand{\Msun}{{\rm M_{\odot}} }

\newcommand{\Mpc}{ {\rm Mpc} }
\newcommand{\kpc}{ {\rm kpc} }
\newcommand{\kms}{\,{\rm km/s}}

\newcommand{\Gadget}{{\small GADGET-3} }
\newcommand{\Anarchy}{\textquotedblleft Anarchy\textquotedblright}

\newcommand{\CLOUDY}{{\small CLOUDY} }

\newcommand{\apjl}{{ApJ}}
\newcommand{\aj}{{AJ}}
\newcommand{\apj}{{ApJ}}
\newcommand{\apjs}{{ApJS}}

\newcommand{\mnras}{{MNRAS}}

\begin{document}
\title[Highly ionized metals in the EAGLE simulations]{Cosmic distribution of highly ionized metals and their physical conditions in the EAGLE simulations}
\author[A.~Rahmati et al.]
  {Alireza~Rahmati$^{1}$\thanks{rahmati@physik.uzh.ch}, Joop~Schaye$^{2}$, Robert A. Crain$^{3}$, Benjamin D. Oppenheimer$^4$,\newauthor Matthieu Schaller$^5$, Tom Theuns$^{5}$\\
  $^1$Institute for Computational Science, University of Z\"urich, Winterthurerstrasse 190, CH-8057 Z\"urich, Switzerland\\	
  $^2$Leiden Observatory, Leiden University, P.O. Box 9513, 2300 RA, Leiden, The Netherlands\\
  $^3$Astrophysics Research Institute, Liverpool John Moores University, 146 Brownlow Hill, Liverpool, L3 5RF, UK\\
  $^4$University of Colorado, Boulder, CO 80309, USA\\
  $^5$Institute for Computational Cosmology, Department of Physics, University of Durham, South Road, Durham, DH1 3LE, UK\\}
\maketitle

\begin{abstract} 
We study the distribution and evolution of highly ionised intergalactic metals in the Evolution and Assembly of Galaxies and their Environment (EAGLE) cosmological, hydrodynamical simulations. EAGLE has been shown to reproduce a wide range of galaxy properties while its subgrid feedback was calibrated without considering gas properties. We compare the predictions for the column density distribution functions (CDDFs) and cosmic densities of $\SiIV$, $\CIV$, $\NV$, $\OVI$ and $\NeVIII$ absorbers with observations at redshift $z=0$ to $\sim 6$ and find reasonable agreement, although there are some differences. We show that the typical physical densities of the absorbing gas increase with column density and redshift, but decrease with the ionization energy of the absorbing ion. The typical metallicity increases with both column density and time. The fraction of collisionally ionized metal absorbers increases with time and ionization energy. While our results show little sensitivity to the presence or absence of AGN feedback, increasing/decreasing the efficiency of stellar feedback by a factor of two substantially decreases/increases the CDDFs and the cosmic densities of the metal ions. We show that the impact of the efficiency of stellar feedback on the CDDFs and cosmic densities is largely due to its effect on the metal production rate. However, the temperatures of the metal absorbers, particularly those of strong $\OVI$, are directly sensitive to the strength of the feedback.

\end{abstract}

\begin{keywords}
  methods: numerical -- galaxies: formation -- galaxies: high-redshift -- galaxies: absorption lines -- quasars: absorption lines -- intergalactic medium
\end{keywords}

\section{Introduction}

The bulk of elements heavier than helium, i.e., metals, is produced by stars. Stellar winds and supernova explosions enrich the gas around stars and distribute metals in and around galaxies. Metals directly influence the gas dynamics by affecting its cooling rate. Furthermore metals can act as tracers of the complex dynamics and evolution of the gas that is \textquotedblleft colored\textquotedblright~by metals in the vicinity of stars. The distribution of metals, therefore, provides us with a powerful tool to study the processes that regulate star formation and gas dynamics in and around galaxies and that govern the enrichment history of the intergalactic medium (IGM).

Ionized metal species can be observed either in emission or absorption. The ionization state of the absorbing/emitting ions is sensitive to the gas density, temperature, and to the ionizing radiation field they are exposed to. Therefore, one can use ions to extract valuable information about the microscopic physical conditions and the radiation field in the regions where they can be detected.

Detecting metals in the gas around galaxies in emission is challenging, particularly at cosmological distances. However, various metal species are readily detected through their absorption signatures in the spectra of bright background sources (e.g., quasars, hereafter QSOs) out to cosmic distances \citep[see e.g.,][]{Young82,Sargent88,Steidel92}. Absorption by metals has been used to measure the cosmic distribution of metals in the IGM \citep[e.g.,][]{Cowie95,Schaye03,Simcoe06,Aguirre08,Tripp08,Thom08,Cooksey11,Danforth14} and around galaxies \citep[e.g.,][]{Chen01,Adelberger03,Steidel10,Tumlinson11,Turner14}. 

While feedback regulated star formation controls directly the production rate of metals, large-scale outflows, which are likely to stem from feedback processes, can transport metals away from stars and into the IGM \citep[e.g.,][]{Aguirre01,Theuns02,Dave07,Cen11}. Modelling the complex and non-linear interaction between star-formation and gas dynamics becomes even more complicated because of the significant impact of metals on the cooling properties of gas, and therefore on its hydrodynamics. Modelling the distribution of ions also requires accurate ionization corrections which can be achieved by incorporating the evolution of the gas density, temperature and the metagalactic ultra-violet background (UVB) radiation. However, the effects of non-equilibrium ionization and local sources of ionizing radiation on the distribution of different ions may not be negligible \cite[e.g.,][]{Schaye06,Oppenheimer13a,Oppenheimer13b,Rahmati13b}

Modelling the transport of metals self-consistently from star forming regions into the IGM through galactic outflows is challenging. Simulations often do not capture the evolution of outflows self-consistently, e.g., because they temporarily turn off radiative cooling \citep[e.g.,][]{Theuns02,Shen10,Shen13} and/or decouple the outflows hydrodynamically \citep[e.g.,][]{Oppenheimer06,Oppenheimer08,Oppenheimer12,Vogelsberger14}, mainly to increase the feedback efficiency and to improve the numerical convergence. Even simulations which generate outflows more self-consistently require subgrid models and those prescriptions come with parameters whose values require calibration \cite[e.g.,][]{Schaye15}. All those approaches can have a significant impact on the hydrodynamics of the gas and therefore on the distribution of metals. Moreover, they can significantly affect the temperature distribution of gas which can drastically change the ionization state of observable ions.

In this work we investigate the cosmic distribution of metals in the Evolution and Assembly of Galaxies and their Environment (EAGLE) simulations (\citealp{Schaye15}, hereafter S15, \citealp{Crain15}). The EAGLE reference simulation has a large cosmological volume with sufficient resolution to capture the evolution of baryons in and around galaxies over a wide ranges of galaxy masses and spacial scales. The galaxy stellar mass function and the star formation rate density of the Universe are reproduced over a wide range of redshifts \citep{Furlong14} which is critical for achieving reasonable metal production rates. The feedback implementation allows the galactic winds to develop naturally, without predetermined mass loading factors or velocities, and without disabling hydrodynamics or radiative cooling. The abundances of eleven elements are followed using stellar evolution models and are used to calculate the radiative cooling/heating rates in the presence of an evolving UVB. This allows ion fractions to be calculated self-consistently with the hydrodynamical evolution of baryons in the simulation. It is important to note that gas properties were not considered in calibrating the simulations and hence provide predictions that can be used to test the simulation.

As we showed in \citet{Rahmati15}, EAGLE successfully reproduces both the observed global column density distribution function of $\HI$ and the observed radial covering fraction profiles of strong $\HI$ absorbers around Lyman Break galaxies and bright quasars. This shows that the EAGLE reference model which is calibrated based on present day stellar content of galaxies is also capable of reproducing the observed cosmic distribution of gas and its connection with galaxies. In this work, we continue studying the gas by analysing the properties of $\CIV$, $\SiIV$, $\NV$, $\OVI$ and $\NeVIII$ ions in EAGLE. We chose to restrict our study to those ions with high ionization potentials (compared to $\HI$) which are commonly detected in observational studies. Moreover, the relatively high ionization energy of those species, and the typical densities in which they appear, make them largely insensitive to the details of the self-shielding correction which is a necessity for simulating species with lower ionization potentials, such as $\HI$ \citep[e.g.,][]{Rahmati13a, Altay11}. 
 
The structure of this paper is as follows. In $\S$\ref{sec:ingredients} we introduce our cosmological simulations and discuss the photoionization corrections required for obtaining the column densities of different ions and calculating their distributions. We present our predictions for the column density distribution function of different ions in $\S$\ref{sec:results} and show the physical conditions different metal absorbers at different redshifts represent. We discuss the impact of feedback variations on our results in $\S$\ref{sec:fdbk} and conclude in $\S$\ref{sec:conclusions}.

\section{Simulation techniques}
\label{sec:ingredients}

In this section we briefly describe our hydrodynamical simulations and methods. Interested readers can find full descriptions of the simulations in S15 and \citet{Crain15}.
\begin{table*}
\caption{List of EAGLE simulations used in this work. The first four simulations use model ingredients identical to the reference simulation of \citet{Schaye15} while the higher-resolution \emph{Recal-L025N0752} has been re-calibrated to the observed present-day galaxy mass function. Model \emph{NoAGN} does not have AGN feedback, \emph{WeakFB} and \emph{StrongFB} use half and twice as much stellar feedback as the reference simulation, respectively (see \citealp{Crain15}).
 From left to right the columns show: simulation identifier; comoving box size; number of particles (there are equally many baryonic and dark matter particles); initial baryonic particle mass; dark matter particle mass; comoving (Plummer-equivalent) gravitational softening; maximum physical softening and a brief description.} 
\begin{tabular}{lccccccl}
\hline
Simulation & $L$       & $N$ & $m_{\rm b}$ & $m_{\rm dm}$ & $\epsilon_{\rm com}$ & $\epsilon_{\rm prop}$ &  Remarks\\  
                & (cMpc) &       & $(\Msun)$     & $(\Msun)$        & (ckpc)          &    (pkpc)                              &  \\
\hline 
\bf{\emph{Ref-L100N1504}} &    \bf{100} & $\mathbf{2\times1504^3}$ & $\mathbf{1.81 \times 10^6}$ & $\mathbf{ 9.70 \times 10^6}$ & \bf{2.66} & \bf{0.70} & \bf{ref. stellar \& ref. AGN feedback}\\
\emph{Ref-L050N0752} &    50  & $2\times752^3$ & $1.81 \times 10^6$ & $ 9.70 \times 10^6$ & 2.66 & 0.70 & ,,  \\
\emph{Ref-L025N0376} &    25  & $2\times376^3$ & $1.81 \times 10^6$ & $ 9.70 \times 10^6$ & 2.66 & 0.70 & ,, \\
\emph{Ref-L025N0752} &    25  & $2\times752^3$ & $2.26 \times 10^5$ & $ 1.21 \times 10^6$ & 1.33 & 0.35 & ,, \\
\emph{Recal-L025N0752} &    25  & $2\times752^3$ & $2.26 \times 10^5$ & $ 1.21 \times 10^6$ & 1.33 & 0.35 & recalibrated stellar \& AGN feedback\\
\emph{NoAGN} &    25  & $2\times376^3$ & $1.81 \times 10^6$ & $ 9.70 \times 10^6$ & 2.66 & 0.70 & ref. stellar \& no AGN feedback \\
\emph{WeakFB} &    25  & $2\times376^3$ & $1.81 \times 10^6$ & $ 9.70 \times 10^6$ & 2.66 & 0.70 & weak stellar \& ref. AGN feedback \\
\emph{StrongFB} &    25  & $2\times376^3$ & $1.81 \times 10^6$ & $ 9.70 \times 10^6$ & 2.66 & 0.70 & strong stellar \& ref. AGN feedback \\
\hline
\end{tabular}
\label{tbl:sims}
\end{table*}

\subsection{The EAGLE hydrodynamical simulations}
\label{sec:hydro}
The simulations we use in this work are part of the Evolution and Assembly of GaLaxies and their Environments (EAGLE) project (S15; \citealp{Crain15}). The cosmological simulations were performed using  the smoothed particle hydrodynamics (SPH) code \Gadget (last described in \citealp{Springel05}) after implementing significant modifications. In particular, EAGLE uses a modified hydrodynamics solver and new subgrid models. 

We use the pressure-entropy formulation of SPH (see \citealp{Hopkins13}) and the time-step limiter of \citet{Durier12}. Those modifications are part of a new hydrodynamics algorithm called \Anarchy (Dalla Vecchio in prep.) which is used in the EAGLE simulations (see Appendix A of S15 and \citealp{Schaller15}).

Our reference model (described below) has a subgrid model for star formation which uses the metallicity-dependent density threshold of \citet{Schaye04} to model the transition from the warm atomic to the cold, molecular interstellar gas phase and which follows the pressure-dependent star formation prescription of \citet{Schaye08}. Galactic winds develop naturally and without turning off radiative cooling or the hydrodynamics thanks to stellar and AGN feedback implementations based on the stochastic, thermal prescription of \citet{DV12}. The adopted stellar feedback efficiency depends on both metallicity and density to account, respectively, for greater thermal losses at higher metallicities and for residual spurious resolution dependent radiative losses \citep{DV12,Crain15}. Supermassive black holes grow through gas accretion and mergers (\citealp{Springel05b, Booth09}; S15) and the subgrid model for gas accretion accounts for angular momentum \citep{Rosas13}. The implementation of metal enrichment is described in \citet{Wiersma09b}, modified as described in S15 and follows the abundances of the eleven elements that are found to be important for radiative cooling and photoheating: H, He, C, N, O, Ne, Mg, Si, S, Ca, and Fe, assuming a \citet{Chabrier03} initial mass function. Element-by-element metal abundances are used to calculate the radiative cooling/heating rates in the presence of the uniform cosmic microwave background and the \citet{HM01} (hereafter HM01) UVB model which includes galaxies and quasars, assuming the gas to be optically thin and in ionization equilibrium \citep{Wiersma09a}. We use the same ion fractions that are used in the cooling/heating rates calculations to compute the column density of ions.

 The subgrid models for energetic feedback used in EAGLE are calibrated based on observed present day galaxy stellar mass function and galaxy sizes, which are reproduced with unprecedented accuracy for a hydrodynamical simulation of this kind (S15;~\citealp{Crain15}). The same simulation also shows very good agreement with a large number of other observed galaxy properties which were not considered in the calibration (S15;~\citealp{Crain15,Furlong14,Schaller14,Trayford15,Rahmati15}).

The adopted cosmological parameters are: $\{\Om=0.307,\ \Ob=0.04825,\ \Ol=0.693,\ \sigeight=0.8288,\ \ns=0.9611,\ h=0.6777\} $ \citep{Planck13}. The reference simulation, \emph{Ref-L100N1504}, has a periodic box of $L = 100$ comoving $\Mpc$ (cMpc) and contains $1504^3$ dark matter particles with mass $9.7 \times 10^6~\Msun$ and initially an equal number of baryonic particles with mass $1.81 \times 10^6~\Msun$. The Plummer equivalent gravitational softening length is set to $\epsilon_{\rm{com}} = 2.66~\kpc$ and is limited to a maximum physical scale of $\epsilon_{\rm{prop}} = 0.7$ proper kpc (pkpc). Throughout this work, we also use simulations with different box sizes (e.g., $L = 25$ or $50$ cMpc), different resolutions and different feedback models to test the impact of those factors on our results. All the simulations used in this work are listed in Table \ref{tbl:sims}.

\subsection{Calculating ion abundances and column densities}
\label{sec:ions_method}
To obtain the ionization states of the elements, we follow \citet{Wiersma09a} and use \CLOUDY version 07.02 \citep{Ferland98} to calculate ion fractions. Assuming the gas is in ionization equilibrium and exposed to the cosmic microwave background and the HM01 model for the evolving UV/X-ray background from galaxies and quasars, the ion fractions are tabulated as a function of density, temperature and redshift. We use these tables to calculate the ion fractions. We stress, that identical ion fractions were used in our cosmological simulations to obtain the heating/cooling rates, which makes our analysis self-consistent.

\begin{figure*}
\centerline{\hbox{{\includegraphics[width=0.5\textwidth]
              {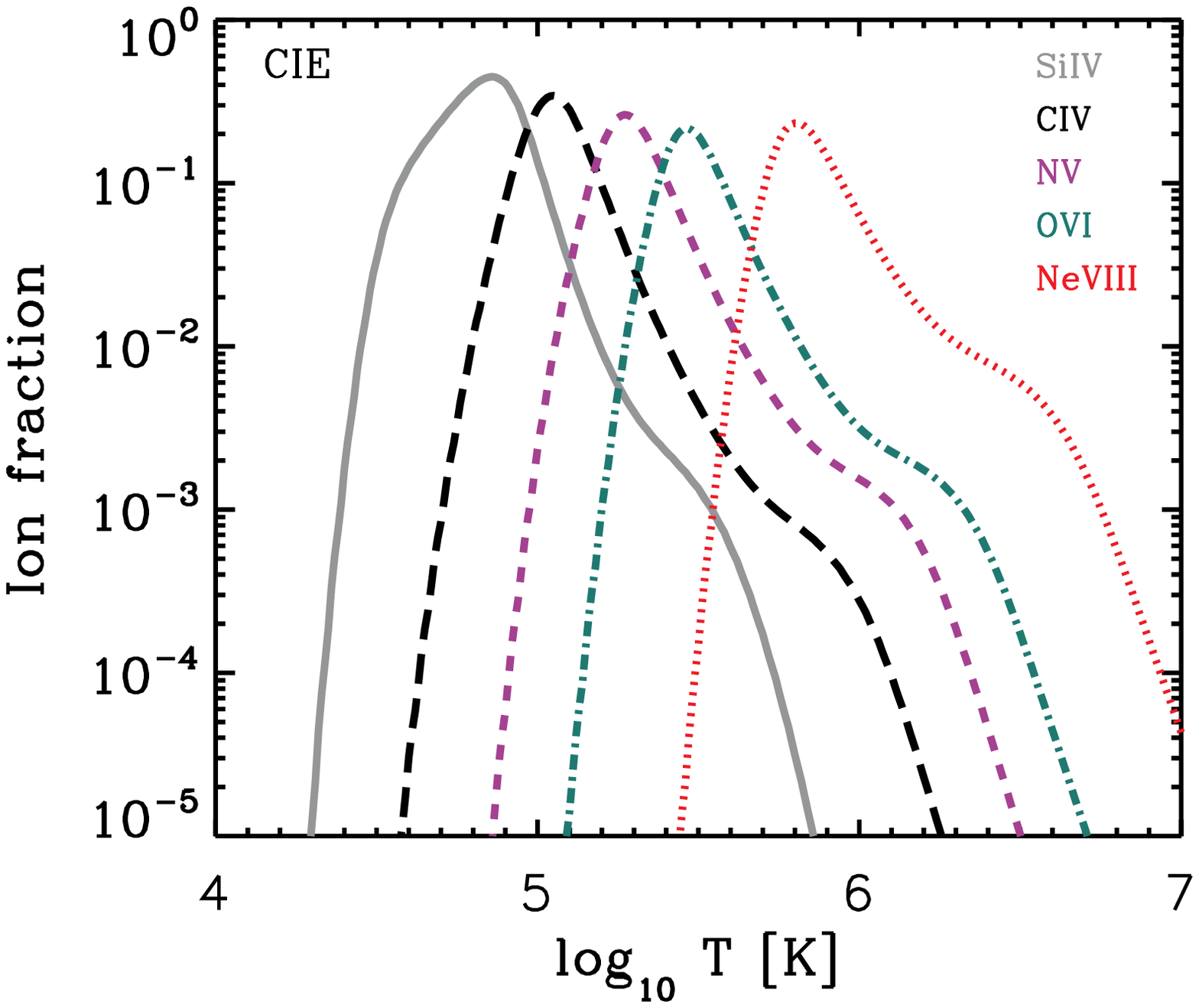}}}
                 \hbox{{\includegraphics[width=0.5\textwidth]	
              {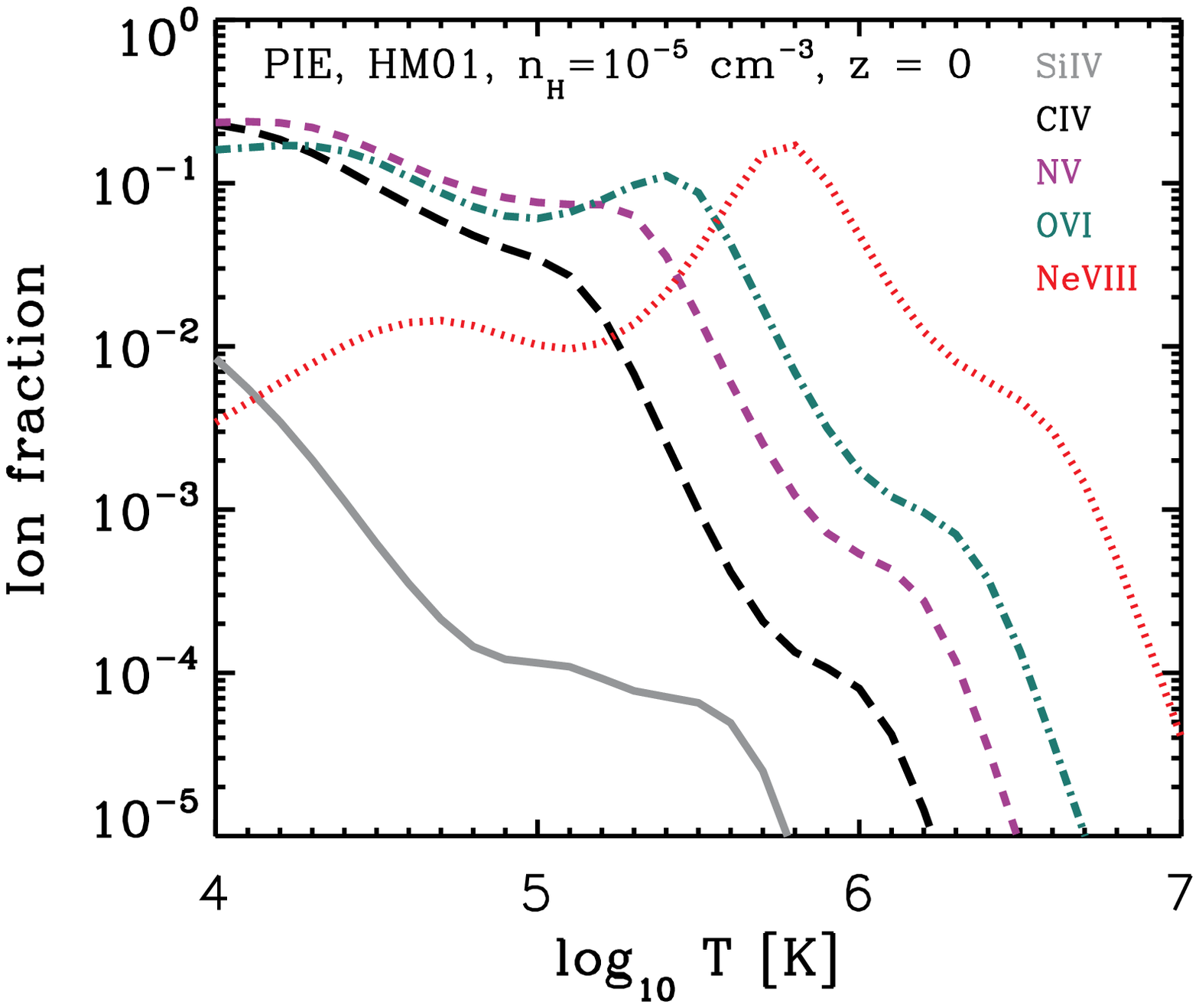}}}}
\caption{Ion fraction of different species as a function of temperature in collisional ionization equilibrium (CIE; left) and in photoionization equilibrium (PIE; right) with the HM01 UVB model for gas density $\nH = 10^{-5} \cmcb$ at $z = 0$. Solid, long-dashed, dashed, dot-dashed and dotted curves show ion fractions for $\SiIV$, $\CIV$, $\NV$, $\OVI$ and $\NeVIII$, respectively. In CIE, the temperature at which ion fractions peak increases with the ionization energy of the species. Photoionization changes the CIE ion fractions by pumping species into or out of a specific ionization state, depending on density and temperature, but becomes unimportant at sufficiently high temperatures.}
\label{fig:CIE}
\end{figure*}
\begin{table}
\begin{center}
\caption{List of ionization energies and peak collisional ionization temperatures for the ions we study in this work. For each ion both the energy required to bring the ion from one level down to the desired level ($\Delta E_{i-1 \rightarrow i}$) and the energy required to ionize it further to the next level ($\Delta E_{i \rightarrow i+1}$) are indicated. The peak collisional ionization temperature, $T_{\rm{CIE,max}}$, is defined as the temperature at which the ion fractions are maximum in collisional ionization equilibrium.} 
\begin{tabular}{lccc}
\hline
Ion &  $\Delta E_{i-1 \rightarrow i}$        & $\Delta E_{i \rightarrow i+1}$ & $T_{\rm{CIE,max}}$ \\  
      &    eV                     &    eV                   &             K \\
\hline 
\SiIV &     33.49           &     45.14            &      $10^{4.9}$           \\
\CIV &     47.89           &     64.49            &       $10^{5.1}$           \\
\NV  &      77.47           &     97.89            &      $10^{5.3}$              \\
\OVI &     113.90          &    138.12           &      $10^{5.5}$              \\
\NeVIII &  207.27         &    239.09            &     $10^{5.8}$               \\
\hline
\end{tabular}
\label{tbl:energies}
\end{center}
\end{table}

Using the tabulated ion fractions as a function of density, temperature (see Fig. \ref{fig:CIE}) and redshift, we calculate the total ion mass in each SPH particle using the smoothed elemental abundances for that particle (i.e., for element $X$ we use $\rho_X / \rho_{\rm{g}}$ instead of $m_X / m_{\rm{g}}$ which is its particle abundance). We calculate the total cosmic density of different ions in our simulation by summing over all particles. Since we use a polytropic equation of state to limit the Jeans mass of the star-forming gas in our simulations, the temperatures of star-forming SPH particles are not physical. Therefore, before calculating the ion fractions we set the temperature of the interstellar medium (ISM) particles to $T_{\rm{ISM}} = 10^4$ K which is the typical temperature of the warm-neutral ISM. Noting, however, that all the ions we study in this work arise almost entirely from densities much lower than that of the ISM gas (see $\S$\ref{sec:physicalproperties}), our results are insensitive to this correction.

Furthermore, we calculate the column densities for different ions using SPH interpolation and by projecting the ionic content of the full simulation box onto a 2-D grid, using several slices along the projection axis. We found for the  \emph{Ref-L100N1504} simulation that using a grid with $10,000^2 = 10^8$ pixels and 8-16 slices (depending on redshift) results in converged column density distribution functions in the range of column densities we show in the present work. We use the same projection technique to calculate the ion-weighted physical properties in each pixel, such as the density, temperature and metallicity. We use those physical quantities to study the physical state of the gas that is represented by different ions at different column densities, and at different redshifts (see $\S$\ref{sec:physicalproperties}).

By adopting a uniform UVB model for calculating the ionization states, our calculations ignore the effect of self-shielding. While self-shielding is expected to play a role for $\nH \gtrsim 10^{-2}-10^{-3} \cmcb$ and $T \lesssim 10^4$ K \citep{Schaye01, Rahmati13a}, the high ionization species we study in this work (i.e., $\SiIV$, $\CIV$, $\OVI$, $\NV$ and $\NeVIII$) are mainly found at much lower densities and/or higher temperatures where it is safe to assume that the gas is optically-thin (see $\S$\ref{sec:physicalproperties}). For ions with lower ionization energies (e.g., $\HI $, $\CII$, $\MgII$), however, it is essential to account for self-shielding and recombination radiation \citep{Raicevic14}, for which fitting functions based on accurate radiative transfer simulations can be used \citep[e.g.,][]{Rahmati13a}.

We also note that our assumption of ionization equilibrium may introduce systematic uncertainties in our ion fractions and cooling rates that are used in the simulations \citep[e.g.,][]{Oppenheimer13a}. In addition, the neglect of inhomogeneity in the UVB radiation close to local sources of ionization radiation can be considered as another caveat. We note, however, that local sources are only thought to be important for absorbers as rare as Lyman limit systems \citep[e.g.,][]{Schaye06,Rahmati13b} and will therefore possibly only affect absorbers with the highest ion column densities. However, as \citet{Oppenheimer13b} showed, non-equilibrium effects in the proximity of the fossil AGNs can enhance the ion fraction of higher ionization stages such as $\OVI$ and $\NeVIII$ while reducing the abundances of lower ionization ions such as $\CIV$. Accounting for the aforementioned effects requires to perform cosmological hydrodynamical simulations with non-equilibrium ionization calculations \citep{Oppenheimer13b,Richings14a, Richings14b} coupled with accurate radiative transfer, which is beyond the scope of this study.

\section{Results}
\label{sec:results}
In this section we present our predictions for the column density distribution functions of different ions at different redshifts ($\S$\ref{sec:ionCDDF}), the evolution of the cosmic density of ions ($\S$\ref{sec:iondensity}) and the physical conditions of the gas traced by different ions for different column densities and epochs ($\S$\ref{sec:physicalproperties}). Furthermore, we compare the CDDFs with observations.
\begin{figure*}
\centerline{\hbox{{\includegraphics[width=0.45\textwidth]
              {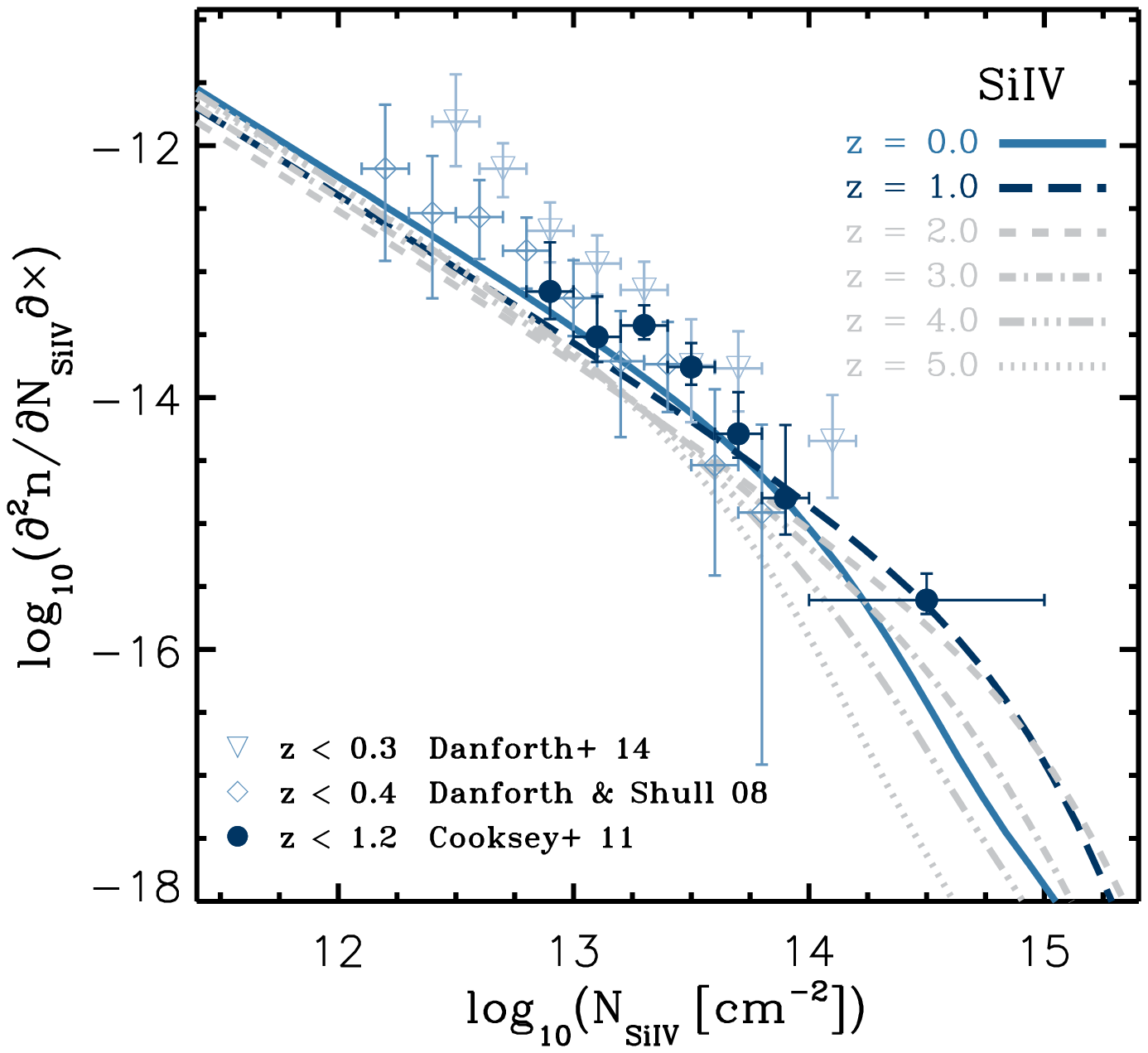}}}
             \hbox{{\includegraphics[width=0.45\textwidth]	
             {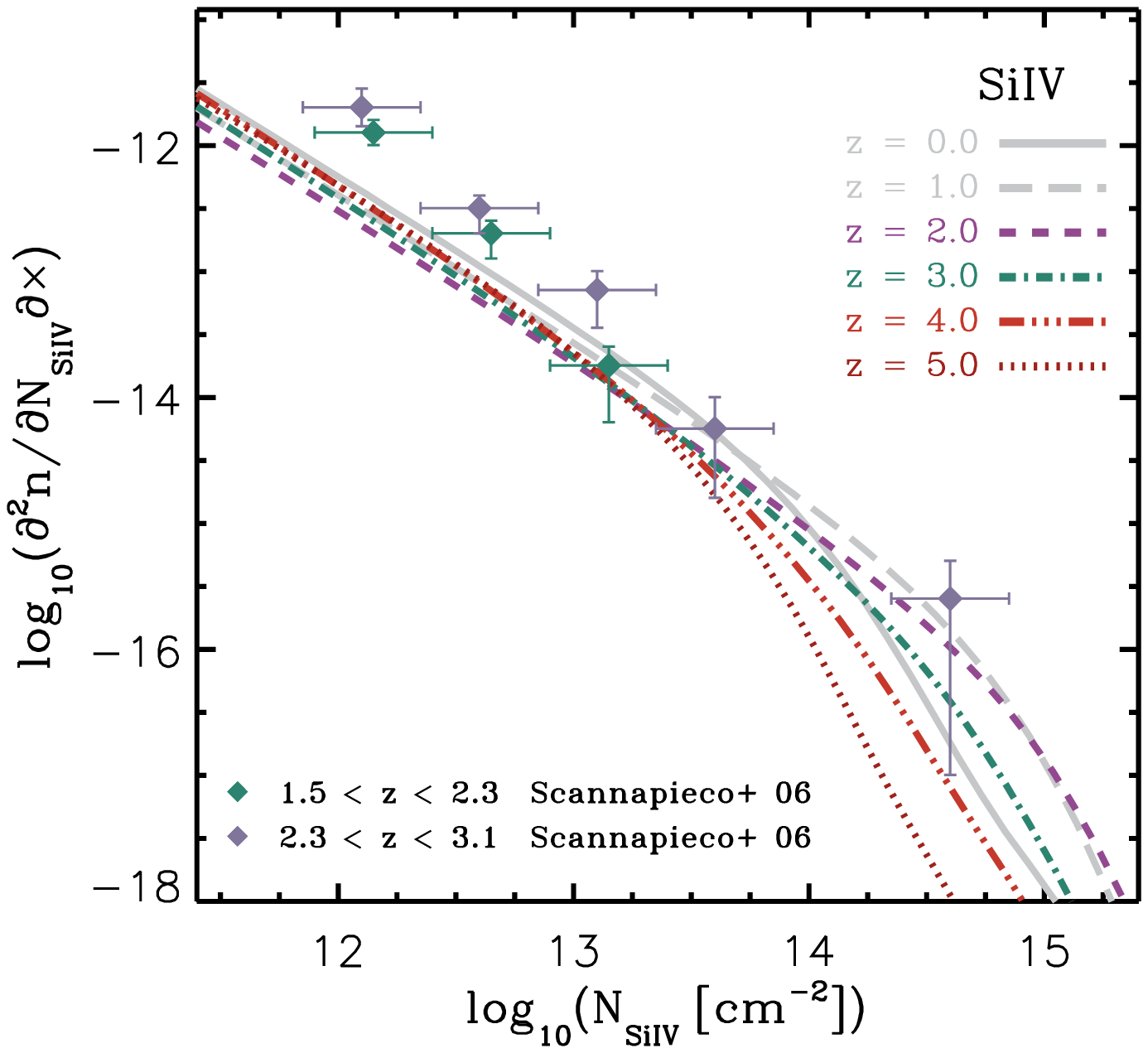}}}}
\centerline{\hbox{{\includegraphics[width=0.45\textwidth]
              {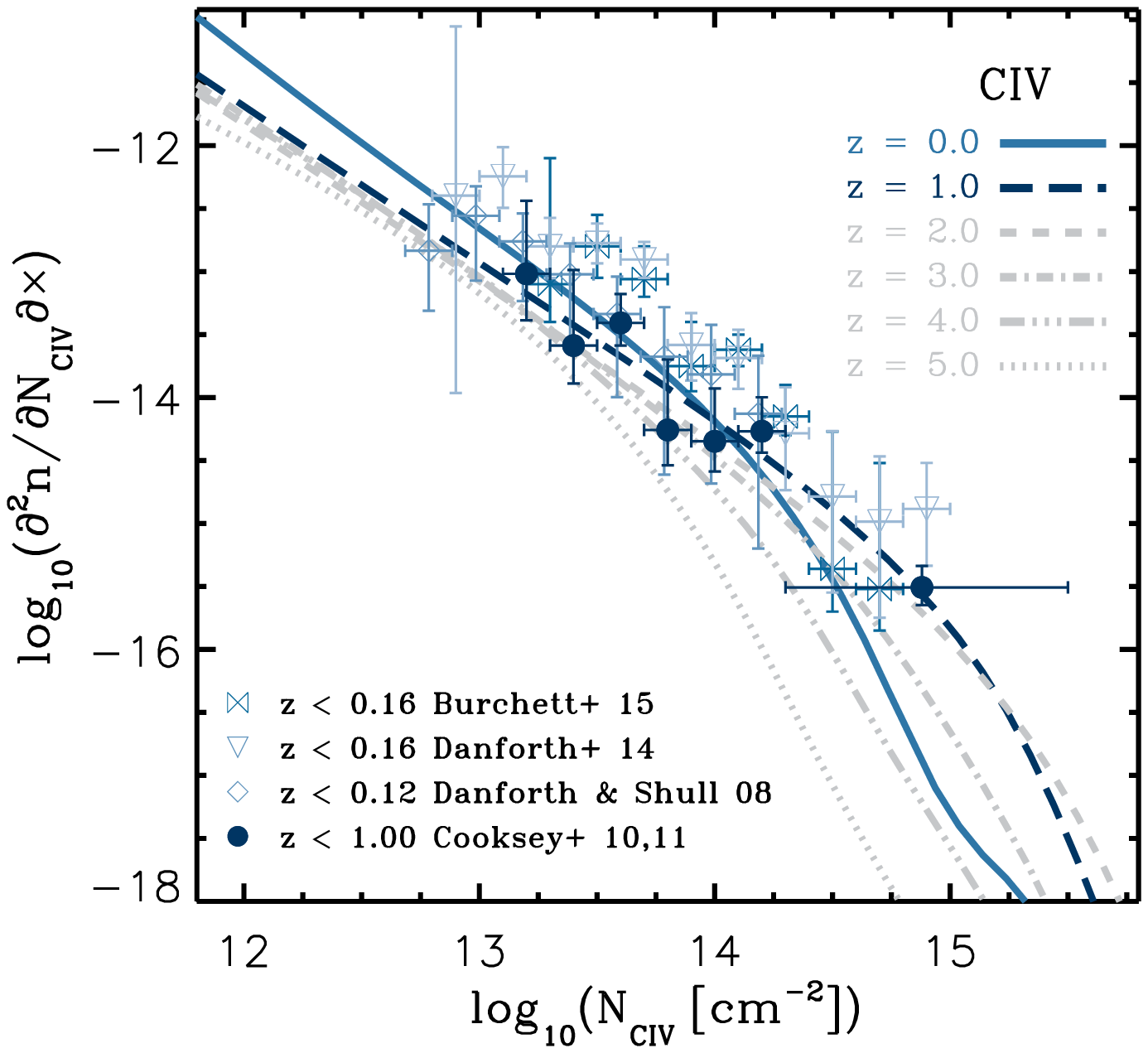}}}
             \hbox{{\includegraphics[width=0.45\textwidth]	
             {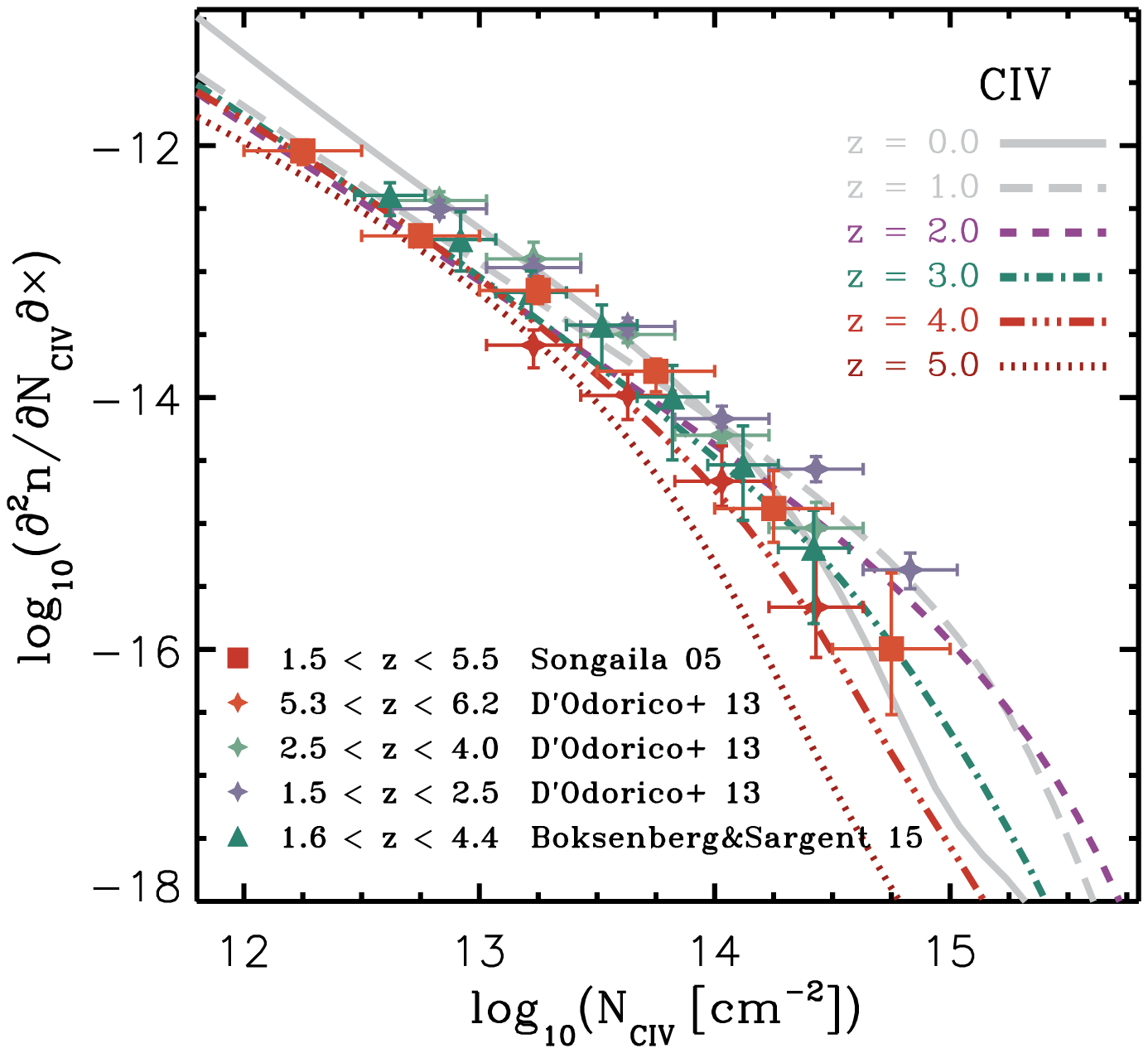}}}}
\centerline{\hbox{{\includegraphics[width=0.45\textwidth]
              {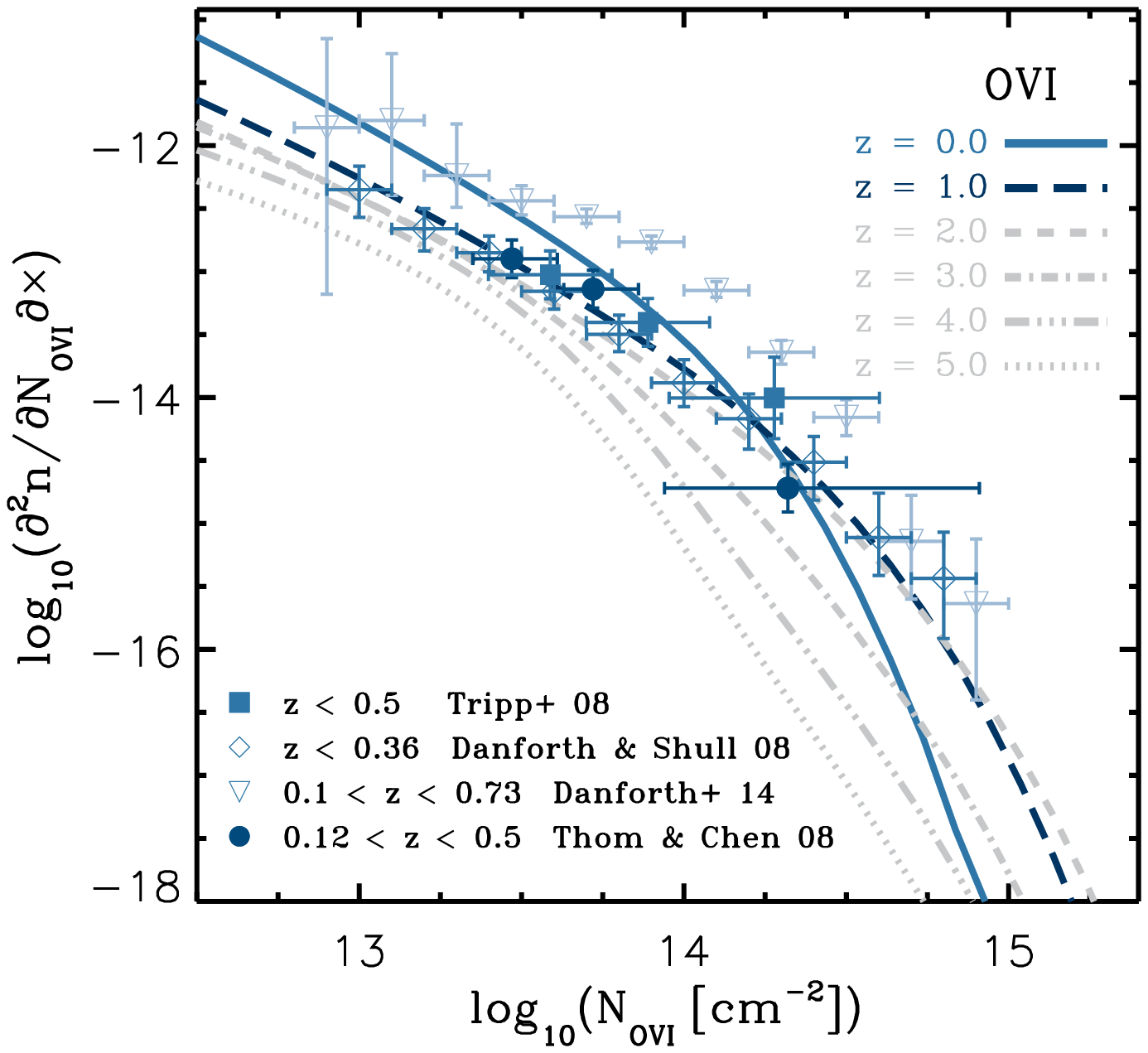}}
             \hbox{{\includegraphics[width=0.45\textwidth]	
             {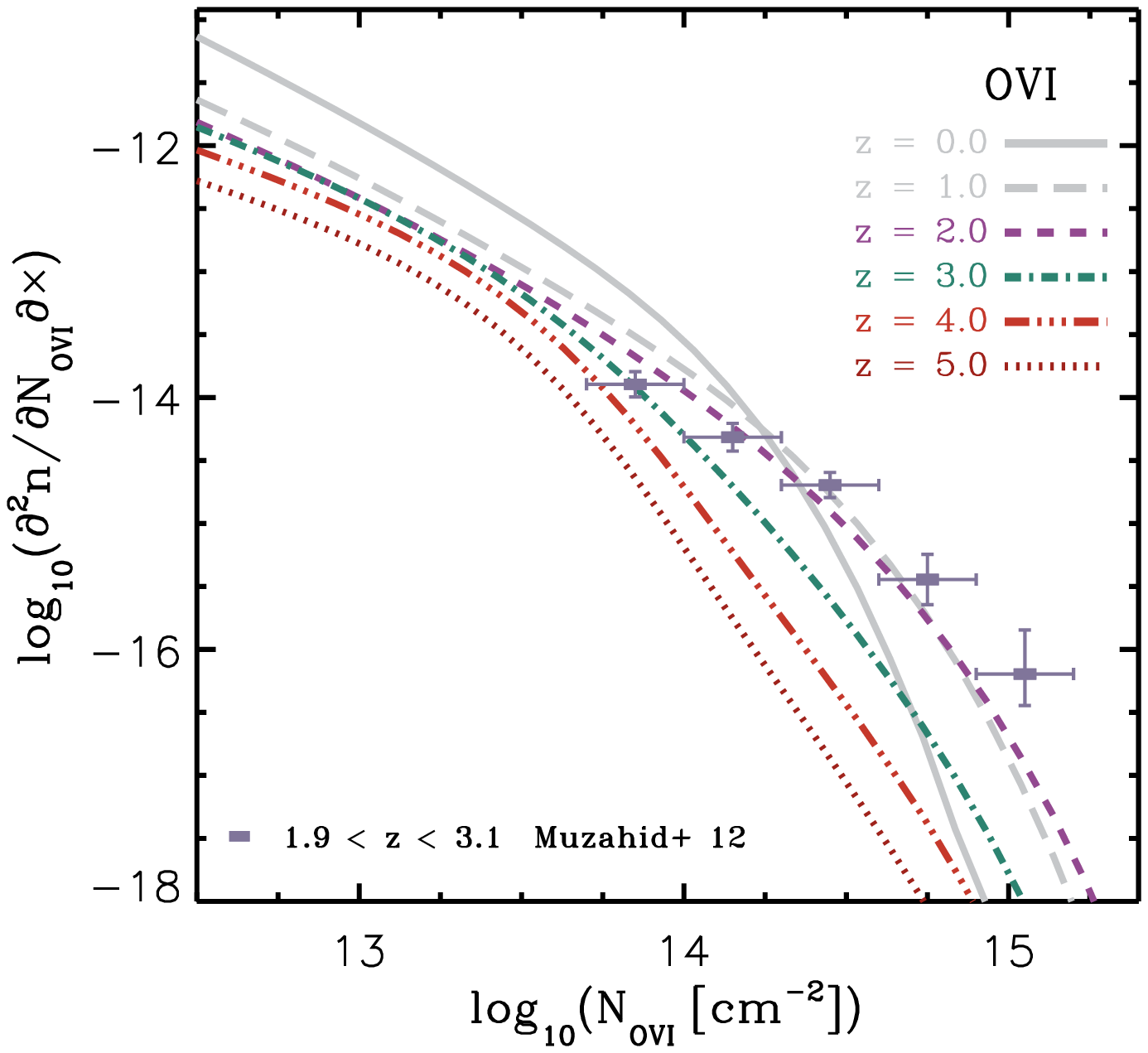}}}}}

\caption{Column density distribution functions of $\SiIV$ (top), $\CIV$ (middle), and $\OVI$ (bottom) at different redshifts in the EAGLE \emph{Ref-L100N1504} simulation. Solid (light blue), long dashed (blue), dashed (purple), dot-dashed (green), dot-dot-dot-dashed (red), and dotted (dark red) curves correspond to $z = 0$, 1, 2, 3, 4 and 5, respectively. In the left column the CDDFs for $z = 0$ and 1 are bold (colored) while other redshifts are shown using light gray. In the right column, higher redshift CDDF are bold (colored) and the CDDFs for $z \le 1$ are gray. Observational measurements (see the legend on the bottom-left of each panel) are also split into a low-redshift sample with $z \lesssim 1$ (left column) and a high-redshift sample (right column) and are shown using symbols with 1-$\sigma$ error bars and have colors matched to curves with appropriate redshifts.}
\label{fig:cddf-allz}
\end{figure*}
\begin{figure*}
\centerline{\hbox{{\includegraphics[width=0.45\textwidth]
              {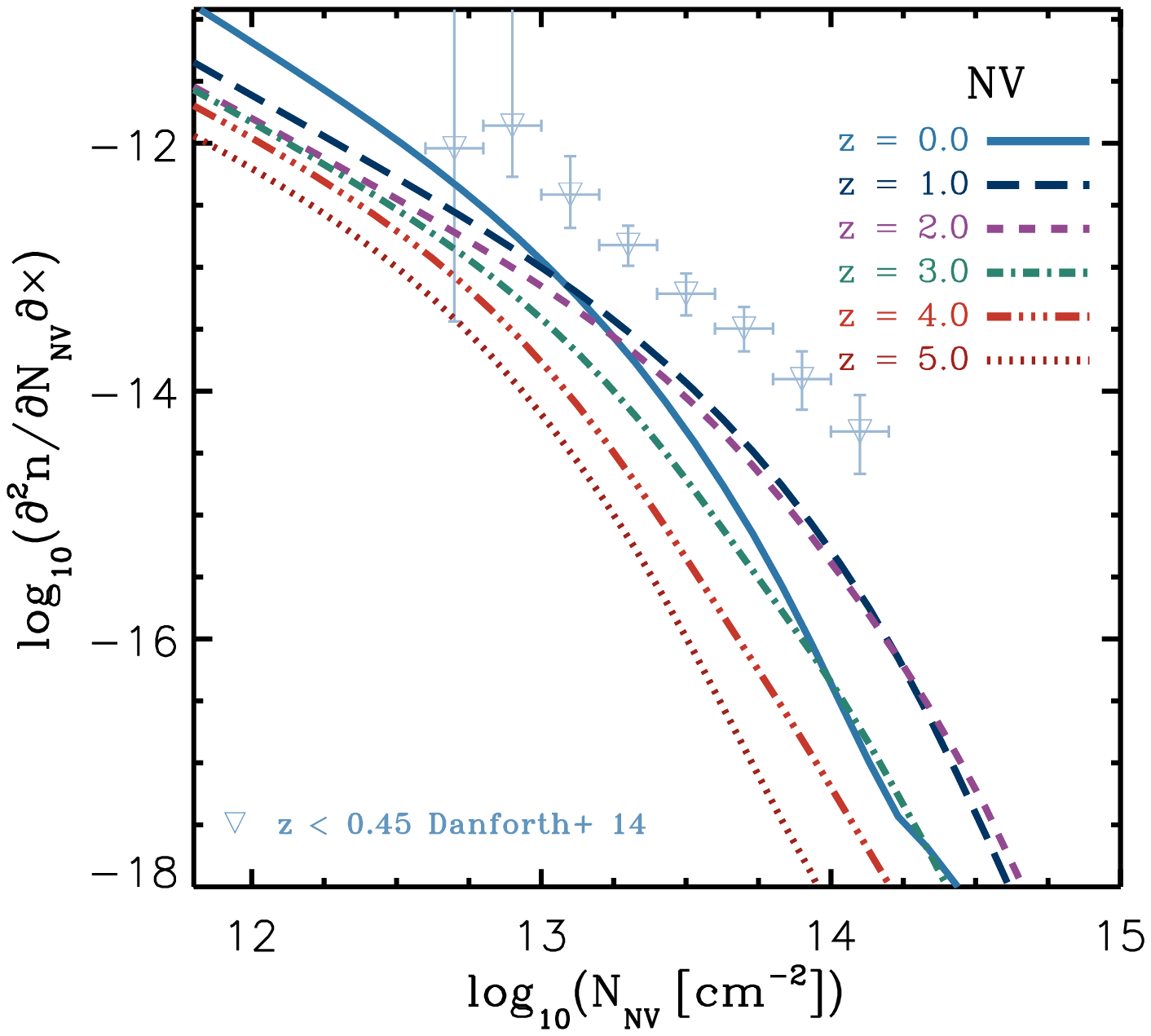}}}
             \hbox{{\includegraphics[width=0.45\textwidth]	
             {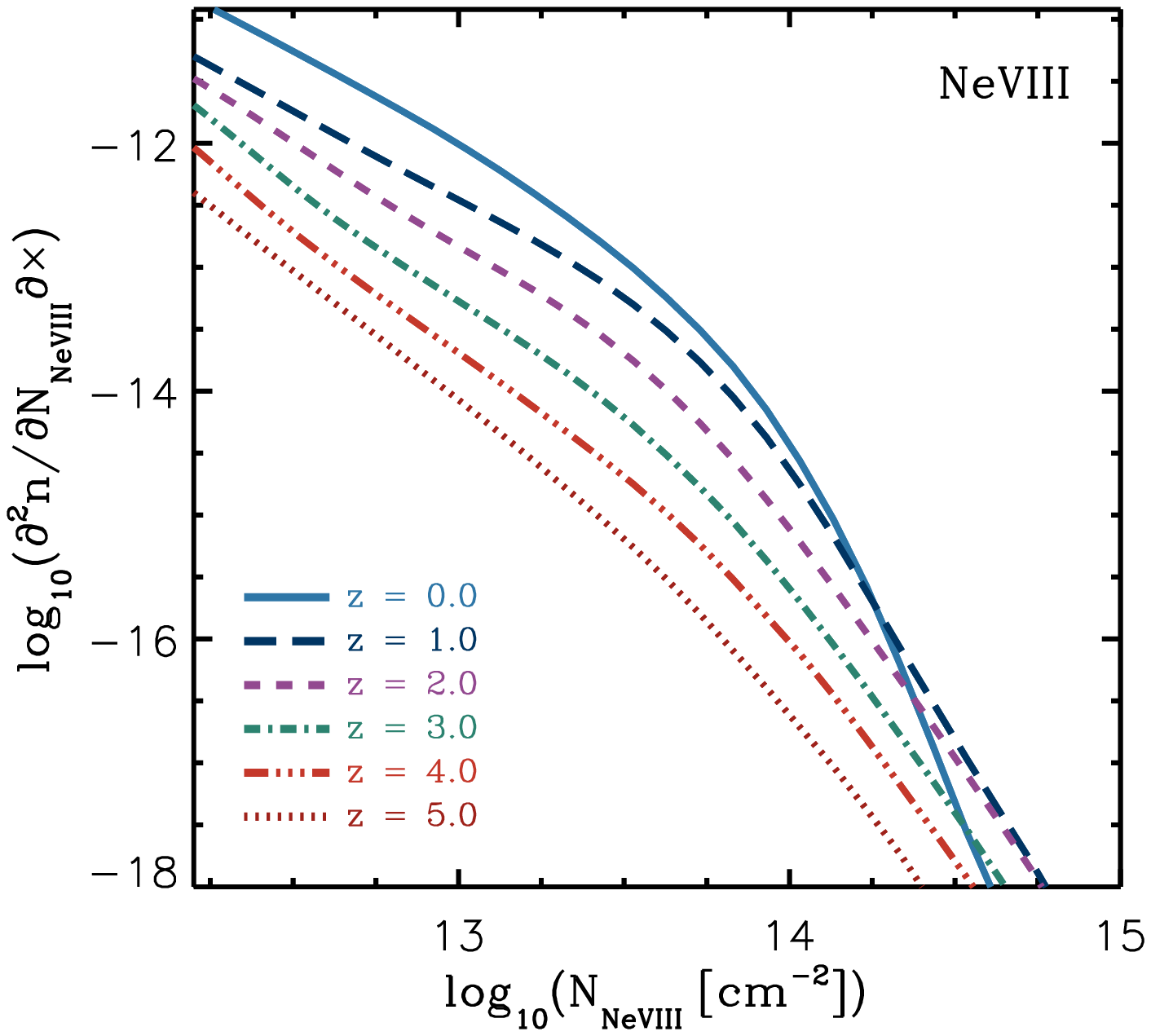}}}}
             
\caption{Column density distribution functions of $\NV$ (left) and $\NeVIII$ (right) at different redshifts in the EAGLE \emph{Ref-L100N1504} simulation. Solid (light blue), long dashed (blue), dashed (purple), dot-dashed (green), dot-dot-dot-dashed (red), and dotted (dark red) curves correspond to $z = 0$, 1, 2, 3, 4 and 5, respectively. $\NV$ observational measurements by \citet{Danforth14} are shown using triangles with 1-$\sigma$ error bars. Note that the CDDFs measured by \citet{Danforth14} for other ions are systematically higher than both our predictions and other available measurements (see Fig. \ref{fig:cddf-allz}) and the significance of their deviation from our results therefore remains to be tested with more observational measurements.}
\label{fig:cddf-allz2}
\end{figure*}
\begin{figure*}
\centerline{\hbox{{\includegraphics[width=0.45\textwidth]
              {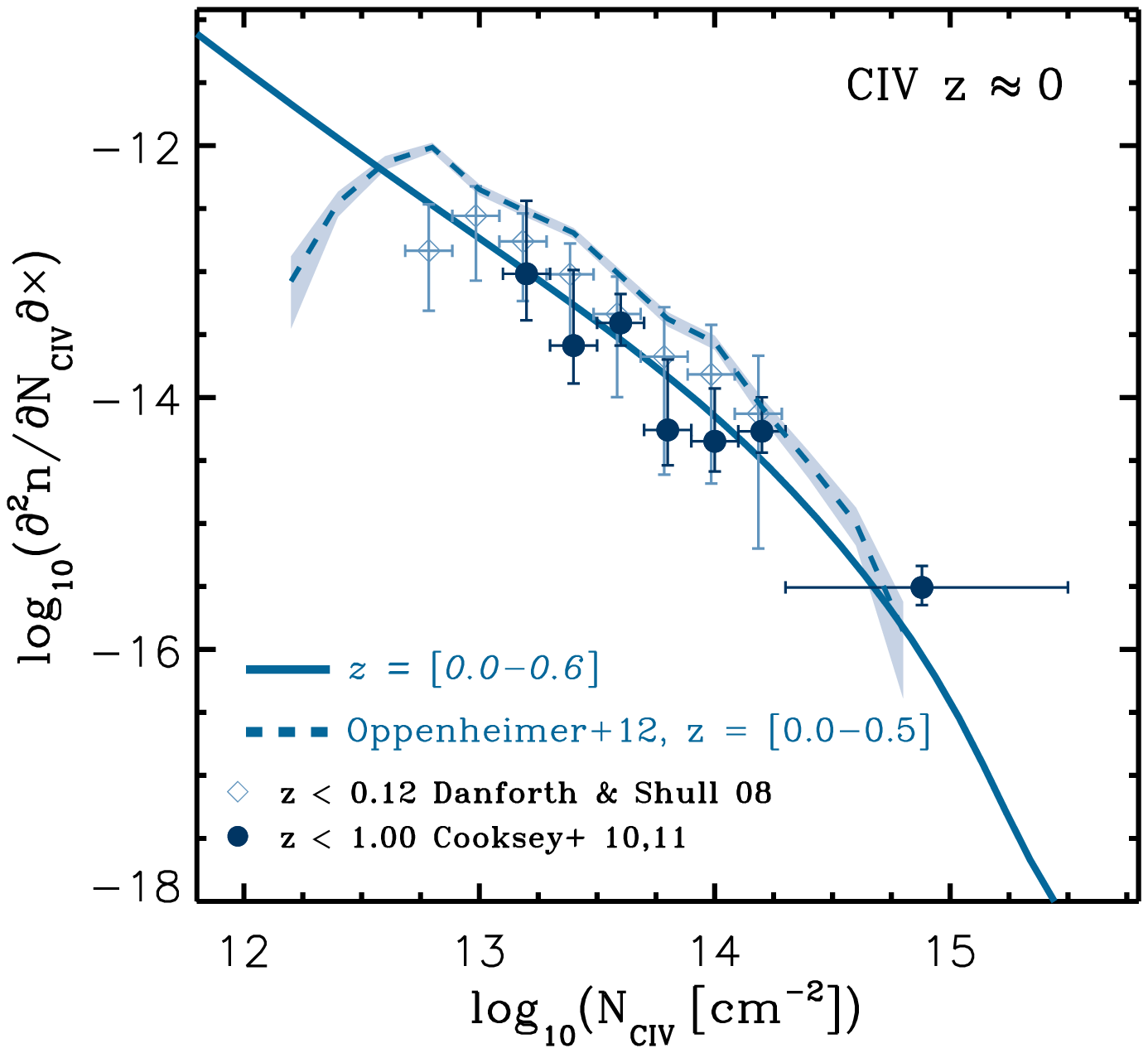}}}
             \hbox{{\includegraphics[width=0.45\textwidth]	
             {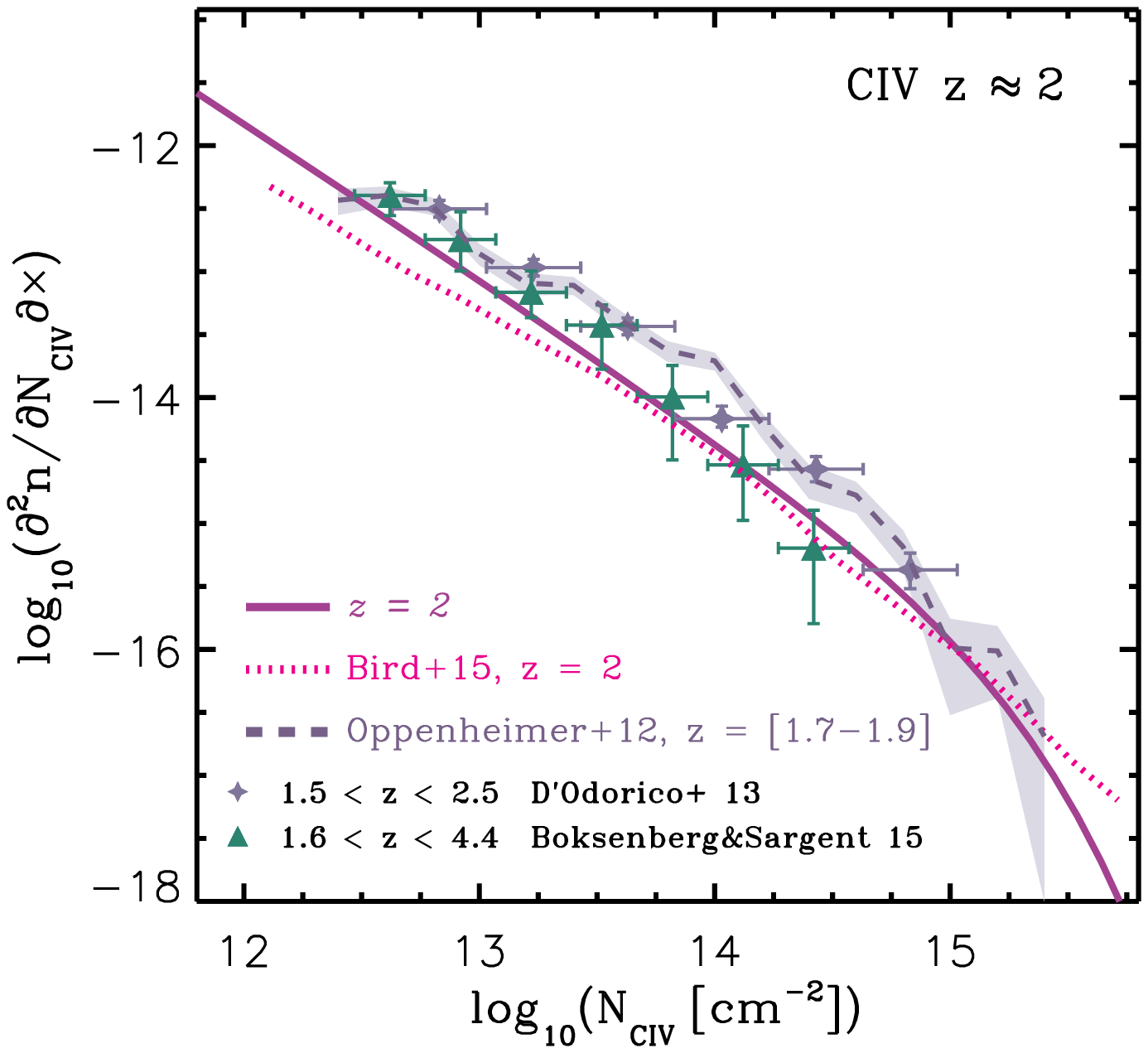}}}}
\centerline{\hbox{{\includegraphics[width=0.45\textwidth]
              {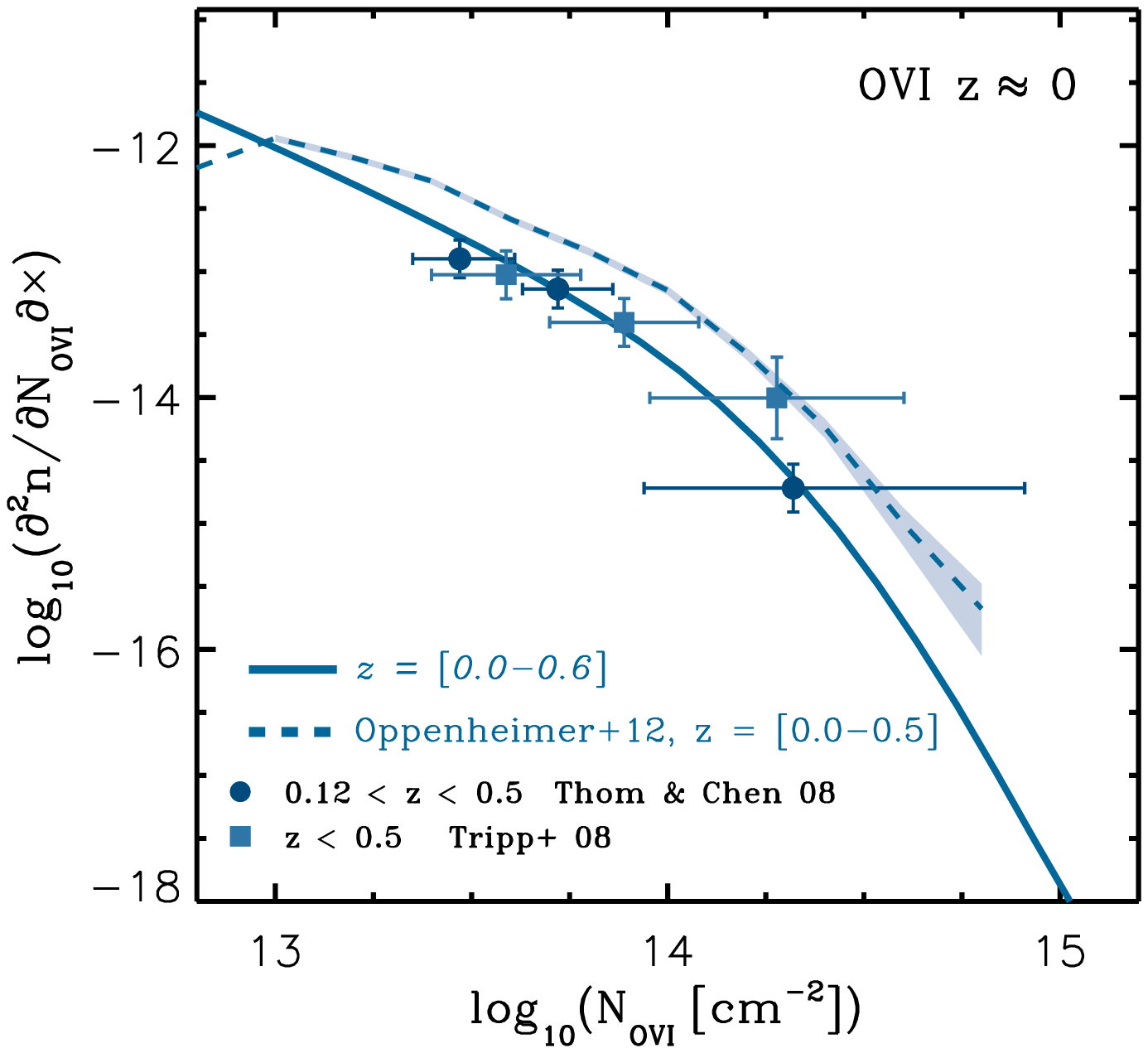}}}
             \hbox{{\includegraphics[width=0.45\textwidth]	
             {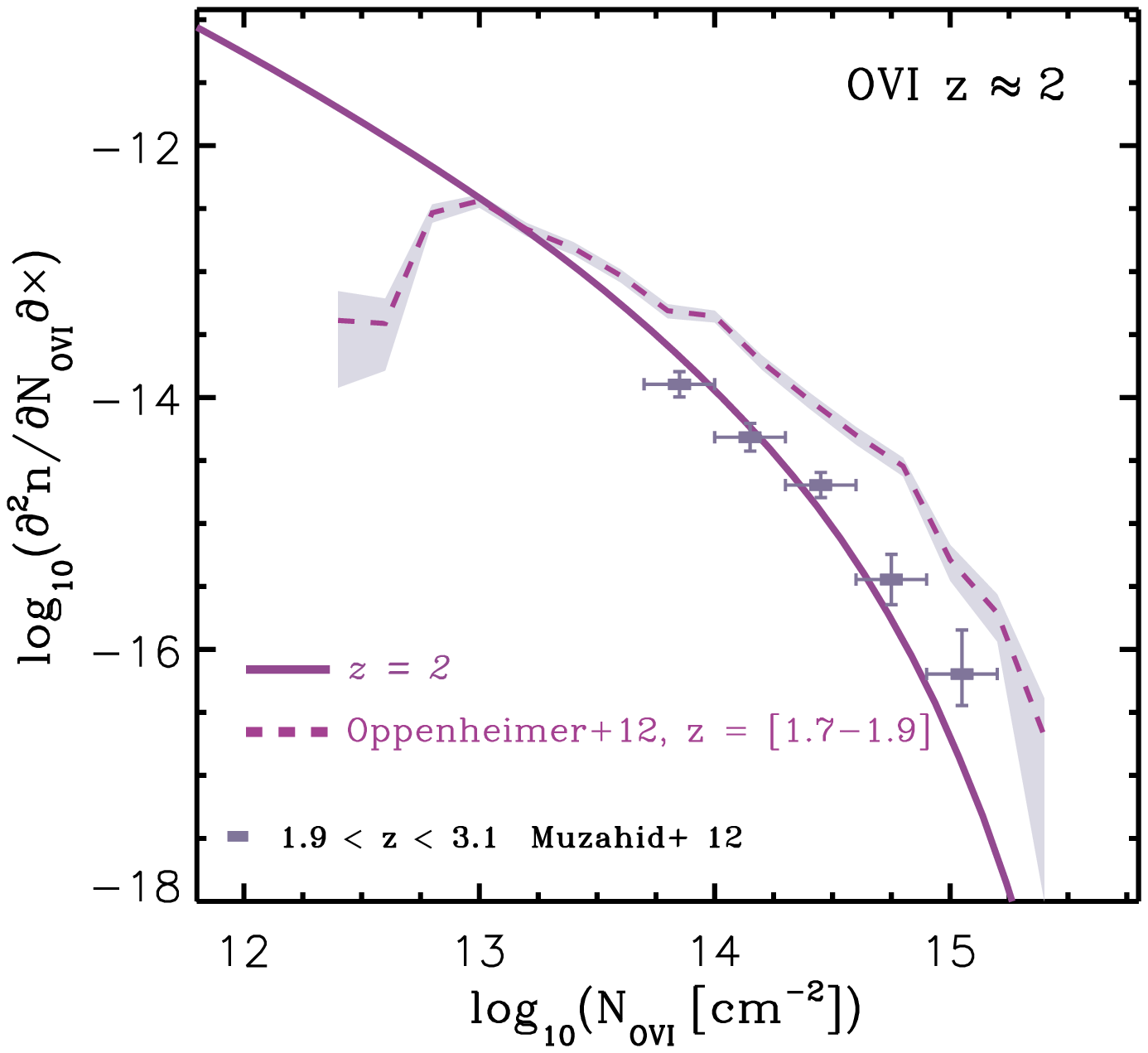}}}}
\caption{Column density distribution functions of $\CIV$ (top) and $\OVI$ (bottom) at $z \approx 0$ (left) and $z \approx 2$ (right) in simulations performed by different groups. Solid, dashed and dotted curves show the results presented in this work, in \citet{Oppenheimer12} and in \citet{Bird15}, respectively. The shaded area around the dashed curve shows the 1-$\sigma$ poisson error associated with the reported CDDFs in \citet{Oppenheimer12}. The $z \approx 2$ results from \citet{Oppenheimer12} represents the mean of snapshots at $z = 1.7$ and $z = 1.9$ while our results and the Illustris $\CIV$ CDDF presented in \citet{Bird15} are calculated at $z = 2$. For $z \approx 0$, and in analogy to \citet{Oppenheimer12} who used the mean of $z = 0$ and $z = 0.5$, we show the mean CDDF of EAGLE at $z = 0$ and $z = 0.6$. Symbols show compilations of observational measurements relevant to each redshift. Despite significant methodological/modelling differences in the different simulations, their predicted CDDFs are in reasonable agreement. Although the predicted $\CIV$ CDDF by \citet{Oppenheimer12} agrees better with the measurements presented in \citet{Dodorico13} at $z \approx 2$, the EAGLE predictions show a better agreement with observational data for $\CIV$ at low redshifts and for $\OVI$ at both low and high redshifts.}
\label{fig:CDDF-sims}
\end{figure*}
\subsection{Column density distribution functions}
\label{sec:ionCDDF}
The rate of incidence of absorbers as a function of column density is a useful statistical quantity which is often measured in observational studies. The column density distribution function (CDDF) for a given ion is conventionally defined as the number of absorbers, $n$, per unit column density, $N_{\rm{ion}}$, and per unit absorption length, $d X =  {d z}~ (H_0 / H(z)) (1+z)^2$:
\begin{eqnarray}
f(N_{\rm{ion}},z) \equiv \frac{d^2n}{d N_{\rm{ion}} d X}
\equiv \frac{d^2n}{d N_{\rm{ion}} d z} \frac{H(z)}{H_0} \frac{1}{(1+z)^2}.
\label{eq:CDDF}
\end{eqnarray}

Fig. \ref{fig:cddf-allz} shows the predicted CDDFs of $\SiIV$ (top), $\CIV$ (middle) and $\OVI$ (bottom) at different redshifts in the EAGLE \emph{Ref-L100N1504} simulation, while $\NV$ and $\NeVIII$ are shown in Fig. \ref{fig:cddf-allz2} (the left and right panels, respectively). In each panel, the solid (light blue), long dashed (blue), dashed (green), dot-dashed (orange), dot-dot-dot-dashed (red), and dotted (dark red) curves correspond to $z = 0$, 1, 2, 3, 4 and 5, respectively.

To compare the predicted CDDFs with observations, different compilations of the observed CDDF for different ions are shown with symbols in figures \ref{fig:cddf-allz} and \ref{fig:cddf-allz2}. To facilitate the comparison for $\SiIV$ (top), $\CIV$ (middle) and $\OVI$, for which there are several measurements, we split the observations into a low-redshift group (with $z \lesssim 1$; shown in the left column) and a high-redshift sample (shown in the right column). Furthermore, we chose a color for each set of data-points similar to the color of the curve with the closest redshift and show the predictions with higher/lower redshifts in gray. All the observational measurements are corrected to make them follow the Planck cosmology which has been used in our simulations. Our compilation of the observed data is collected from the following papers:  \citet{Songaila05, Scannapieco06, Tripp08, Thom08, Danforth08,Dodorico13,Cooksey10,Cooksey11,Muzahid12,Danforth14,Boksenberg03,Burchett15}, as indicated in the bottom-left of each panel. Wherever the reported CDDFs were available for both absorption \emph{components} and \emph{systems} (see below), we showed the CDDF calculated by counting absorption \emph{systems} \cite[e.g.,][]{Boksenberg03}. It is important to note that different data sets at the same redshift do not always agree within the error bars. As an example, this is evident by comparing the low-redshift $\OVI$ data points from \citet{Danforth14} (blue triangles) with the other low-redshift data points, e.g., the data points from \citet{Danforth08} (open diamonds), in the bottom-left panel of Fig. \ref{fig:cddf-allz}, particularly at $N_{\OVI} \lesssim 10^{14}\cmsq$. In other words, the differences between different observational data sets sometimes exceed the reported statistical uncertainties which suggests that the presence of significant systematic errors in the reported observations is highly plausible.

When comparing the predicted ion CDDFs and observations, it is important to keep in mind that the column densities are not measured using the same method. Observationally, column densities are measured by decomposing the (normalized) spectra into Voigt profile components which are characterized by their redshifts, width and column density. It is worth noting, however, that the column densities measured using Voigt profile fit is not necessarily equal to the true column density since the observed absorption lines are not expected to be perfect Voigt profiles (e.g., because part of the broadening is due to bulk motions) and because of noise, continuum fitting errors and contamination. By using the projection technique to obtain the simulated column densities, on the other hand, we effectively group together nearby absorption systems that are along the projection direction and measure their true combined column densities. The velocity width corresponding to the typical slice width we use is $\Delta v \sim 200~ \kms$ (i.e., at $z \ge 2$ and $\Delta v \sim 400~ \kms$ at lower redshifts). This velocity width is comparable to the velocity interval over which individual metal absorption \emph{components} are found to be strongly clustered and are grouped into \emph{systems} \citep[e.g.,][]{Boksenberg03}. Therefore, the predicted CDDFs shown in figures \ref{fig:cddf-allz} and \ref{fig:cddf-allz2} should be compared with observed CDDFs that considered absorption systems by grouping absorption components using similar velocity widths. Different observational studies use different (and often unspecified) criteria for grouping absorption systems before calculating the CDDFs. This makes it difficult to account for different grouping schemes. However, we note that using a velocity width different by a factor of a few compared to our typical value of $\Delta v \sim 200~ \kms$ does not change the CDDFs significantly (the result only starts to change significantly for $\Delta v \lesssim 50~ \kms$). Using absorption \emph{components} instead of \emph{systems} in CDDF calculations would decrease the high-$N_{\rm{ion}}$ end of the CDDF at the expense of boosting the low-$N_{\rm{ion}}$ end of the CDDF \cite[e.g.][]{Boksenberg03,Tripp08,Oppenheimer12}. We note that line saturation in the observed spectra makes the CDDFs quite uncertain at the high-$N_{\rm{ion}}$ end. Assuming a signal-to-noise of 50 and a $b = 10$ km/s the Voigt profiles already saturate at $N_{ion}/\cmsq = 10^{13.5},~10^{13.9},~10^{14.1},~10^{14.3}$ and $10^{14.5}$ for $\SiIV$, $\CIV$, $\NV$, $\OVI$ and $\NeVIII$, respectively. Also, the relatively wide redshift ranges used for measuring the observed ion CDDFs complicate comparison between different observations and between the observations and simulations. Given the aforementioned large uncertainties in measuring the observed column densities of ions, the distinction between absorption \emph{components} and \emph{systems} is not critical for our purpose and is not expected to change our main results.

Fig. \ref{fig:cddf-allz} shows broad agreement between the predicted CDDFs and the observed data. This extends the good agreement which was shown in S15 between the EAGLE reference simulation and observed low-redshift $\CIV$ and $\OVI$ CDDFs, to higher redshifts and other ions like $\SiIV$. 

Despite the overall good agreement between EAGLE and the observational data, we note that the high column density end of the predicted CDDFs slightly underproduces the observed values. This discrepancy is largest for the highest column density $\OVI$ measurements. As mentioned earlier, different systematics such as line blending introduce large uncertainties in identifying high column density absorbers and in measuring their column densities. Therefore, the significance of the difference between our predictions and observations at high $\OVI$ column densities is not clear. We note, however, that increasing the resolution (see Appendix \ref{ap:res}) and/or using a stronger UVB at wavelengths relevant for the photoionization of $\OVI$ (see Appendix \ref{ap:UVB}) would increase the modelled $\OVI$ CDDF at $N_{\rm{OVI}} \gtrsim 10^{13} \cmsq$.

Moreover, the comparison between the predicted and observed CDDFs suggests that a better agreement can be achieved by changing the normalization of the predicted CDDFs. For instance, shifting the predicted CDDFs towards higher column densities (i.e., to the left in figures \ref{fig:cddf-allz} and \ref{fig:cddf-allz2}) by a factor of $\sim 2$ can improve their agreement with the observed data, particularly for $\NV$. Noting that even for a fixed initial mass function (IMF) the stellar nucleosynthetic yields are uncertain by a factor of $\sim 2$ \cite[e.g.,][]{Wiersma09b}, such a shift in our predictions is indeed plausible. Such a shift would particularly improve the agreement between our predictions and the measurements reported by \citet{Danforth14}. However, those observations are systematically higher than all other observational measurements. Indeed, the other observations are in better agreement with our results and do not require a significant shift in the column densities or the normalization of the CDDFs.

As the top panels of Fig. \ref{fig:cddf-allz} suggest, the discrepancy between our predictions and the observed $\SiIV$ CDDF seems to increase with decreasing column density, particularly at higher redshifts. As mentioned above, the predictions are based on absorption systems by grouping \emph{components} that are within $\Delta v \sim 200~ \kms$ (at $z > 2$ and  $\Delta v \sim 400~ \kms$ at lower redshifts) from each other into \emph{systems} of absorbers. The observed $\SiIV$ CDDFs, on the other hand, are often based on counting absorption \emph{components} \cite[e.g.,][]{Scannapieco06}, or use narrower velocity windows than what we use here for identifying \emph{systems} of absorbers \cite[e.g.,][]{Danforth14}. Noting that the low column density end of the observed CDDFs can be reduced by up to a factor $\approx 5$ by grouping absorption  \emph{components} into  \emph{systems} of absorbers without significantly changing the high end of the CDDF \citep{Boksenberg03}, the discrepancy between the predicted and observed CDDFs in the top panels of Fig. \ref{fig:cddf-allz} is reasonable and expected.

As the panels in the top and middle rows of Fig. \ref{fig:cddf-allz} show, the CDDFs of $\SiIV$ and $\CIV$ evolve very weakly at column densities $N_{\rm{ion}} \ll 10^{14} \cmsq$, particularly for $z \le 1$, both in the simulation and in the observations. This trend is similar to the (lack of) evolution of the $\HI$ CDDF for Lyman Limit systems (LLS) with $10^{17} \lesssim \NHI \lesssim 10^{20}~ \cmsq$ \citep{Rahmati13a}. While at lower column densities, the CDDFs of other ions increase monotonically as the total abundance of heavy elements increases with time, at relatively high column densities (i.e., above the knee of the CDDF), the evolution of the normalization of CDDFs closely resembles that of the star formation rate density of the Universe, peaking at $z \sim 1-2$. This is very similar to the evolution of the $\HI$ CDDF for Damped Lyman-$\alpha$ (DLA) systems with $\NHI \gtrsim 10^{20}\cmsq$ \citep{Rahmati13a}. 

The aforementioned trends can be understood if we note that different ions and different column densities represent different physical conditions. As we will show in $\S$\ref{sec:physicalproperties}, at a fixed ion column density, $\SiIV$ and $\CIV$ absorbers trace densities closer to those corresponding to strong $\HI$ absorbers (i.e., LLSs and DLAs; see \citealp{Rahmati13a}), which are higher than those traced by $\OVI$ and $\NeVIII$ absorbers. Moreover, the typical density of the absorbing gas increases with the column density. As a result, higher ion column densities are associated with regions that are typically denser and therefore closer to where star formation takes place. Consequently, the typical physical properties of higher column density systems (e.g., their abundances and metallicities) are closer to those found in the ISM and hence have stronger correlations with the average star formation activity of the Universe. This explains why the ionic CDDFs at high column densities follow qualitatively the cosmic star-formation history.

\begin{figure*}
\centerline{\hbox{{\includegraphics[width=0.47\textwidth]
              {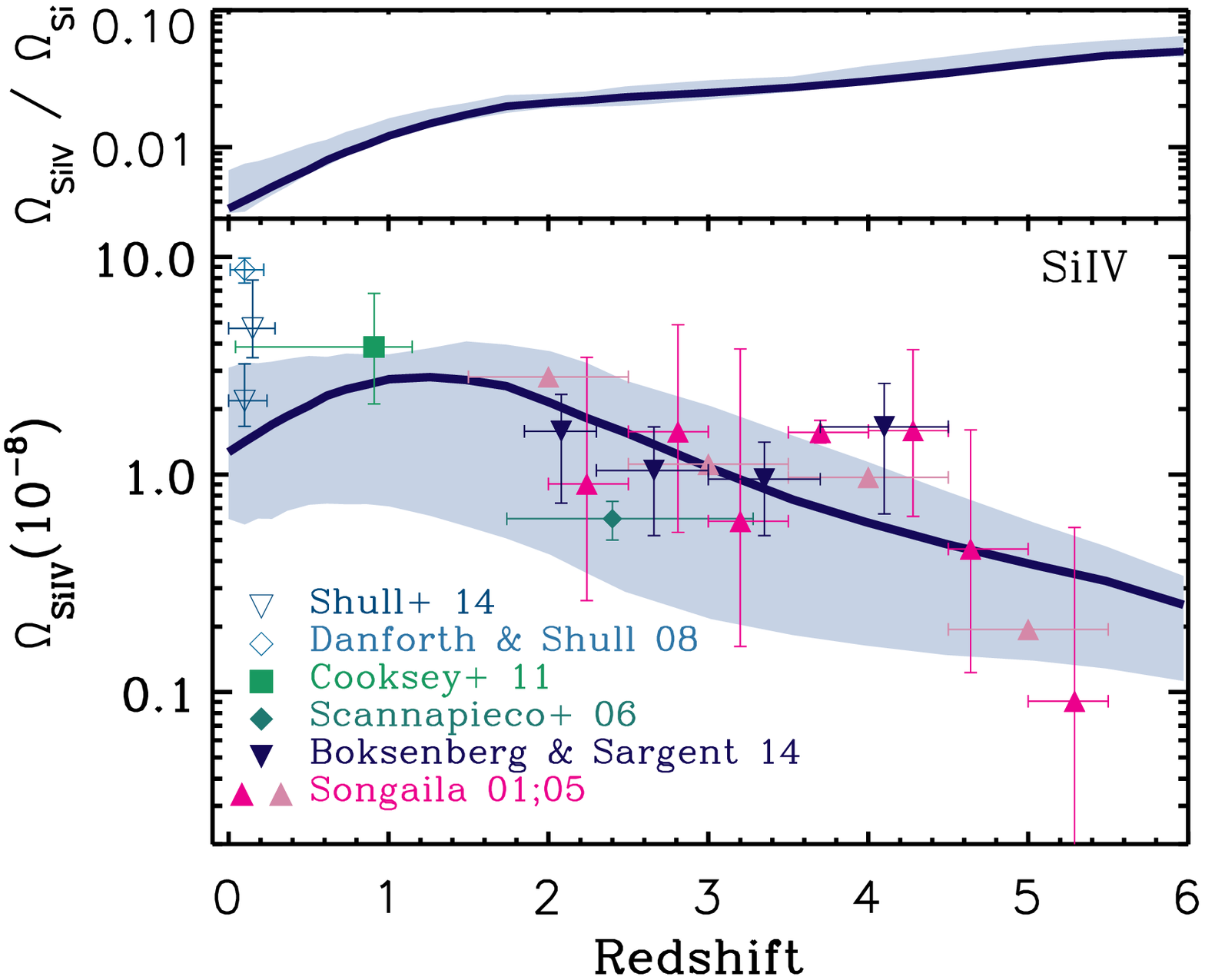}}}
             \hbox{{\includegraphics[width=0.47\textwidth]	
             {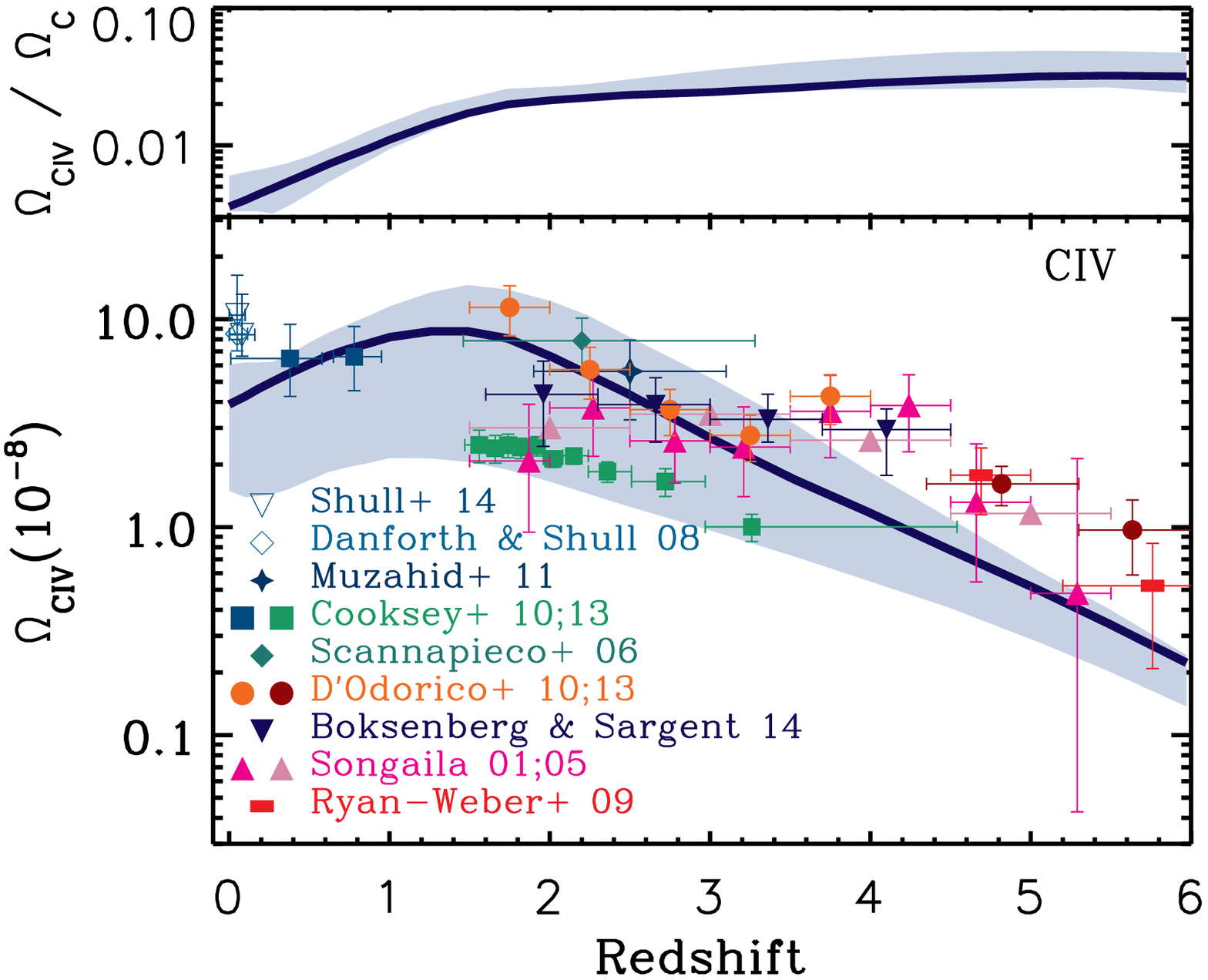}}}}
\centerline{\hbox{{\includegraphics[width=0.47\textwidth]
              {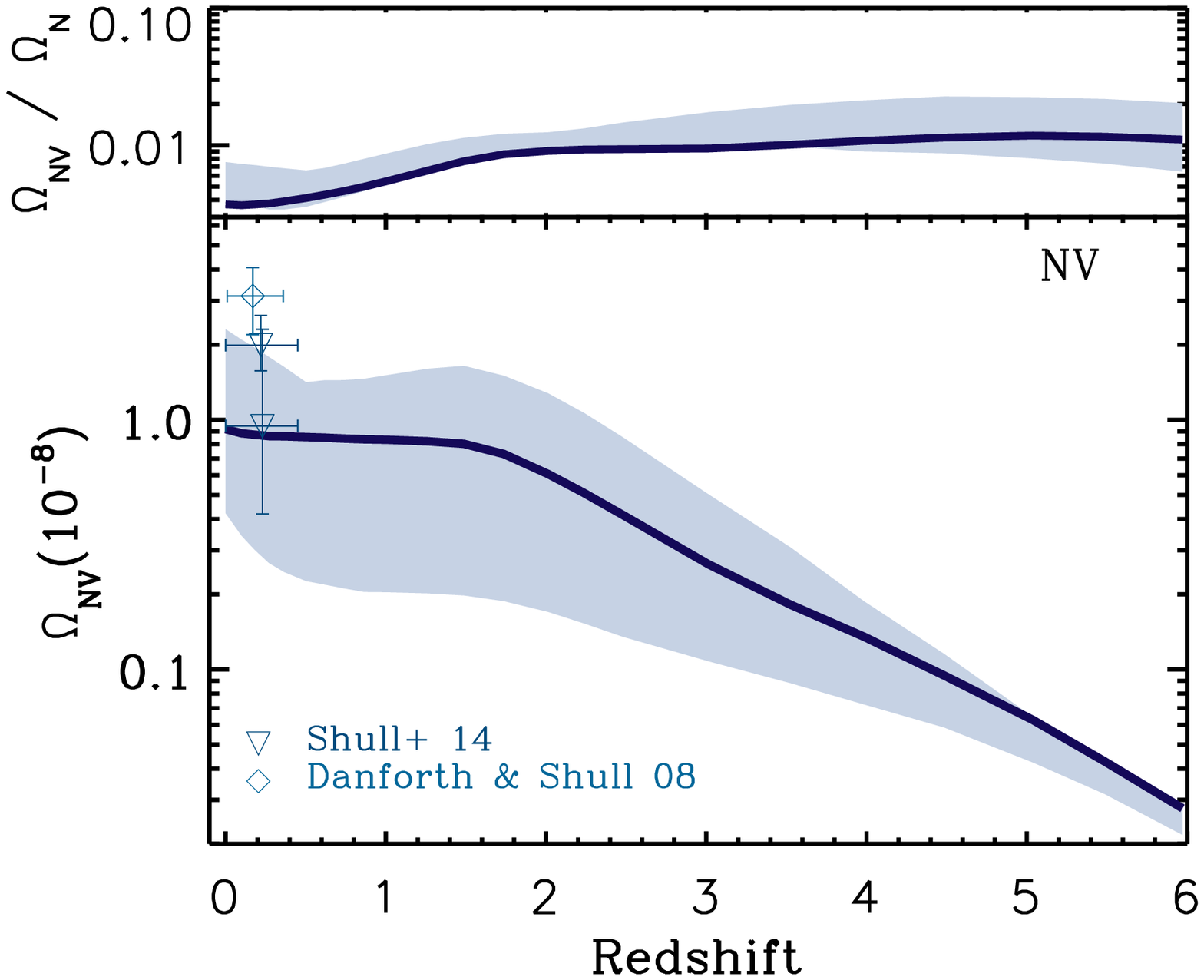}}}
             \hbox{{\includegraphics[width=0.47\textwidth]	
              {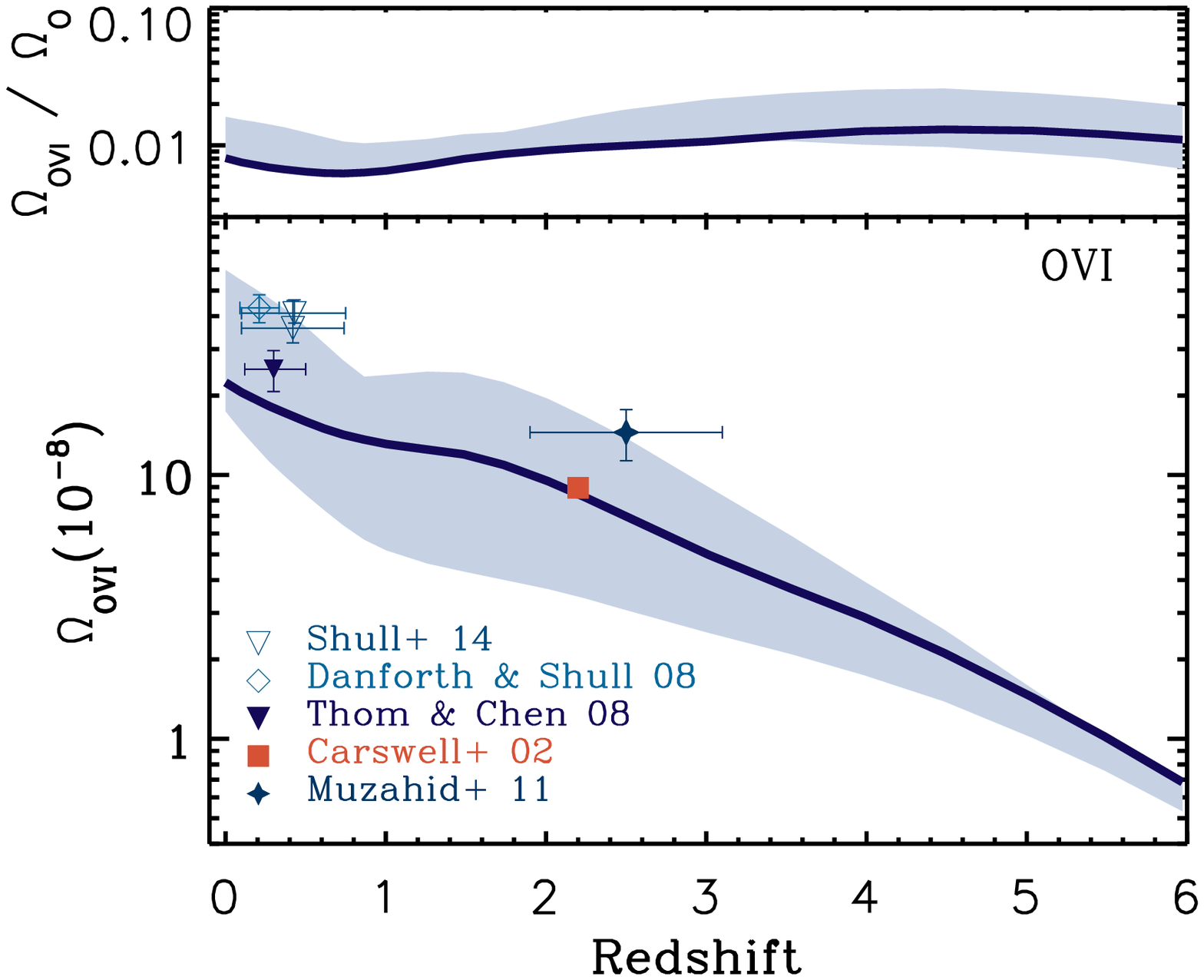}}}}
\centerline{\hbox{{\includegraphics[width=0.47\textwidth]
	      {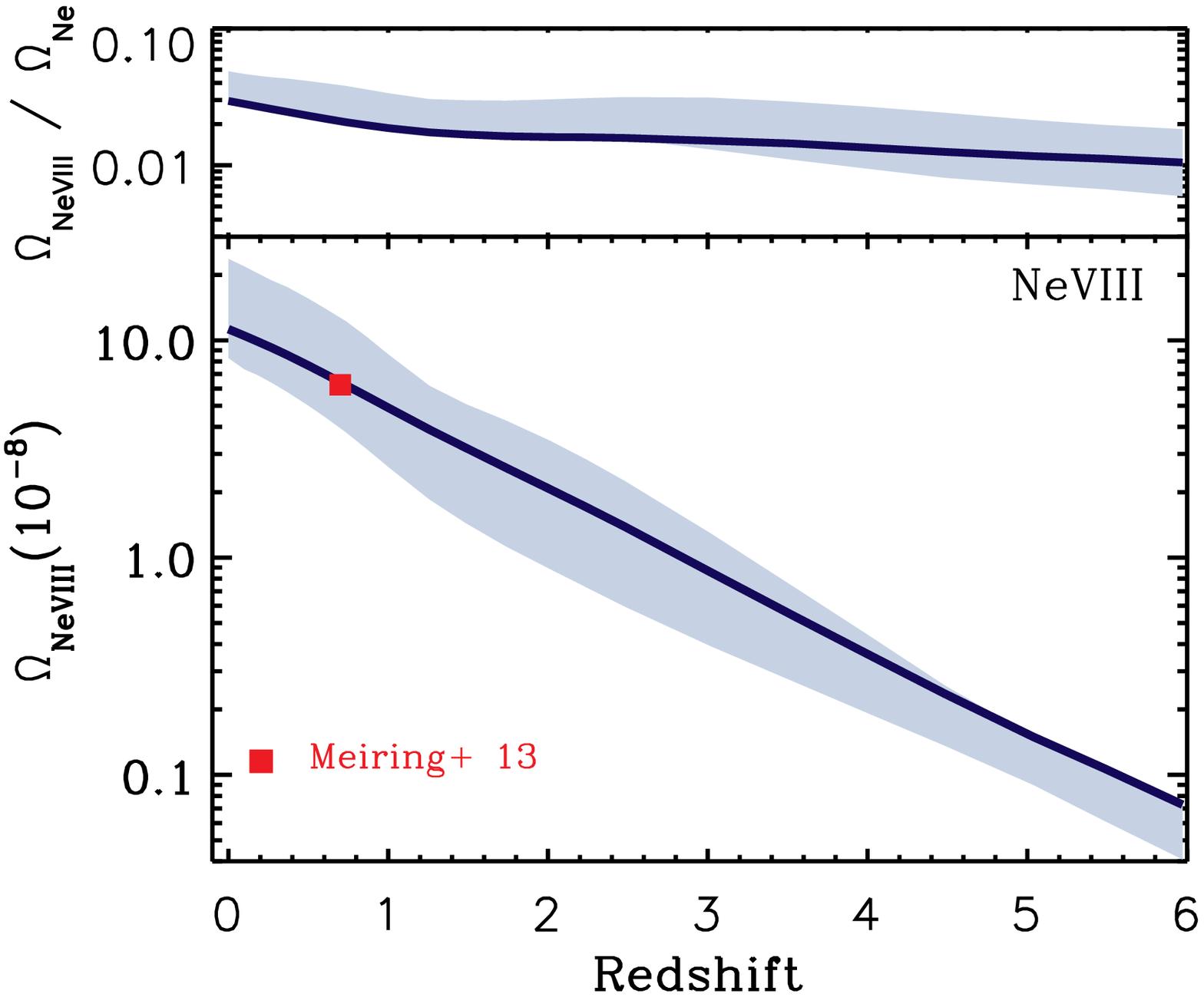}}}}

\caption{Cosmic density of high ionization metals in the EAGLE simulations. Top-left, top-right, middle-left, middle-right and bottom panels show $\SiIV$, $\CIV$, $\NV$, $\OVI$ \& $\NeVIII$, respectively. Solid curves show the predictions from the \emph{Ref-L100N1504} simulation while the shaded areas show the range of predictions from simulations listed in Table \ref{tbl:sims} with different box sizes, resolutions and feedback models. Colored symbols in each panel show observational measurements taken from references indicated in the legends. The top section of each panel shows the ratio between the cosmic density of each ion and the cosmic density of its parent element in gas (e.g., $\Omega_{\rm{C_{IV}}} / \Omega_{\rm{C}}$). The cosmic density of high ionization metals in the EAGLE simulation agrees reasonably well with the observations, although the simulations may under predict the data at $z \simeq 0$ for all ions, at $z \simeq 4$ for $\SiIV$ and at $z \gtrsim 4$ for $\CIV$.}
\label{fig:ion-dens-obs}
\end{figure*}
\begin{figure*}
\centerline{\hbox{{\includegraphics[width=0.5\textwidth]
             {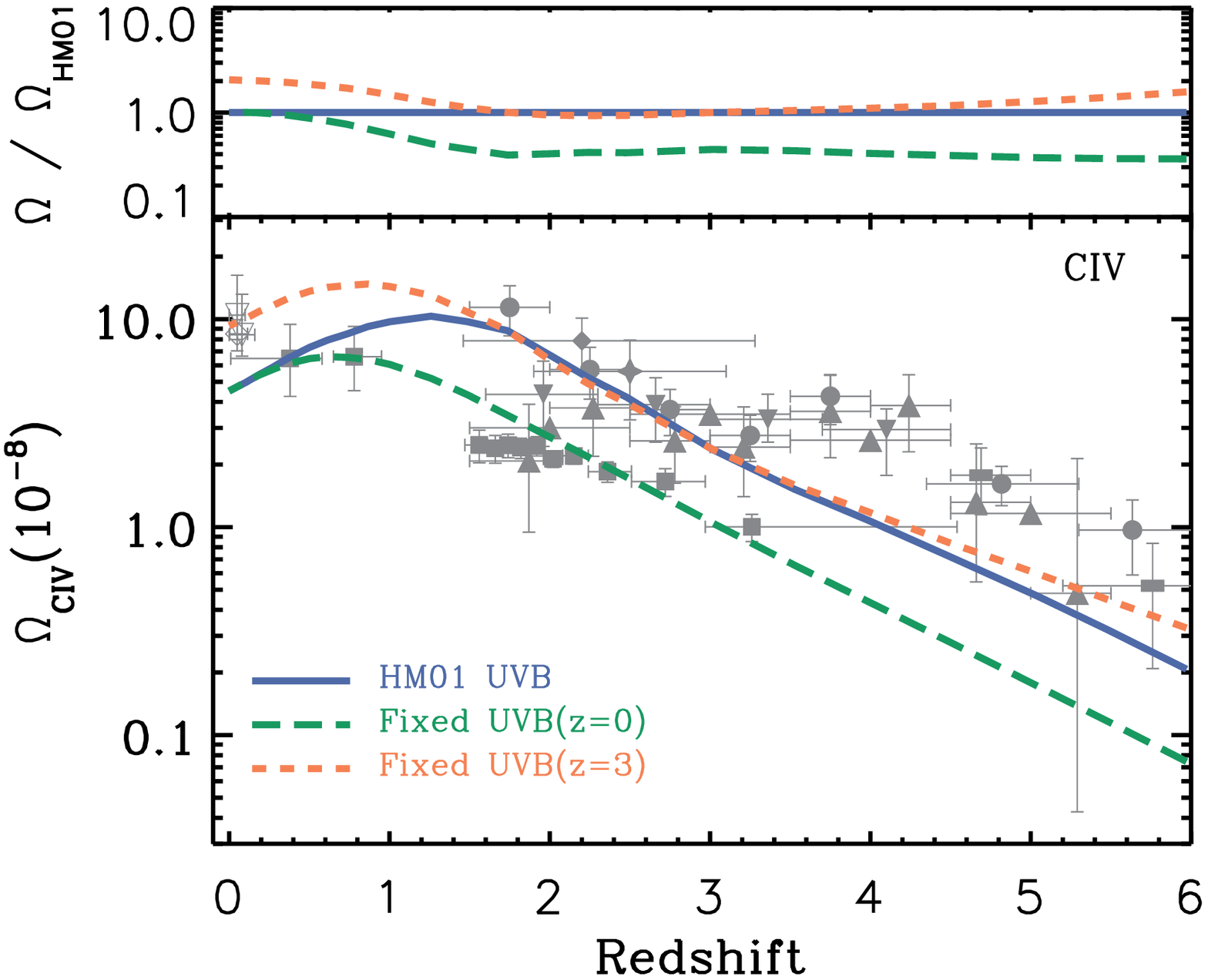}}}
             \hbox{{\includegraphics[width=0.5\textwidth]
             {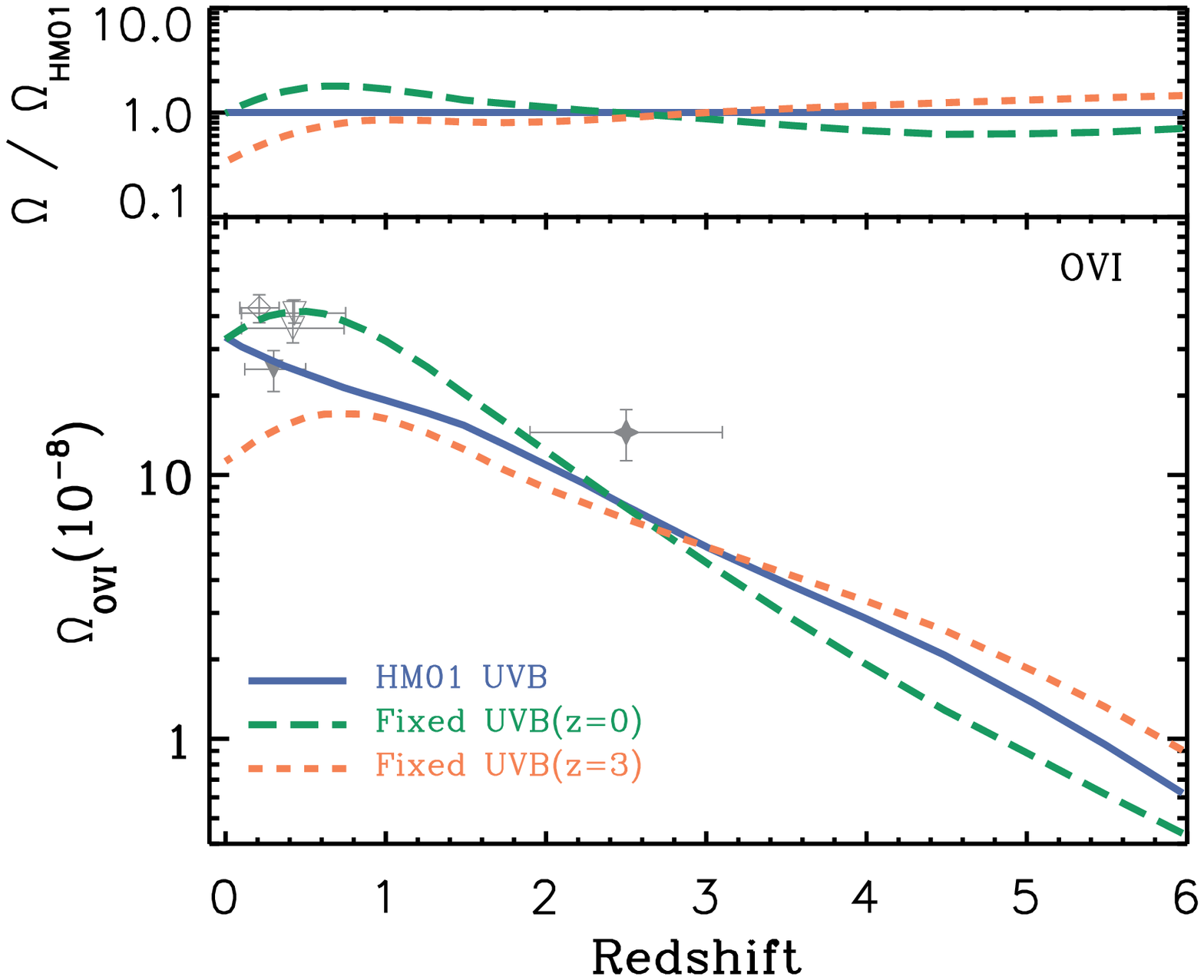}}}}
\caption{Evolution of the cosmic density of $\CIV$ (left) and $\OVI$ (right) in the EAGLE \emph{Ref-L025N0376} simulation for different photoionization scenarios. Blue (solid) shows the fiducial evolving UVB model (i.e., HM01).  Green (long-dashed) and orange (dashed) curves show the results using constant UVB models chosen identical the HM01 UVB at $ z= 0$ and $z = 3$, respectively. The grey symbols show a compilation of observational measurements, identical to the one shown in Fig. \ref{fig:ion-dens-obs}. The top section of each panel shows the ratio between the ion density and that of the evolving HM01 UVB model. }
\label{fig:ion-dens-uvbtest}
\end{figure*}

To compare our results with other simulations which have been used to study the cosmic distribution of some of the metals we studied here, Fig. \ref{fig:CDDF-sims} compares our predictions (solid curves) for the CDDFs of $\CIV$ (top) and $\OVI$ (bottom) at $z \approx 2$ (right) and $z \approx 0$ (left) with those reported in \citet{Oppenheimer12} for their preferred \emph{vzw} model, and in \citet{Bird15} for their reference model (shown using dashed and dotted curves, respectively). The shaded area around the dashed curve shows the 1-$\sigma$ poisson error associated with the reported CDDFs in \citet{Oppenheimer12}. The $z \approx 2$ results from \citet{Oppenheimer12} represent the mean distribution of the $z = 1.7$ and $z = 1.9$ snapshots while our results and the Illustris $\CIV$ CDDF presented in \citet{Bird15} are for $z = 2$. Noting the rather modest evolution in the $\CIV$ CDDF at $z \approx 2$ for EAGLE (see Fig. \ref{fig:cddf-allz}), matching the exact redshifts is not expected to affect the comparison. At low redshifts, however, the evolution is stronger. Therefore, for $z \approx 0$, and in analogy to \citet{Oppenheimer12} who used the mean distributions of two snapshots at $z = 0$ and $z = 0.5$, we also show the mean CDDF of the EAGLE at $z = 0$ and $z = 0.6$. Symbols show compilations of observational measurements relevant to each redshift. 

As shown in Fig. \ref{fig:CDDF-sims}, despite significant differences in the methodologies (e.g., the differences in the hydrodynamics solver, the subgrid wind modelling and the column density calculation) and resolutions (e.g., our reference simulation has a volume $\approx 10$ times larger and a mass resolution more than 20 times better than those used in \citealp{Oppenheimer12}), the predicted CDDFs are in reasonable agreement. Although the $\CIV$ CDDF from \citet{Oppenheimer12} agrees better with the measurements presented in \citet{Dodorico13} at $z \sim 2$, our predictions show a better agreement with observational data for $\CIV$ at low redshifts and for $\OVI$ at both low and high redshifts. The differences in the shapes and normalizations of the CDDFs in the different simulations shown in Fig. \ref{fig:CDDF-sims} result in differences in the cosmic ion densities. For instance, as we show in the next section, we predict a cosmic $\CIV$ density which peaks at $1 < z < 2$ and decreases by more than a factor of $\approx 2$ by $z \sim 0$ while the $\CIV$ cosmic density reported in \citet{Oppenheimer12} remains nearly constant below $z \approx 2$.
\begin{figure*}
\centerline{\hbox{{\includegraphics[width=0.9\textwidth]
              {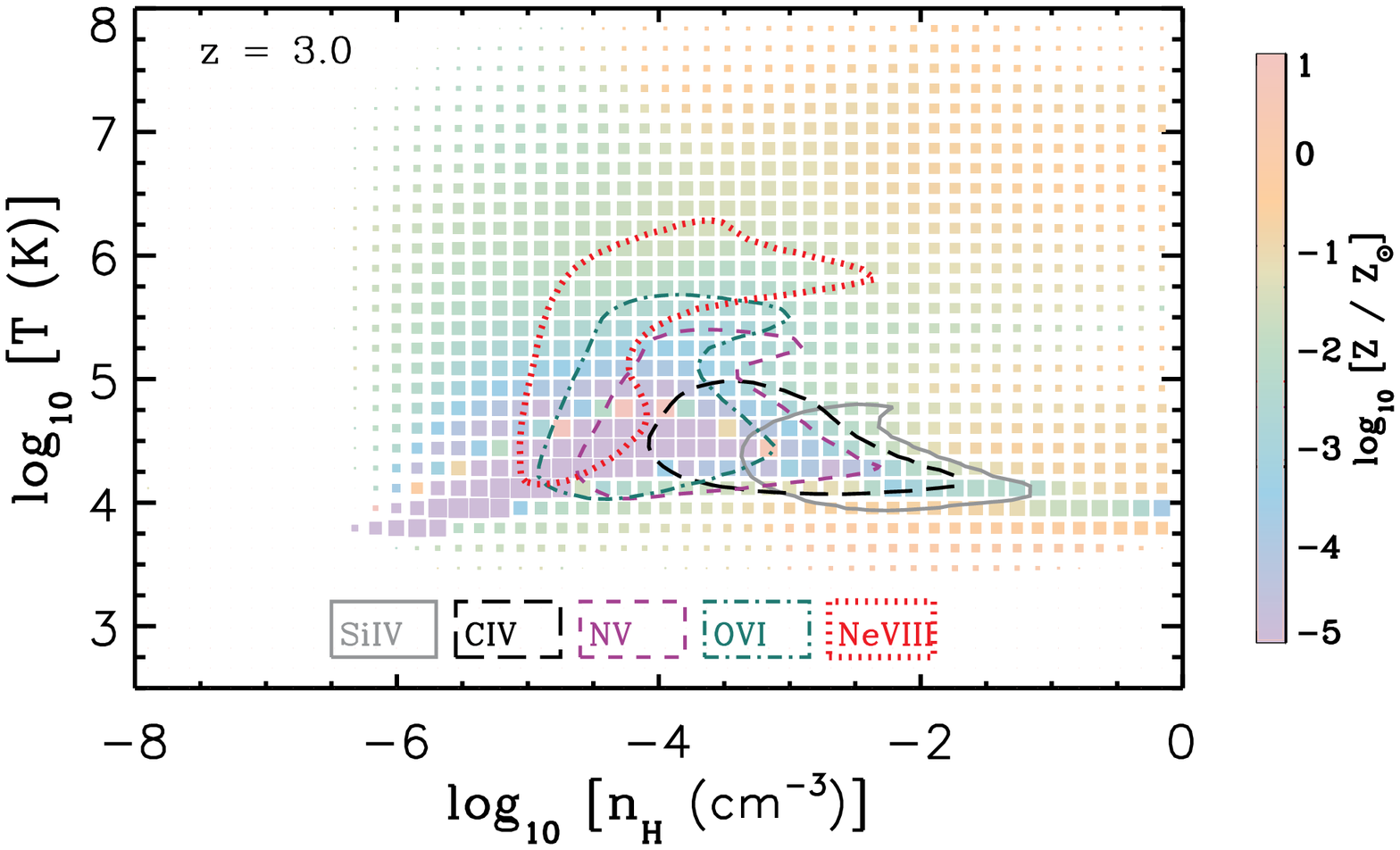}}}}
\centerline{\hbox{{\includegraphics[width=0.9\textwidth]
              {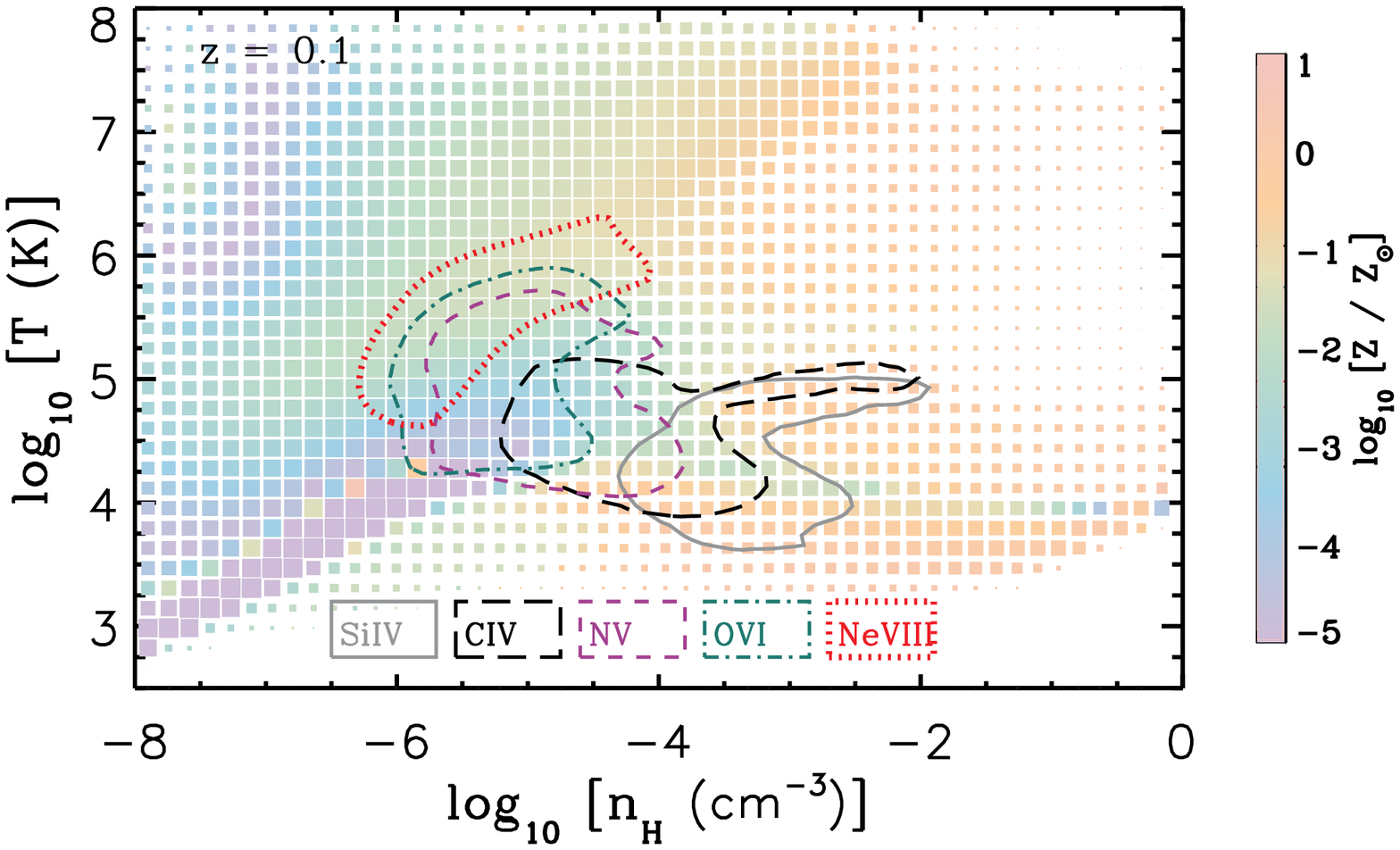}}}}
\caption{Temperature-density distribution of gas in the EAGLE \emph{Ref-L100N1504} simulation at $z = 3$ (top) and $z = 0.1$ (bottom). The size of each cell is proportional to the logarithm of the gas mass enclosed in it and its color shows its median metallicity. The areas enclosed by contours with different colors and line styles show the range of temperature-densities that contain $80\%$ of the mass of the corresponding ions. Gray solid, black long-dashed, purple dashed, green dot-dashed and red dotted lines show $\SiIV$, $\CIV$, $\NV$, $\OVI$ and $\NeVIII$, respectively. The typical temperature (density) of absorbers increases (decreases) with their ionization energy and time.}
\label{fig:4dplot-all-ions}
\end{figure*}
\begin{figure*}
\centerline{\hbox{{\includegraphics[width=0.45\textwidth]
              {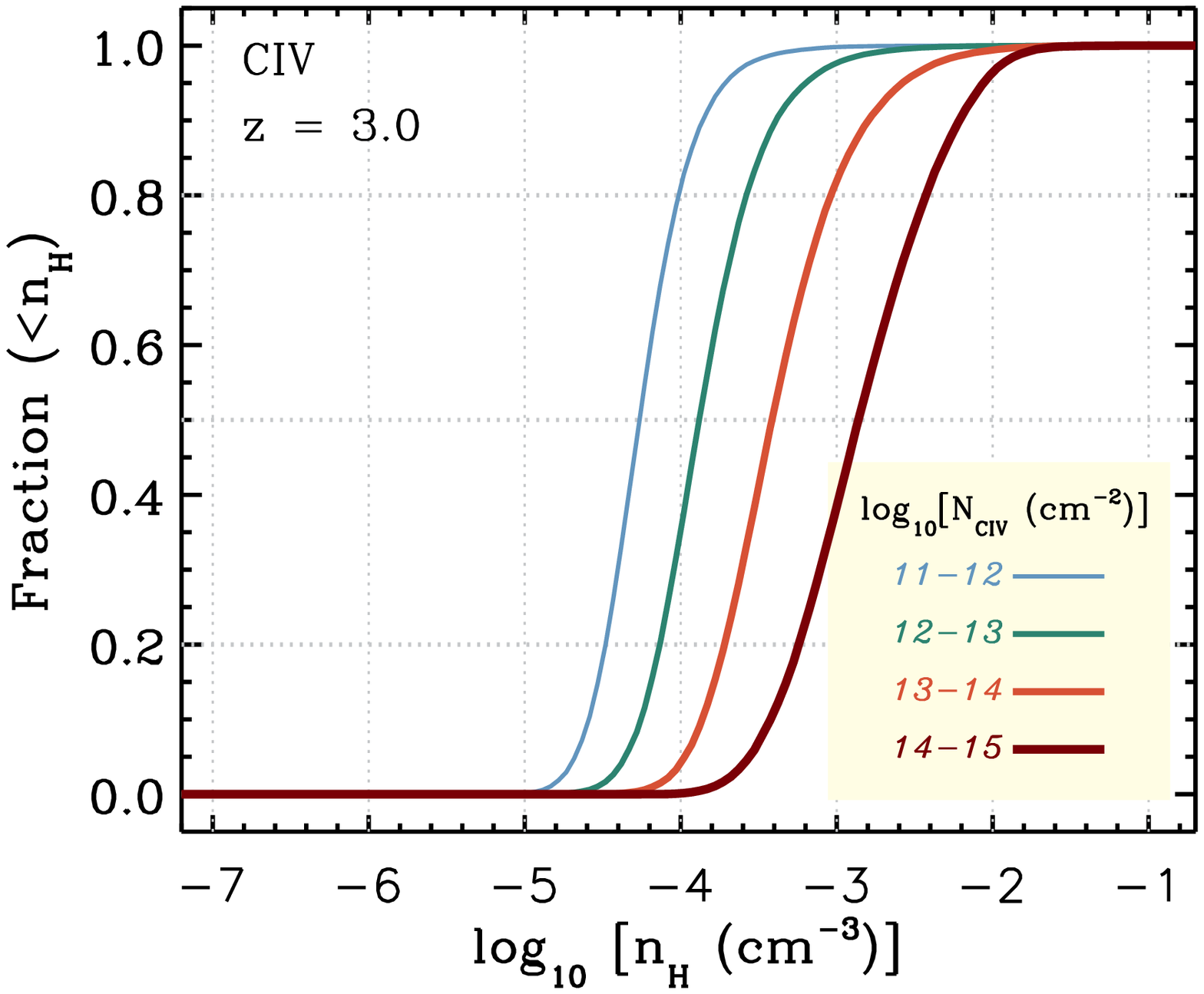}}}
             \hbox{{\includegraphics[width=0.45\textwidth]	
              {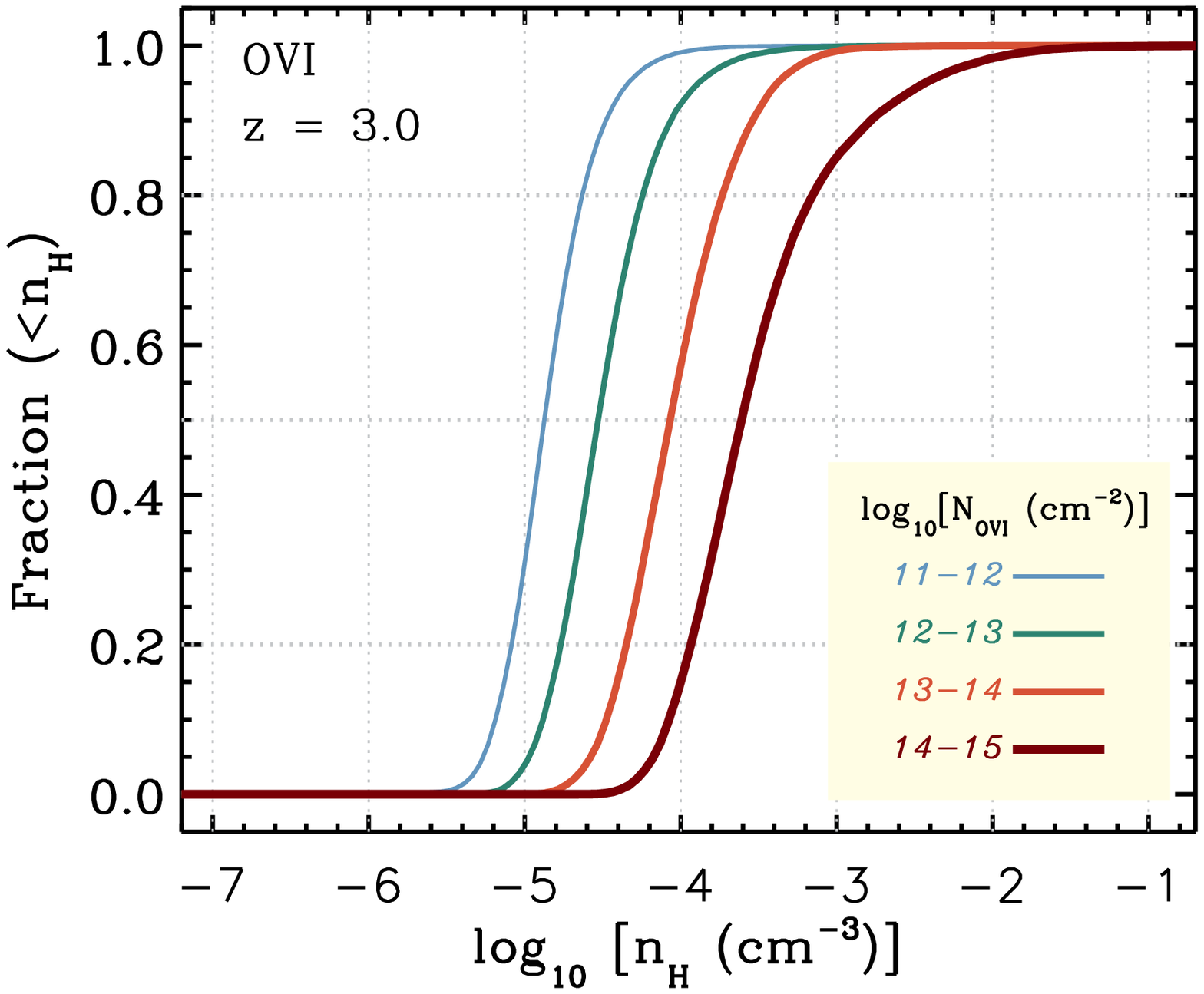}}}}
\centerline{\hbox{{\includegraphics[width=0.45\textwidth]
              {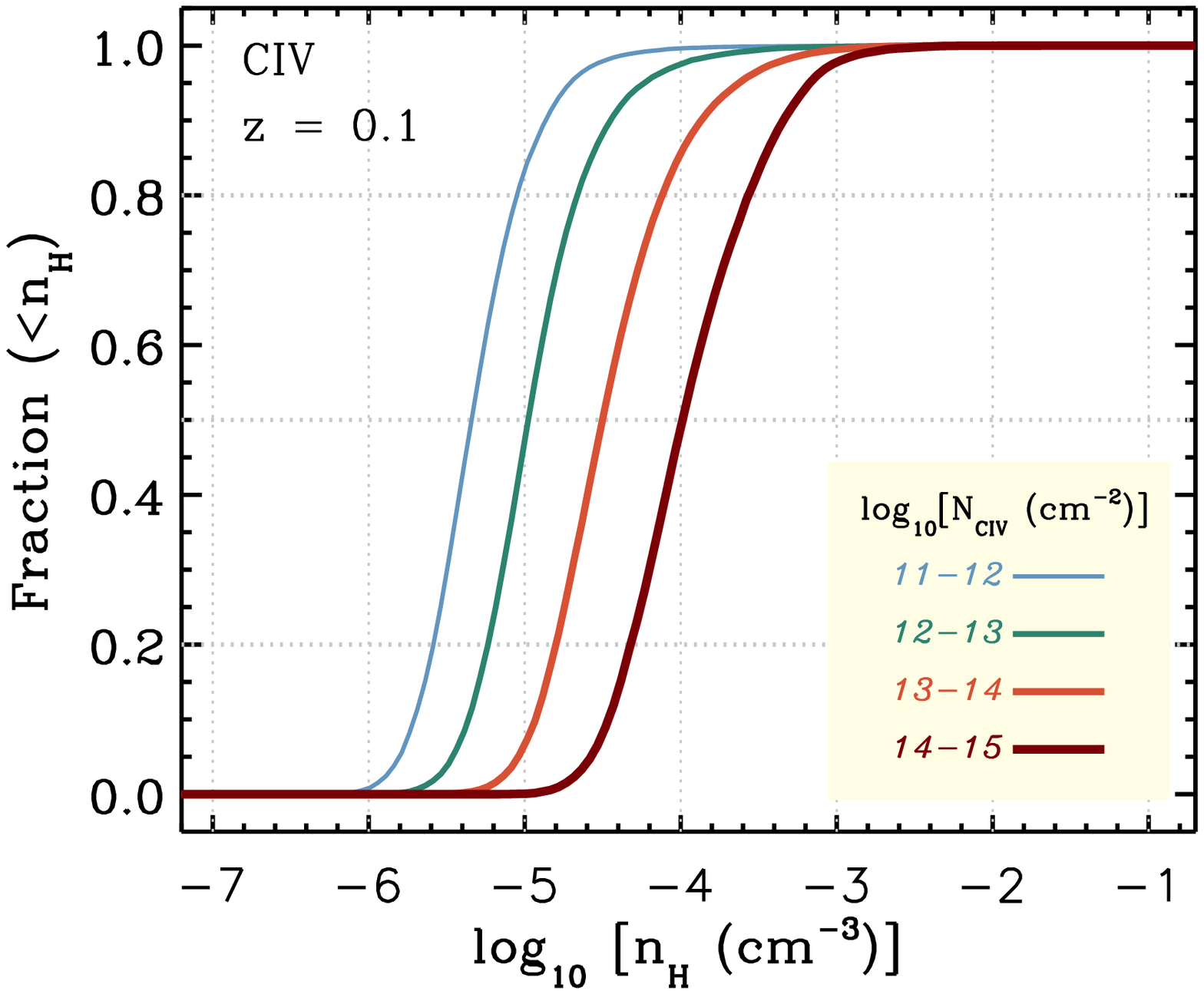}}}
             \hbox{{\includegraphics[width=0.45\textwidth]	
             {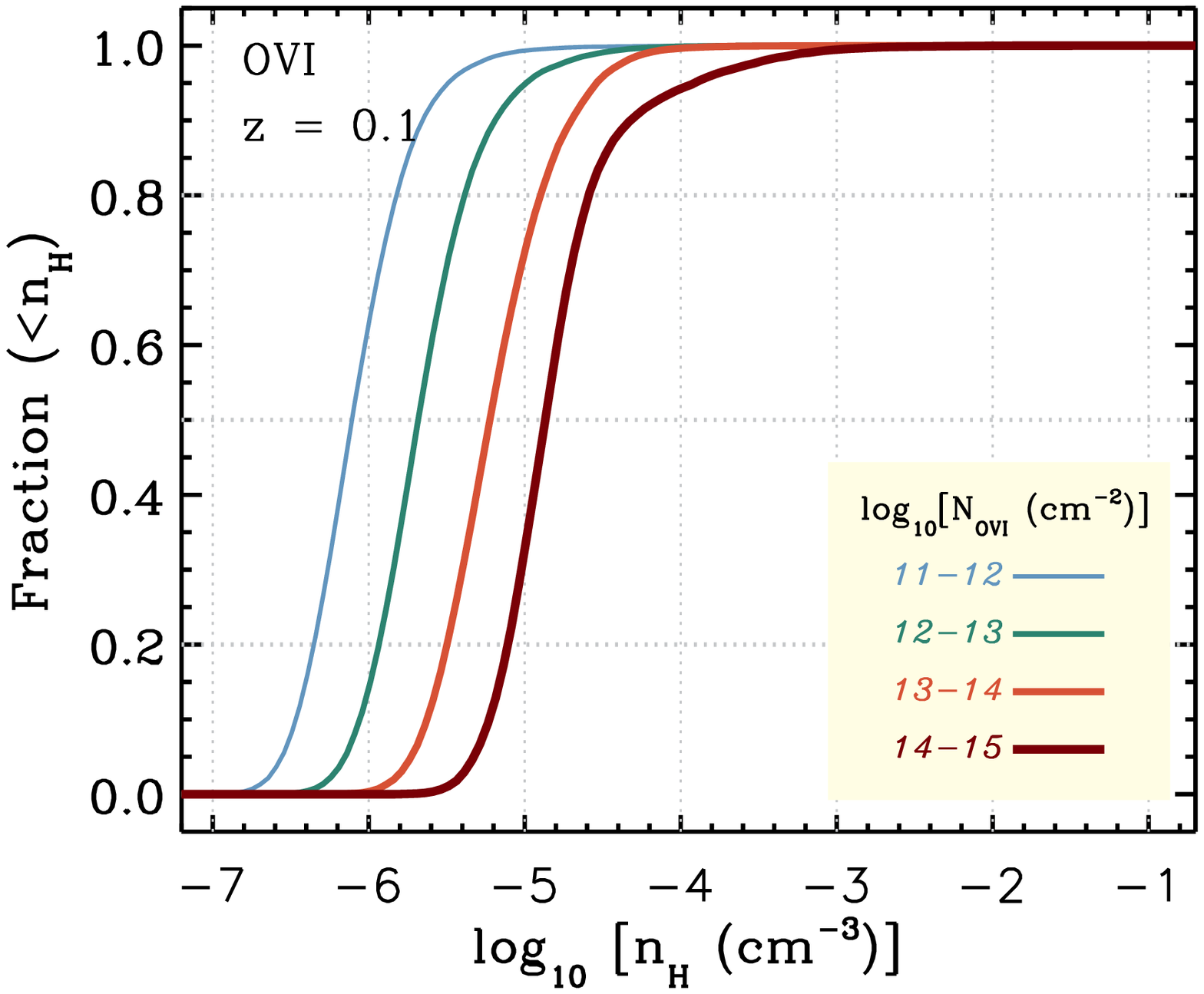}}}}
\caption{Cumulative fractions of $\CIV$ (left) and $\OVI$ (right) absorbers with ion-weighted gas densities lower than $\nH$ as a function of $\nH$ for different column densities at $z = 3$ (top) and $z = 0.1$ (bottom), in the EAGLE \emph{Ref-L100N1504} simulation. Curves from thin to thick show column density bins ranging from $10^{11}  < N_{\rm{ion}} \leq 10^{12} \cmsq$ to $10^{14}  < N_{\rm{ion}} \leq 10^{15} \cmsq$. The typical gas density of absorbers increases with their column density and redshift.}
\label{fig:cum-dist-density-c4o6}
\end{figure*}
\begin{figure*}
\centerline{\hbox{{\includegraphics[width=0.45\textwidth]
              {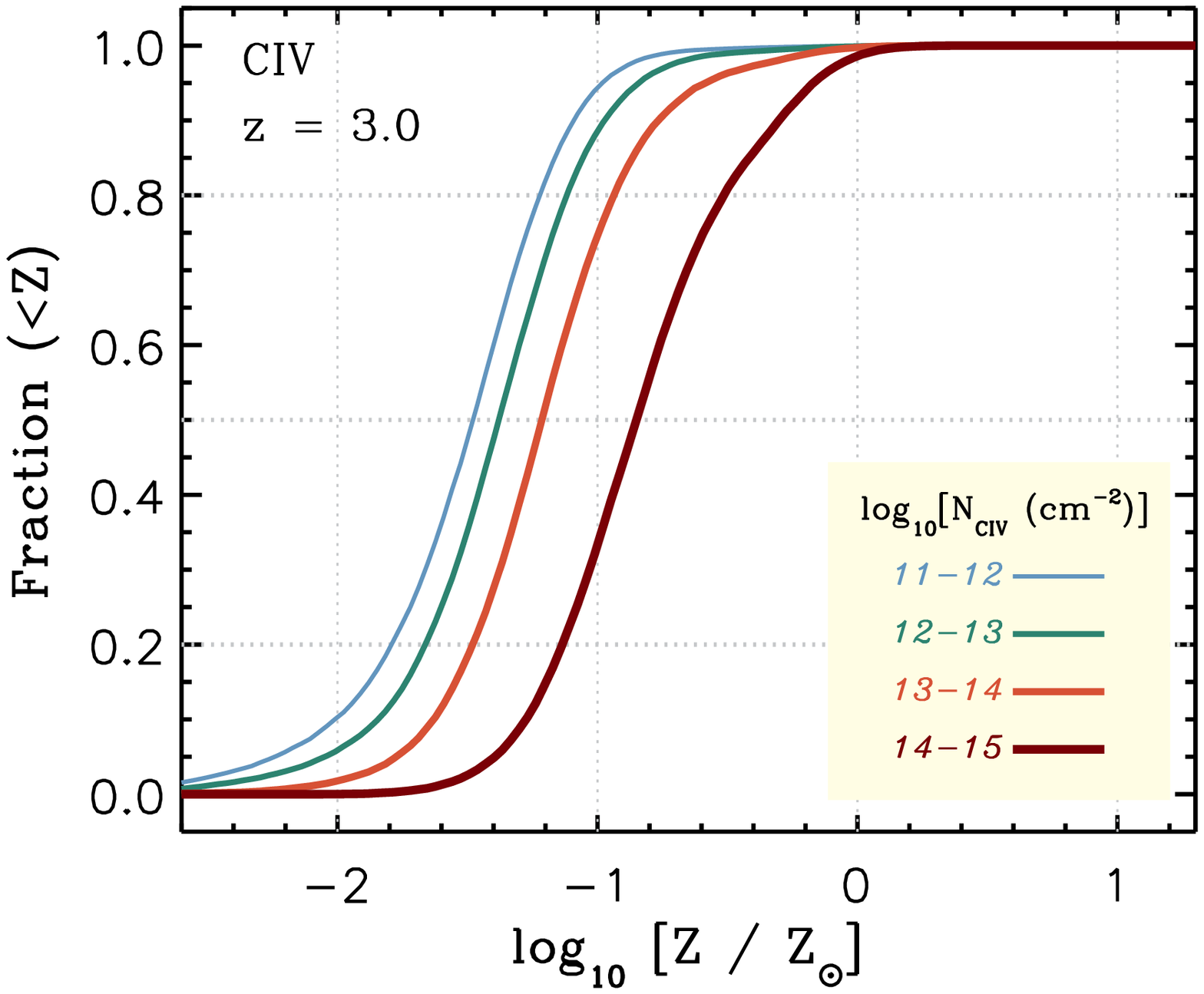}}}
             \hbox{{\includegraphics[width=0.45\textwidth]	
             {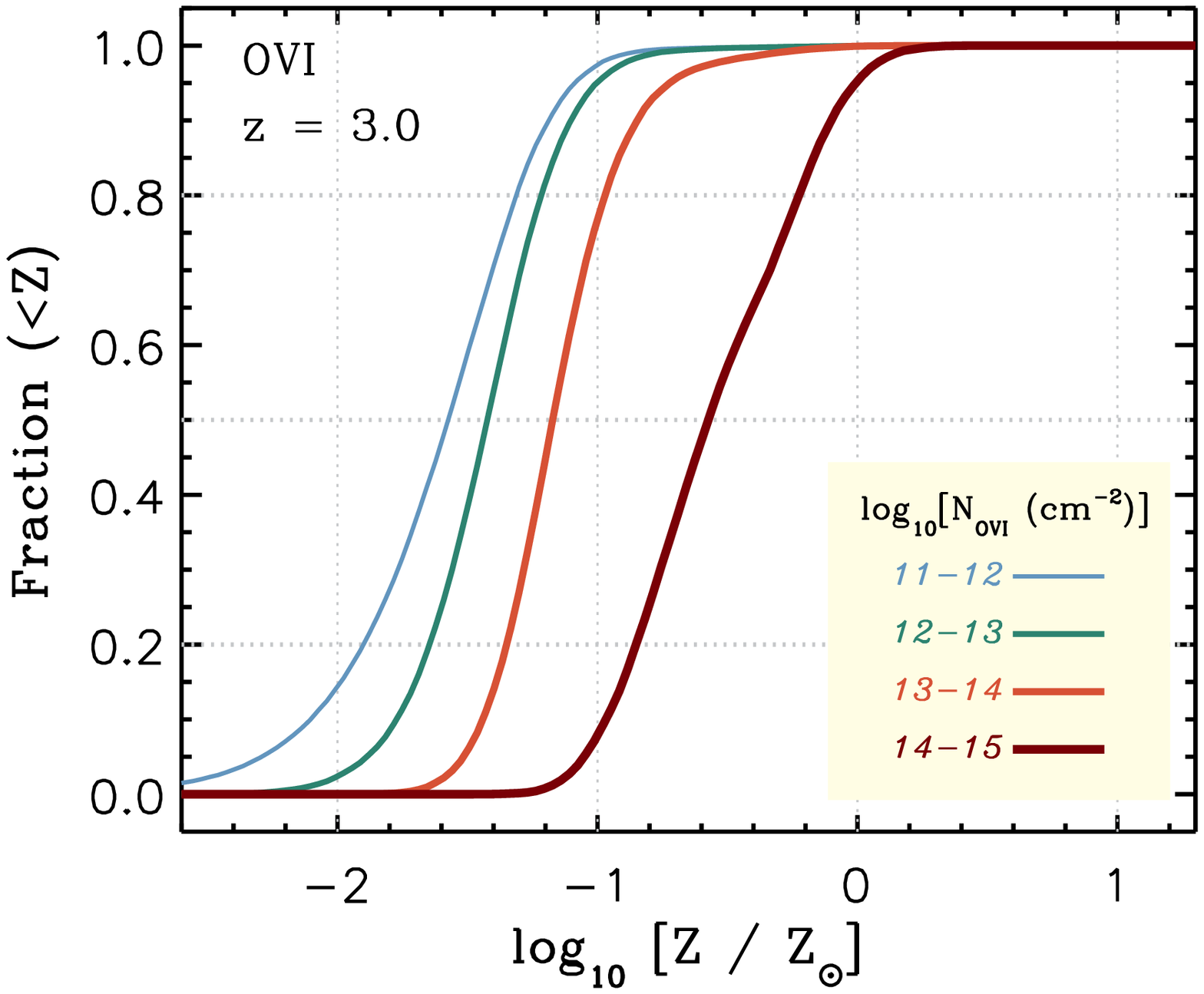}}}}
\centerline{\hbox{{\includegraphics[width=0.45\textwidth]
              {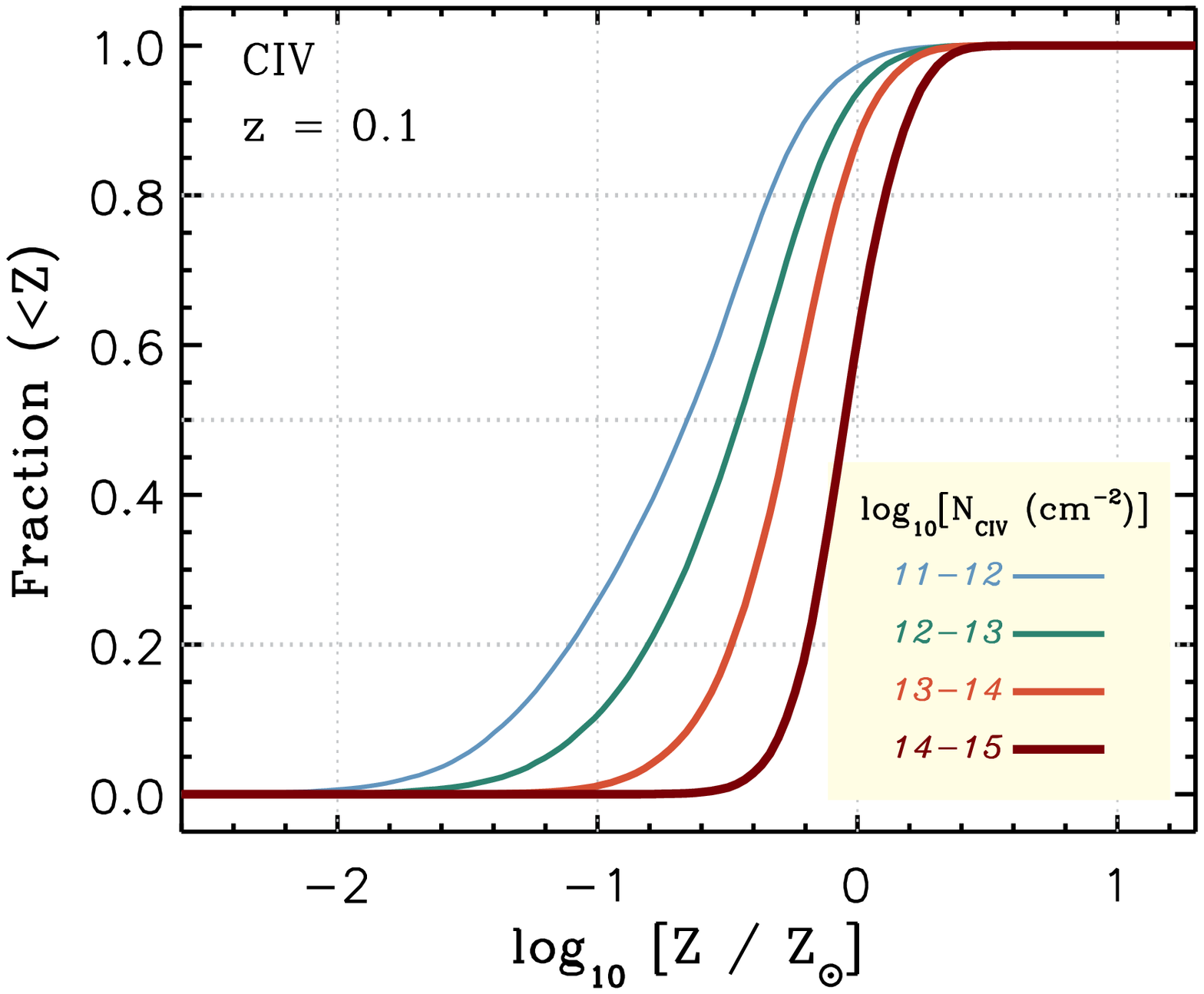}}}
             \hbox{{\includegraphics[width=0.45\textwidth]	
             {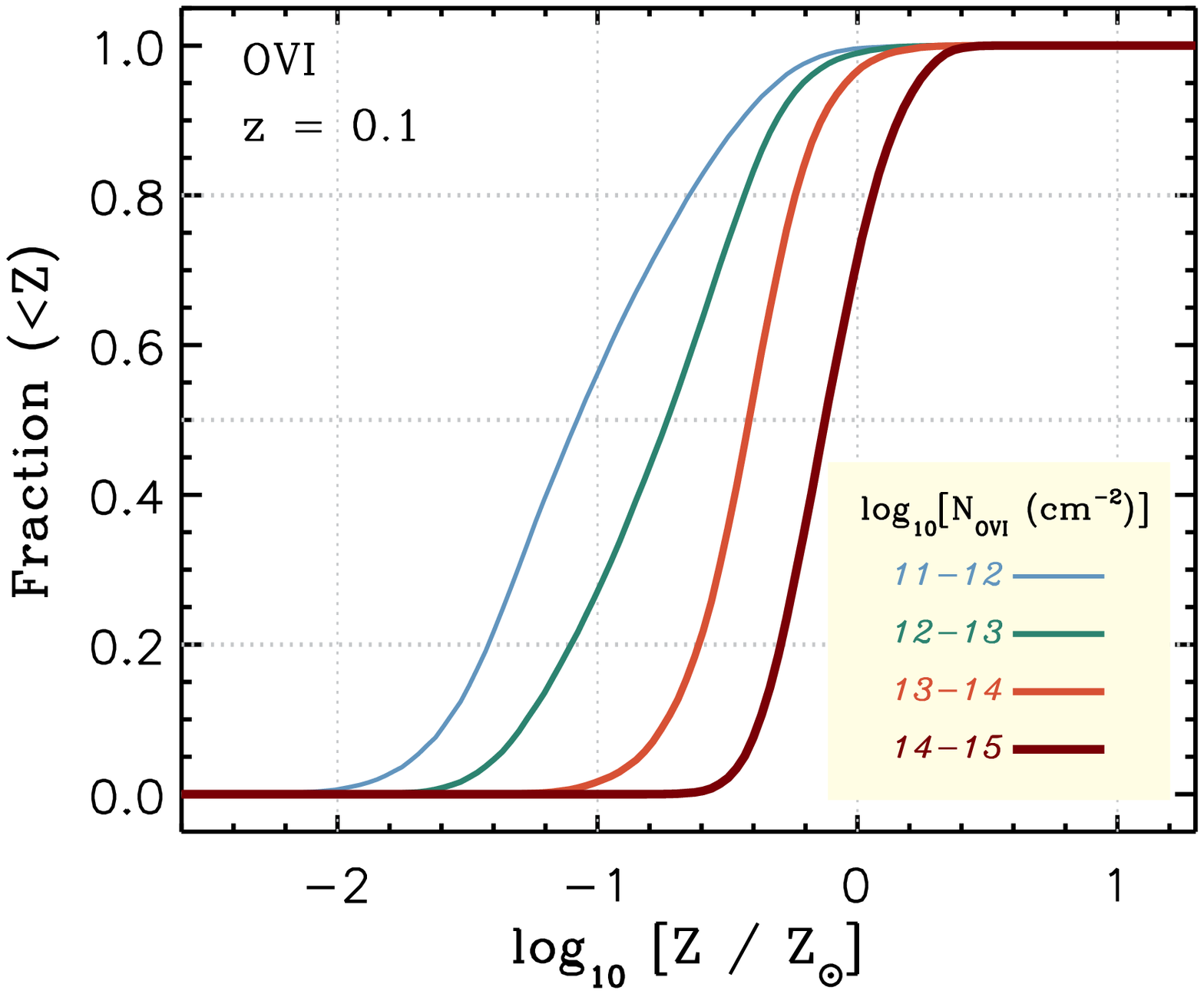}}}}
\caption{Cumulative fraction of $\CIV$ (left) and $\OVI$ (right) absorbers with ion-weighted metallicities less than $Z$ as a function of $Z$ for different column densities at $z = 3$ (top) and $z = 0.1$ (bottom), in the EAGLE \emph{Ref-L100N1504} simulation. Curves from thin to thick show column density bins ranging from $10^{11}  < N_{\rm{ion}} \leq 10^{12} \cmsq$ to $10^{14}  < N_{\rm{ion}} \leq 10^{15} \cmsq$. The typical metallicity of absorbers increases with their column density and time.}
\label{fig:cum-dist-Z-c4o6}
\end{figure*}

\subsection{Cosmic density of ions}
\label{sec:iondensity}
The cosmic density of ions, $\Omega_{\rm{ion}}$, encapsulates the evolution of metal absorbers in the Universe. The cosmic density of an ion can be obtained by integrating the CDDF,
\begin{equation}
\Omega_{\rm{ion}} (z) = \frac{H_0 m_{\rm{ion}}}{c \rho_{\rm{crit}}} \int_0^\infty N_{\rm{ion}} f(N_{\rm{ion}},z)  dN_{\rm{ion}},
\label{eq:Omega}
\end{equation}
where $H_0 = 100~h~\rm{km~s^{-1}~Mpc^{-1}}$ is the Hubble constant, $m_{\rm{ion}}$ is the atomic weight of a given ion, $c$ is the speed of light, $\rho_{\rm{crit}} = 1.89 \times 10^{-29} h^2 \rm{g~cm^{-3}}$ and $f(N_{\rm{ion}}, z)$ is the CDDF as defined in equation \eqref{eq:CDDF}. For the ions we study in this work only a narrow range of column densities, around the ``knee" of the CDDF significantly contribute to $\Omega_{\rm{ion}}$. This is due to an increasing contribution of higher column densities in the integral together with the rapid change in the slope of the CDDF to values below $-2$ around its knee which makes the absorbers with higher column densities too rare to make a significant contribution to $\Omega_{\rm{ion}}$. Due to different CDDFs shapes (see figures \ref{fig:cddf-allz} and \ref{fig:cddf-allz2}), the appropriate range of important column densities varies from one ion to another, in addition to being redshift dependent. For instance, as figures \ref{fig:cddf-allz} and \ref{fig:cddf-allz2} show, at $z \sim 0$ the knees of the CDDFs happen at $N_{ion}\sim 10^{14} \cmsq$ for $\SiIV$, $\CIV$ and $\OVI$, at $N_{\NV}\sim 10^{13}$ for $\NV$ and at $N_{\NeVIII} \sim 10^{13.5}\cmsq$ for $\NeVIII$. Note that those values vary by $\sim \pm 0.5$ dex depending on redshift. 

The evolution of the cosmic ion densities for $\SiIV$, $\CIV$, $\NV$, $\OVI$ and $\NeVIII$ in the EAGLE \emph{Ref-L100N1504} simulation are shown in Fig. \ref{fig:ion-dens-obs} (solid curves). The shaded areas around the curves show the range of predictions from the simulations listed in Table \ref{tbl:sims} with different box sizes, resolutions and feedback models. For each ion, a compilation of observational measurements is shown using different symbols. All the observational measurements have been corrected to the Planck cosmology used by EAGLE.

Overall, the predicted cosmic ion densities agree reasonably well with the observations. One should note, however, that different observational studies use different ranges of column densities to compute the integral of eq. \eqref{eq:Omega}. Therefore, many of the observed ion densities shown in Fig. \ref{fig:ion-dens-obs} are lower limits and should be corrected for incompleteness due to missing some column densities. Such corrections are not simple and require knowing the true CDDFs at different redshifts. Therefore, it is not straightforward to compare different observational measurements (even in the same study at different redshifts), or observed data points with modelled ion densities. Moreover, the large uncertainties involved in measuring column densities can contribute significantly to the uncertainty of ion density calculations.

Despite the overall good agreement with the observed cosmic ion densities shown in Fig. \ref{fig:ion-dens-obs}, we note that our results underproduce the observed cosmic densities of $\SiIV$, $\CIV$, $\NV$ and $\OVI$ at very low redshifts ($z \simeq 0$) by a factor of $\sim 2$. Using a higher resolution simulation (see Appendix \ref{ap:res}) and/or a different UVB model (see Appendix \ref{ap:UVB}) improves the agreement between our predictions and the observed cosmic ion densities. While increasing the resolution of simulations does not change the ion densities significantly at higher redshifts ($z \gtrsim 0.5$), using the \citet{HM12} (HM12) UVB model can change them over a wider range of redshifts. For example using the HM12 UVB model increases the $\OVI$ density and improves the agreement between our results and the measurement of \citet{Muzahid12}. However, we note that our fiducial model is consistent with the $\OVI$ measurement of \citet{Carswell02}. Moreover, it is important to note that the contamination from the $\HI$ Ly$\alpha$ forest makes identifying and characterising $\OVI$ absorbers very challenging, particularly at high redshifts ($z \sim 2-3$). As a result, there are not many reported measurements for the statistical properties of $\OVI$ absorbers at those redshifts, and existing measurements could be contaminated. We also note that the simulations may under predict the data at $z \simeq 4$ for $\SiIV$, and at $z \gtrsim 4$ for $\CIV$.

The cosmic density of each ion divided by the cosmic density of the corresponding element (in gas) is shown in the top sections of each panel of Fig. \ref{fig:ion-dens-obs}. Both the sign and amount of evolution in the ion fractions seem to be sensitive to the ionization potential of each ion. The ion fraction of $\SiIV$ is $\approx 0.05$ at $z = 6$, drops by a factor of $\approx 2$ between $z = 6$ and $z = 2$ before reaching $\approx 0.003$ by the present time. The ion fraction of $\CIV$ is very similar to that of $\SiIV$ at $z \lesssim 2$. At higher redshifts, however, it evolves more slowly compared to $\SiIV$ and reaches to a value $\approx 0.03$ at $z = 6$. Showing a similar behaviour, $\NV$ ion fraction decreases with time but its evolution is weaker: it remains nearly constant (at $0.01$) at $z \gtrsim 2$ before dropping to $\approx 0.004$ at $z = 0$. The $\OVI$ ion fraction evolves even more weakly and remains nearly constant at $0.01$ for the full redshift range we consider. It has, however, a mild dip around $z \approx 1$ before recovering by $z=0$. The ion fraction of $\NeVIII$ on the other hand evolves differently from the rest of ions we show and increases monotonically with time from $\approx 0.01$ at $z = 6$ to its present value of $\approx 0.03$. 

The cosmic densities of ions evolve as a result of evolution in the cosmic density of the relevant elements and their evolving ionization fractions. At zeroth order, as the metal content of the Universe increases, the cosmic densities of different heavy elements are also expected to increase. Assuming weakly evolving ion fractions, this translates into monotonic increase in the ionic cosmic densities. As mentioned above, this is indeed what happens at all redshifts for $\OVI$ and $\NeVIII$, and at $z \gtrsim 2$ for all the ions we show in Fig. \ref{fig:ion-dens-obs}. At lower redshifts, however, the ion fractions evolve significantly, which can cancel the cosmic increase in the elemental abundances (e.g., $\NV$), cause a decreasing ion cosmic densities with time (e.g., $\SiIV$ and $\CIV$) or even accelerate the increase in the ion cosmic densities caused by increasing average metal density of the Universe (e.g., $\NeVIII$).

Different processes can change the evolution of the ion fractions. For instance, the physical gas densities of absorbers change with redshift which changes their ion fractions even in the presence of a fixed UVB radiation. The evolving temperature structure of the IGM due to shock heating caused by structure formation also affects the ion fractions by changing collisional ionization and recombination rates, particularly at low redshifts. In addition, the strength and spectral shape of the UVB is evolving which changes the photoionization rates. While we discuss the density and temperature evolution in the next section, we show the impact of the evolving UVB on the cosmic densities of $\CIV$ and $\OVI$ in the left and right panels of Fig. \ref{fig:ion-dens-uvbtest}, respectively. The solid blue curves show the cosmic densities of $\CIV$ and $\OVI$ in the \emph{Ref-L025N0376} simulation using our fiducial HM01 UVB model. The orange dashed and green long-dashed curves, on the other hand, show the cosmic densities where the UVB properties (i.e., normalization and spectral shape) are kept fixed based on the HM01 UVB model at $z=3$ and $z = 0$, respectively. As shown in the left and right panels of Fig. \ref{fig:ion-dens-uvbtest}, the cosmic density of $\CIV$ decreases with time in part due to the evolution of the UVB at $z \lesssim 2$ while the cosmic $\OVI$ density shows the opposite behaviour, implying that other factors compensate for the evolution of the UVB. 

\begin{figure*}
\centerline{\hbox{{\includegraphics[width=0.35\textwidth]
              {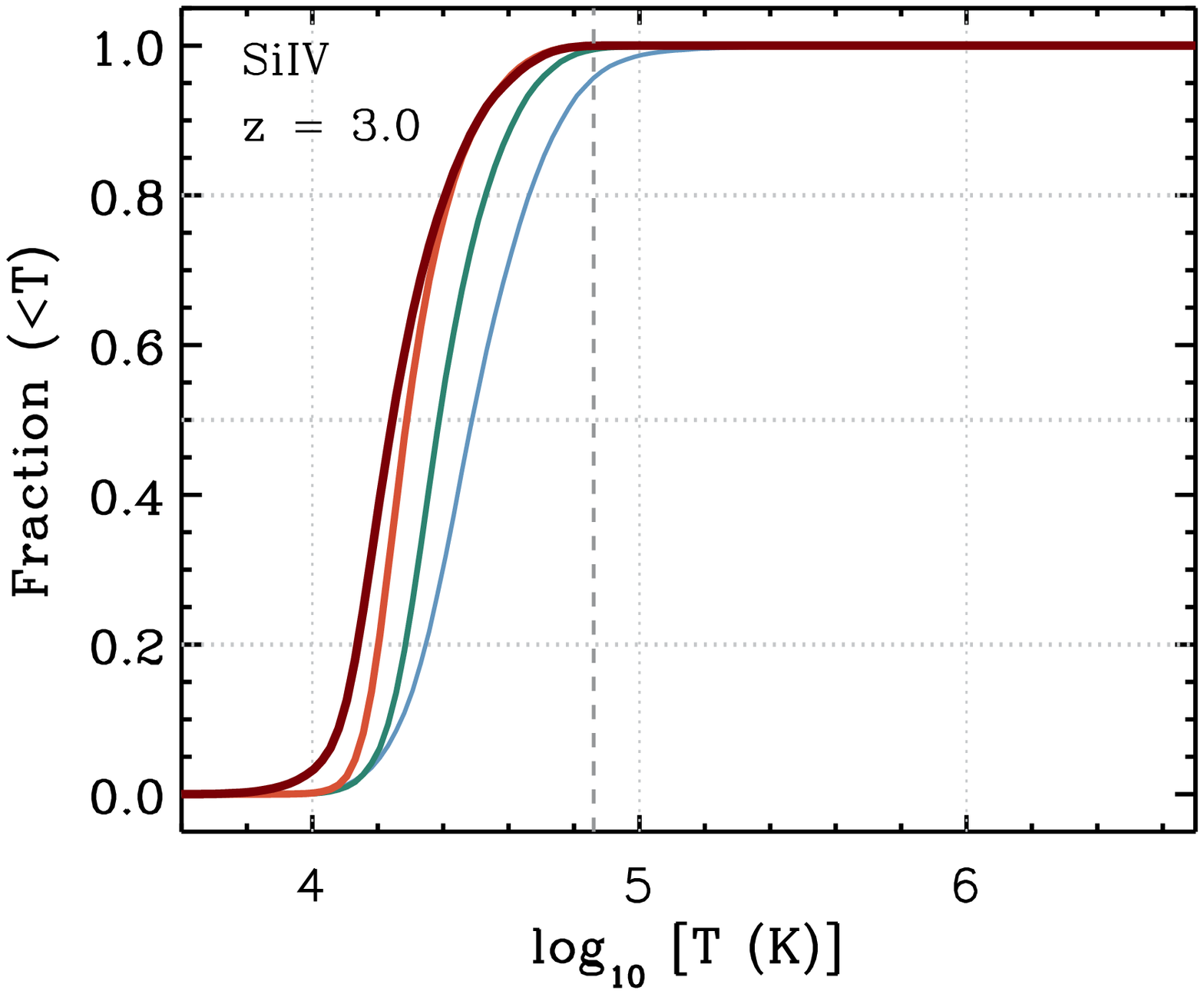}}}
             \hbox{{\includegraphics[width=0.35\textwidth]	
             {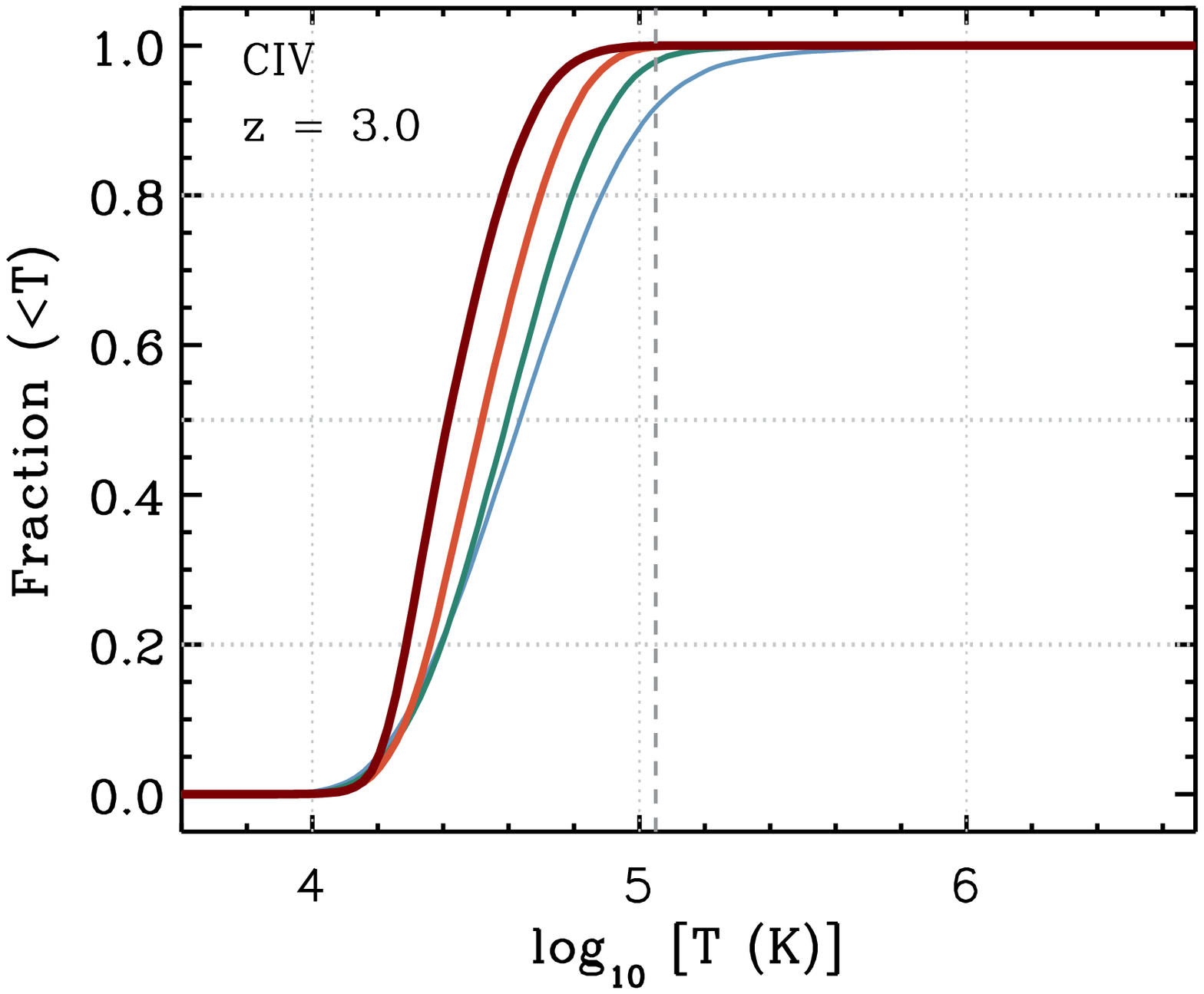}}}
             \hbox{{\includegraphics[width=0.35\textwidth]	
             {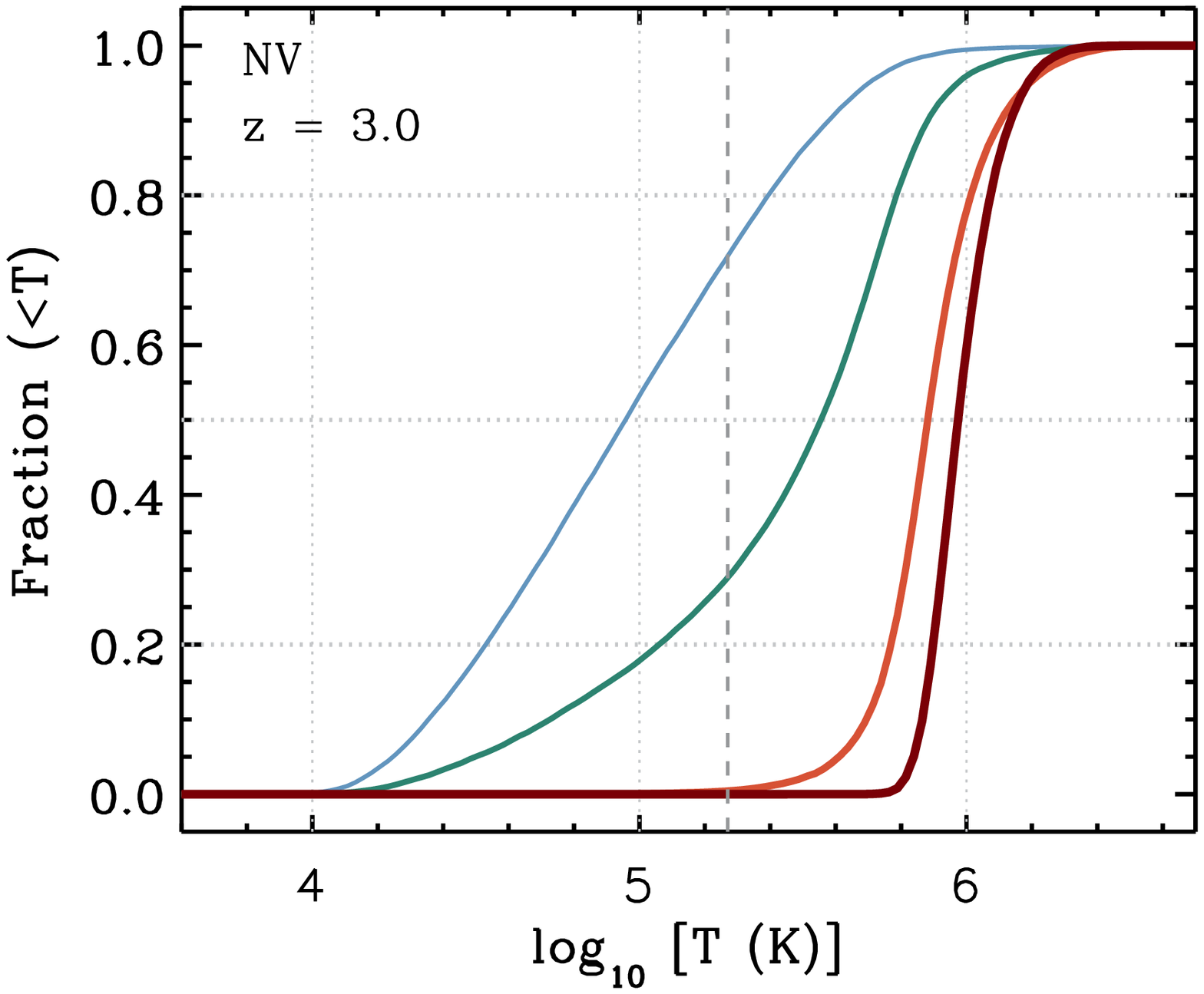}}}}
\centerline{\hbox{{\includegraphics[width=0.35\textwidth]
              {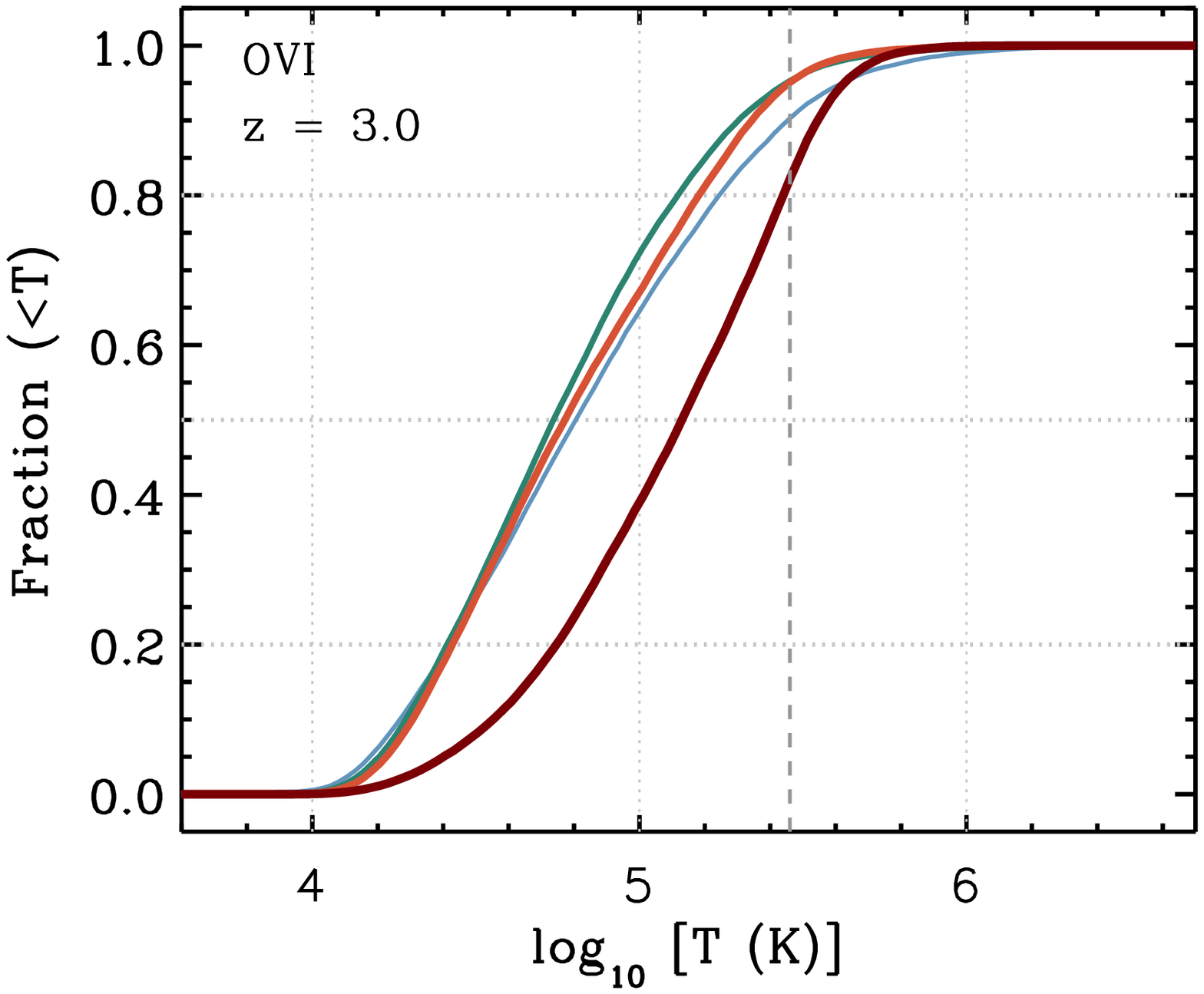}}}
             \hbox{{\includegraphics[width=0.35\textwidth]	
             {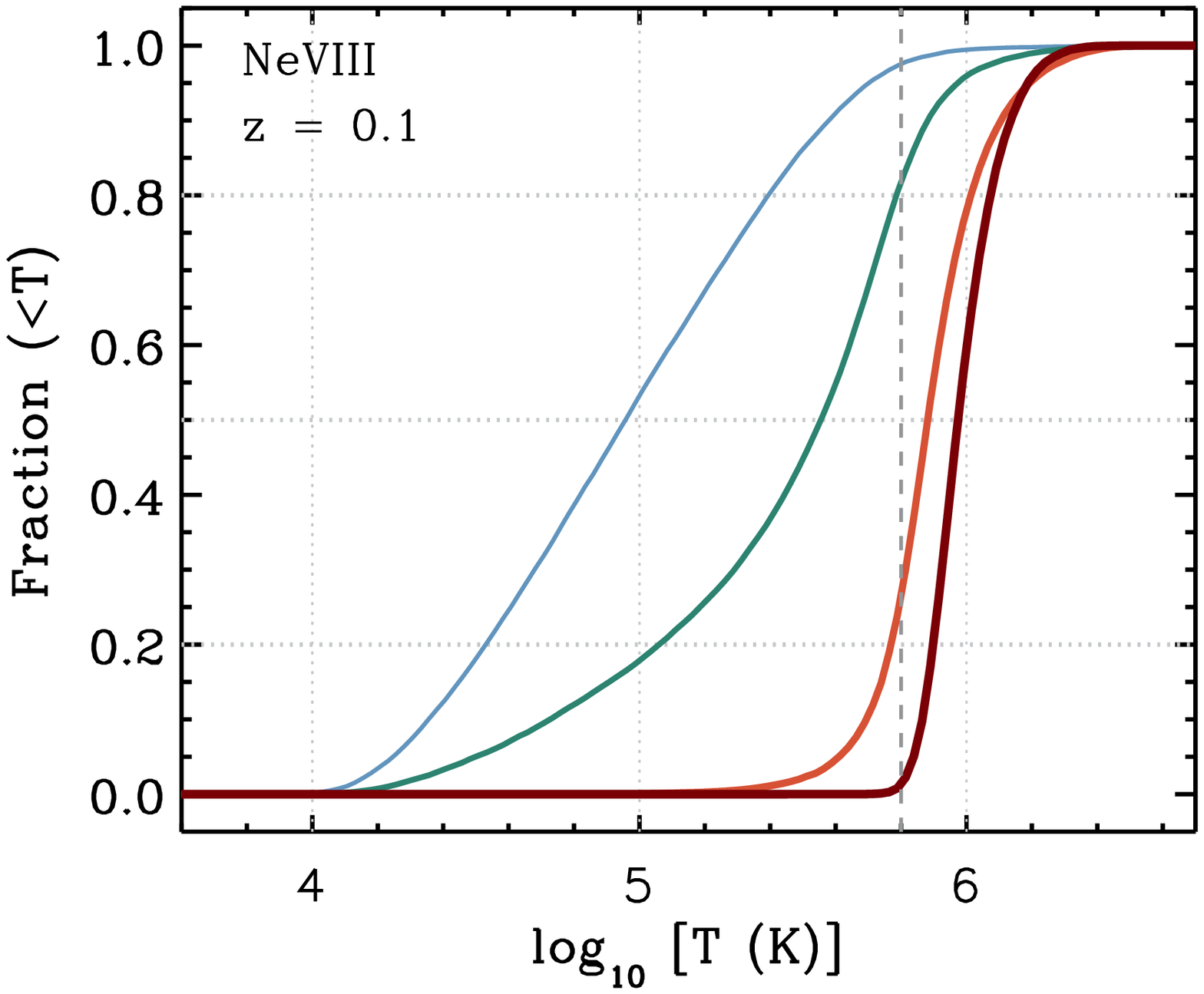}}}
             \hbox{{\includegraphics[width=0.35\textwidth]	
             {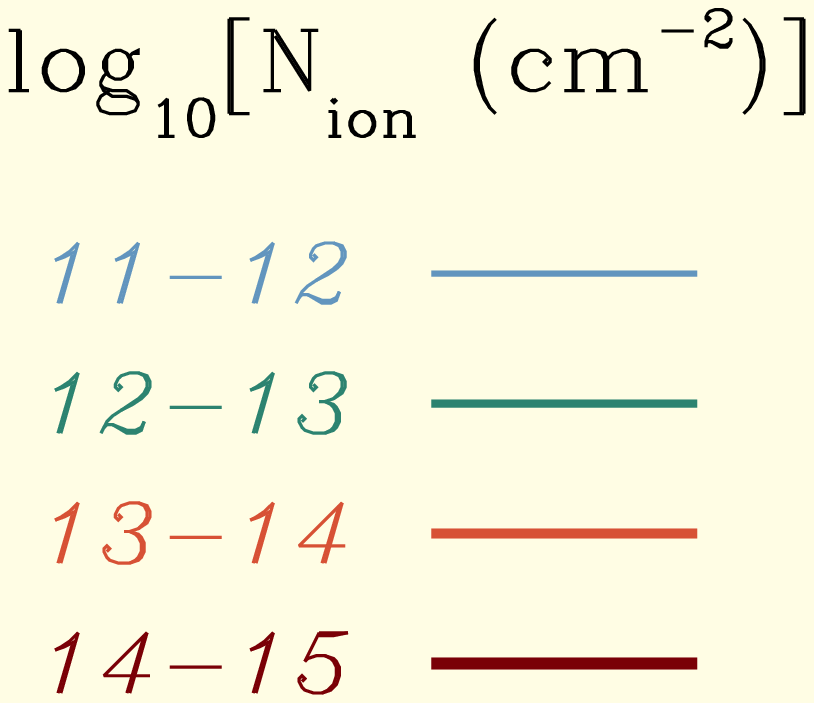}}}}
\caption{Cumulative fraction of $\SiIV$ (top-left), $\CIV$ (top-middle), $\NV$ (top-right), $\OVI$ (bottom-left) and $\NeVIII$ (bottom-right) absorbers with ion-weighted temperatures below $T$ as a function of $T$ for different column densities at $z = 3$ in the EAGLE \emph{Ref-L100N1504} simulation. Curves from thin to thick show column density bins ranging from $10^{11}  < {\rm{N_{ion}}} \leq 10^{12} \cmsq$ to $10^{14}  < {\rm{N_{ion}}} \leq 10^{15} \cmsq$. The vertical dashed lines indicate the temperature at which ion fraction peaks in collisional ionization equilibrium. Higher column density $\SiIV$ and $\CIV$ absorbers have lower temperatures while the opposite holds for $\OVI$ and $\NeVIII$ absorbers. Moreover, while a negligible fraction of $\SiIV$ and $\CIV$ absorbers are collisionally ionized, this fraction increases with ionization potential of the ion and becomes significant for $\OVI$ and $\NeVIII$ absorbers.}
\label{fig:cum-dist-T-z3}
\end{figure*}
\begin{figure*}
\centerline{\hbox{{\includegraphics[width=0.35\textwidth]
              {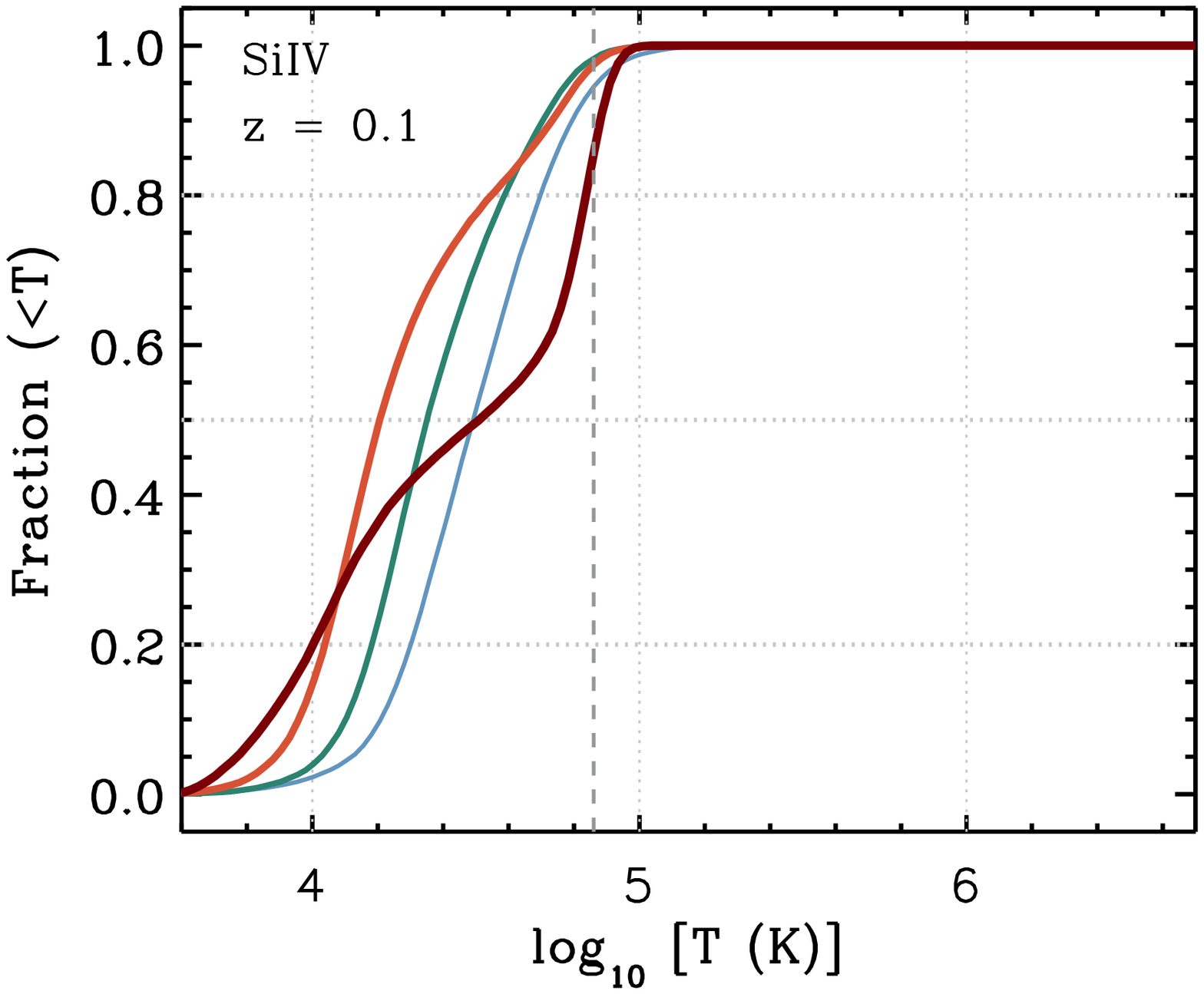}}}
             \hbox{{\includegraphics[width=0.35\textwidth]	
             {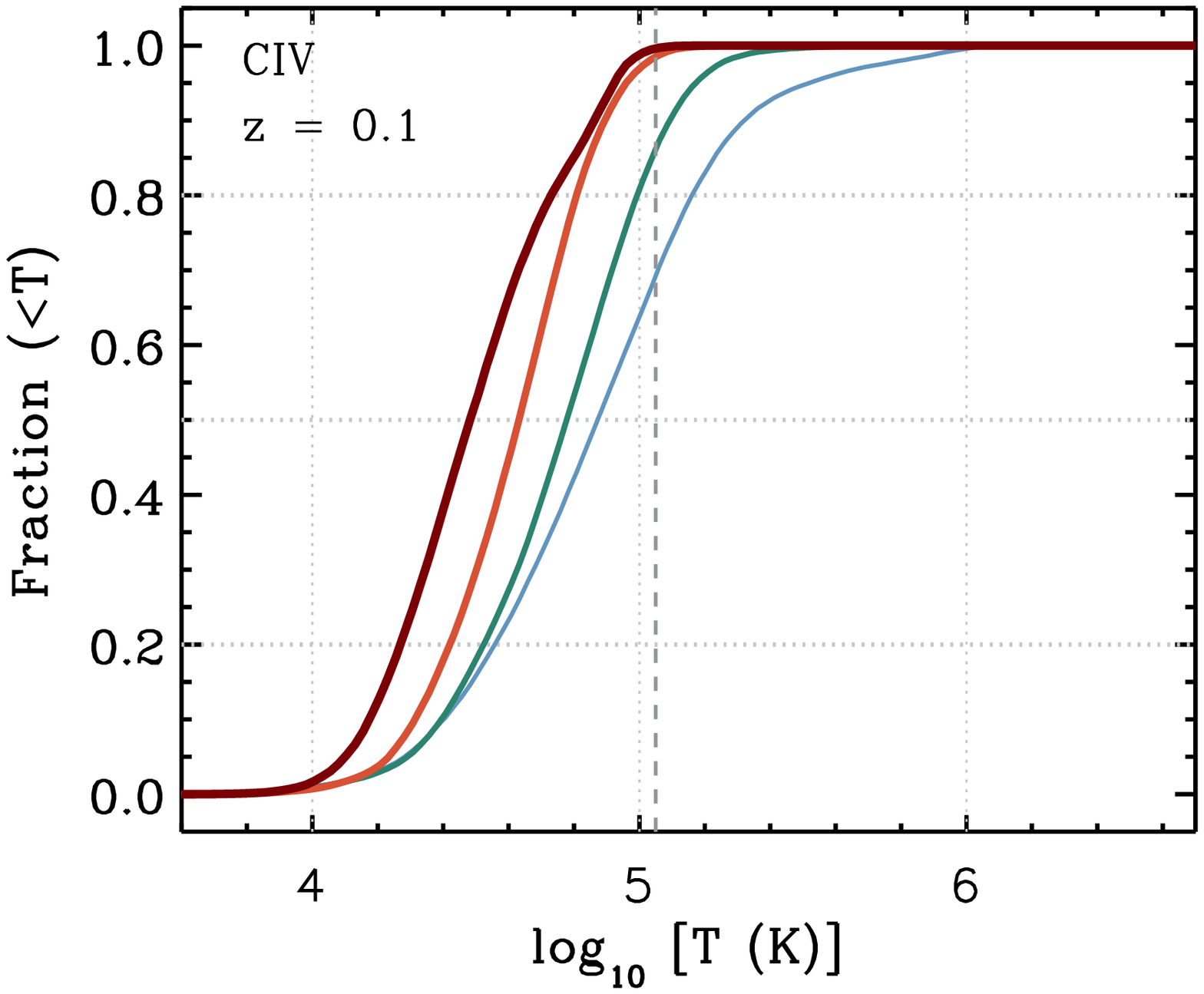}}}
             \hbox{{\includegraphics[width=0.35\textwidth]	
             {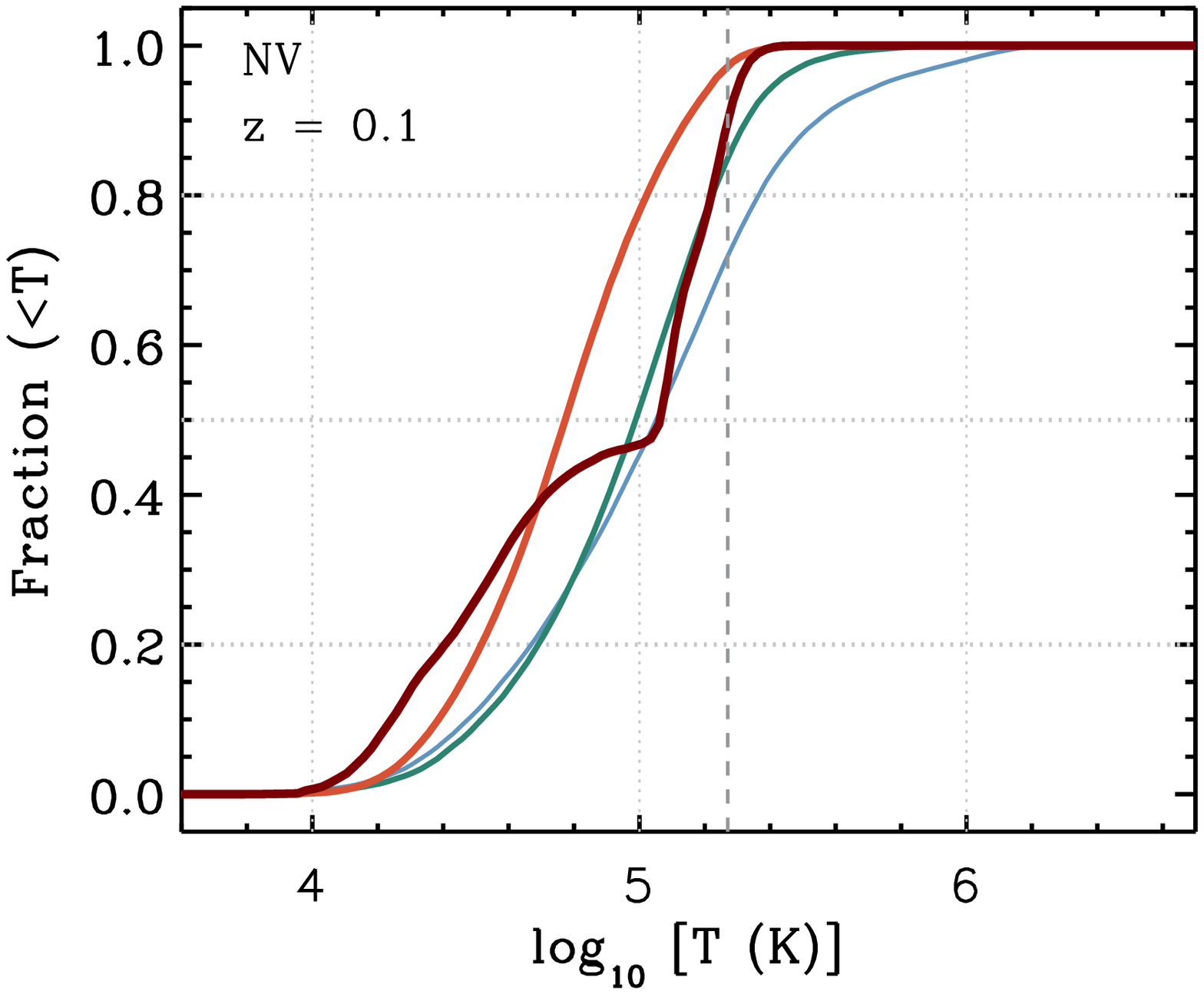}}}}
\centerline{\hbox{{\includegraphics[width=0.35\textwidth]
              {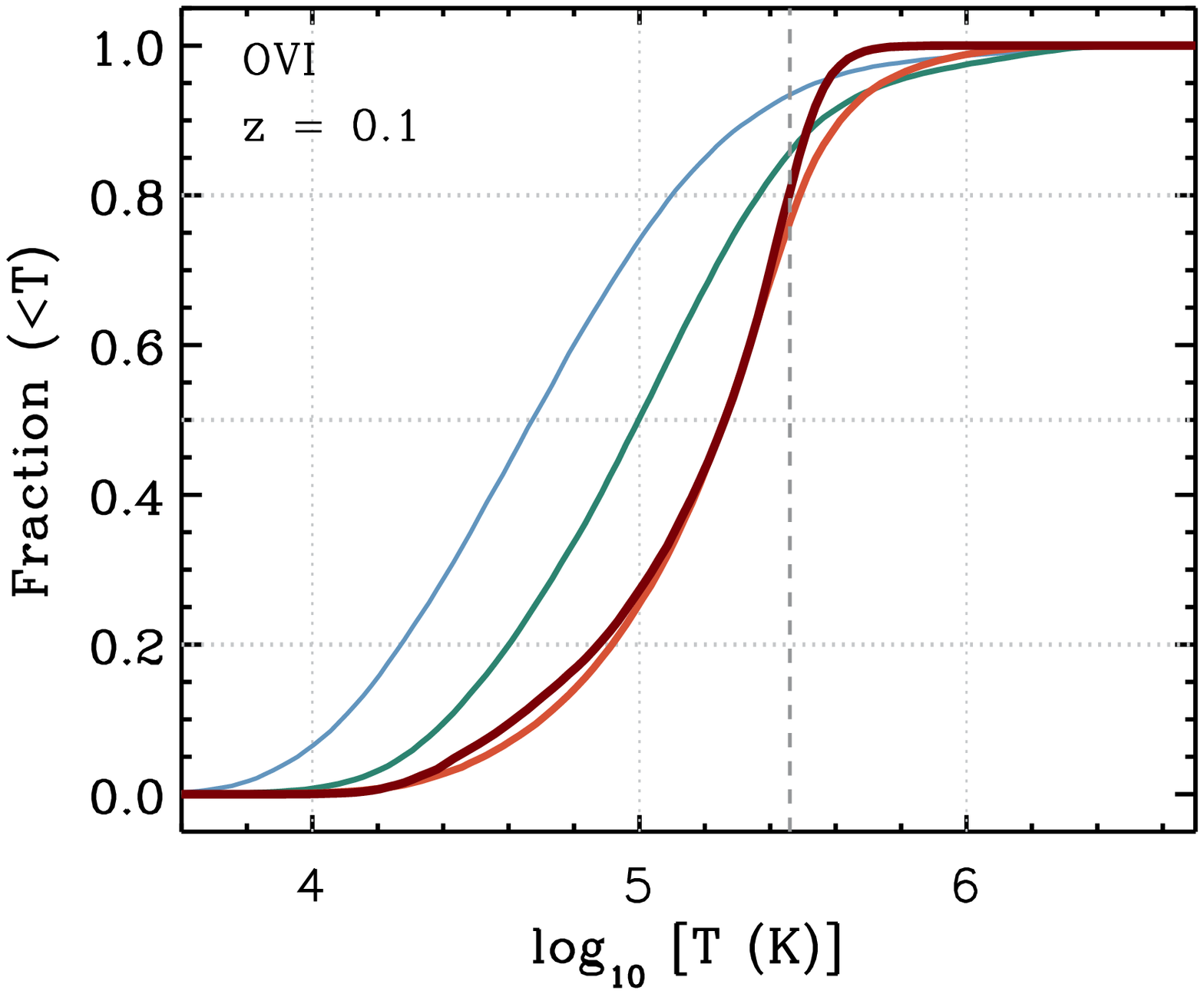}}}
             \hbox{{\includegraphics[width=0.35\textwidth]	
             {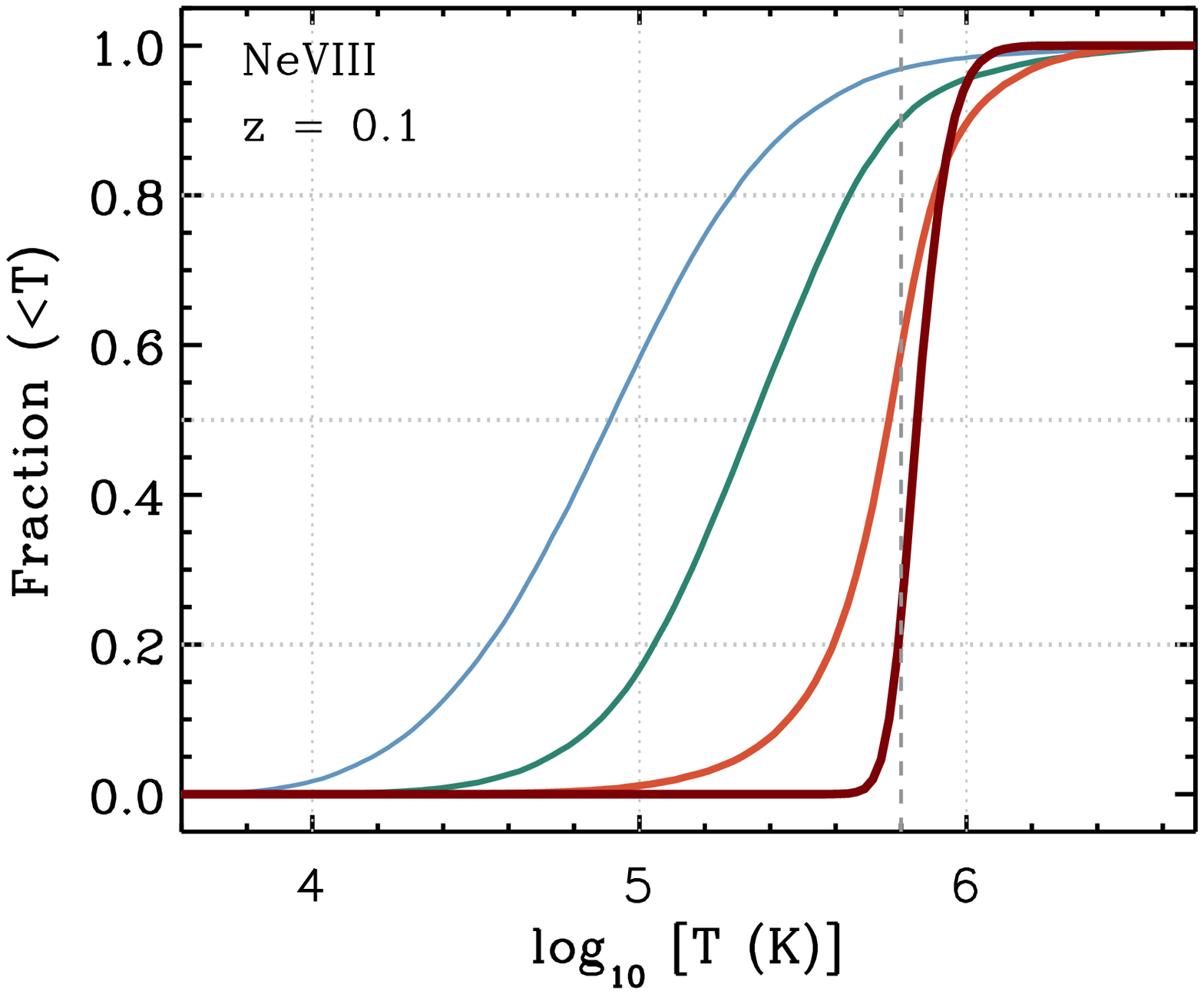}}}
             \hbox{{\includegraphics[width=0.35\textwidth]	
             {./finalplots/Tleg.eps}}}}
\caption{The same as Fig. \ref{fig:cum-dist-T-z3} but at $z = 0.1$. For all ions, the column density dependence of the temperature distribution for $\SiIV$ and $\CIV$ becomes stronger at lower redshifts and, for all ions, the fraction of absorbers that are collisionally ionized increases.}
\label{fig:cum-dist-T-z0p1}
\end{figure*}
\subsection{Physical properties of absorbers}
\label{sec:physicalproperties}

The fraction of atoms found in a given ionization state depends on the gas density and temperature on the one hand, and the properties of the UVB on the other hand. These dependencies are set by atomic energy structures and vary from one species to another, making different ions tracers of different physical conditions (see Fig. \ref{fig:CIE}). Studying those conditions helps us to understand the processes that shape the distribution of different ions (e.g., collisional ionization vs. photoionization) in addition to the physical regimes that they represent. 

The gas density-temperature distribution in the EAGLE \emph{Ref-L100N1504} simulation is shown in Fig. \ref{fig:4dplot-all-ions} for $z = 3$ (top) and $z=0.1$ (bottom). The size of each cell in those diagrams is proportional to the logarithm of the gas mass enclosed in it, while its color shows its median metallicity. The contours with different colors and line-styles show the regions inside which $80\%$ of the total masses of different ions are found. The typical temperature of the absorbers increases with the ionization energy (see Table \ref{tbl:energies}). The typical density of absorbers, on the other hand, decreases with increasing ionization energy. In addition, as a comparison between the top and bottom panels of Fig. \ref{fig:4dplot-all-ions} shows, the typical densities of absorbers decrease with decreasing redshift while their typical temperatures increase towards lower redshifts, at least from $z = 3$ to $z = 0.1$.

For all ions there is a strong correlation between the gas density and the column density. This is shown in Fig. \ref{fig:cum-dist-density-c4o6} for $\CIV$ (left) and $\OVI$ (right) absorbers\footnote{Note that we opted not to include all absorbers in Fig. \ref{fig:cum-dist-density-c4o6} and \ref{fig:cum-dist-Z-c4o6} since all absorbers show similar trends in their density and metallicity distributions while the temperature distribution of each species is different from the others.} at $z = 3$ (top) and $z = 0.1$ (bottom). In each panel, curves with different colors show the cumulative functions of absorbers as a function of the ion-weighted gas density for column density bins ranging from $10^{11}  < N_{\rm{ion}} \leq 10^{12} \cmsq$ to $10^{14}  < N_{\rm{ion}} \leq 10^{15} \cmsq$.

Noting that systems with higher densities are typically closer to stars which are sources of metal production, it is reasonable to expect that the metallicities of absorbers increase with increasing their densities and hence their column densities. Fig. \ref{fig:cum-dist-Z-c4o6} shows that at $z = 3$ the typical metallicity of absorbers increases from $\sim 10^{-1.5}~{\rm{Z}}_{\odot}$ at $N_{\rm{ion}} \sim 10^{12} \cmsq$ to $\sim 10^{-1} Z_{\odot}$ for $N_{\rm{ion}} \sim 10^{14} \cmsq$. The strong correlation between metallicity and column density holds at all epochs but the typical metallicity of the absorbers increases with decreasing redshift. The typical metallicity of absorbers at fixed column density increases by $\sim 1$ dex from $z = 3$ to $z = 0.1$.

The aforementioned correlation between density (column density) and metallicity can also be seen in Fig. \ref{fig:4dplot-all-ions} as the metallicity increases by moving from left to right within each contour. It is important to note, however, that the colors in Fig. \ref{fig:4dplot-all-ions} show the median mass-weighted metallicity of the gas with temperatures and densities within the range indicated by each pixel. While this quantity can be used to deduce the qualitative change in metallicity with density, it does not represent the actual metallicity of the absorption systems. Indeed, comparison with Fig. \ref{fig:cum-dist-Z-c4o6} shows that the metallicities of absorbers are higher than the typical metallicity of gas at temperatures and densities  similar to those of the metal absorbers (\citealp[see also e.g.,][]{Oppenheimer09}). For instance, while the colors in Fig. \ref{fig:4dplot-all-ions} suggest that the bulk of the gas in the temperature and density range similar to that of low column density $\CIV$ absorbers ($N_{\CIV} \sim 10^{12} \cmsq$; $T\sim 10^{4.8}$ K, $\nH \sim 10^{-5} \cmcb$) at $z = 0.1$ has a typical metallicity $\lesssim 10^{-3} \rm{Z}_{\odot}$, the bottom-left panel of Fig. \ref{fig:cum-dist-Z-c4o6} shows that those absorbers have typical metallicities $\approx100$ times higher.

\begin{figure*}
\centerline{\hbox{{\includegraphics[width=0.45\textwidth]
              {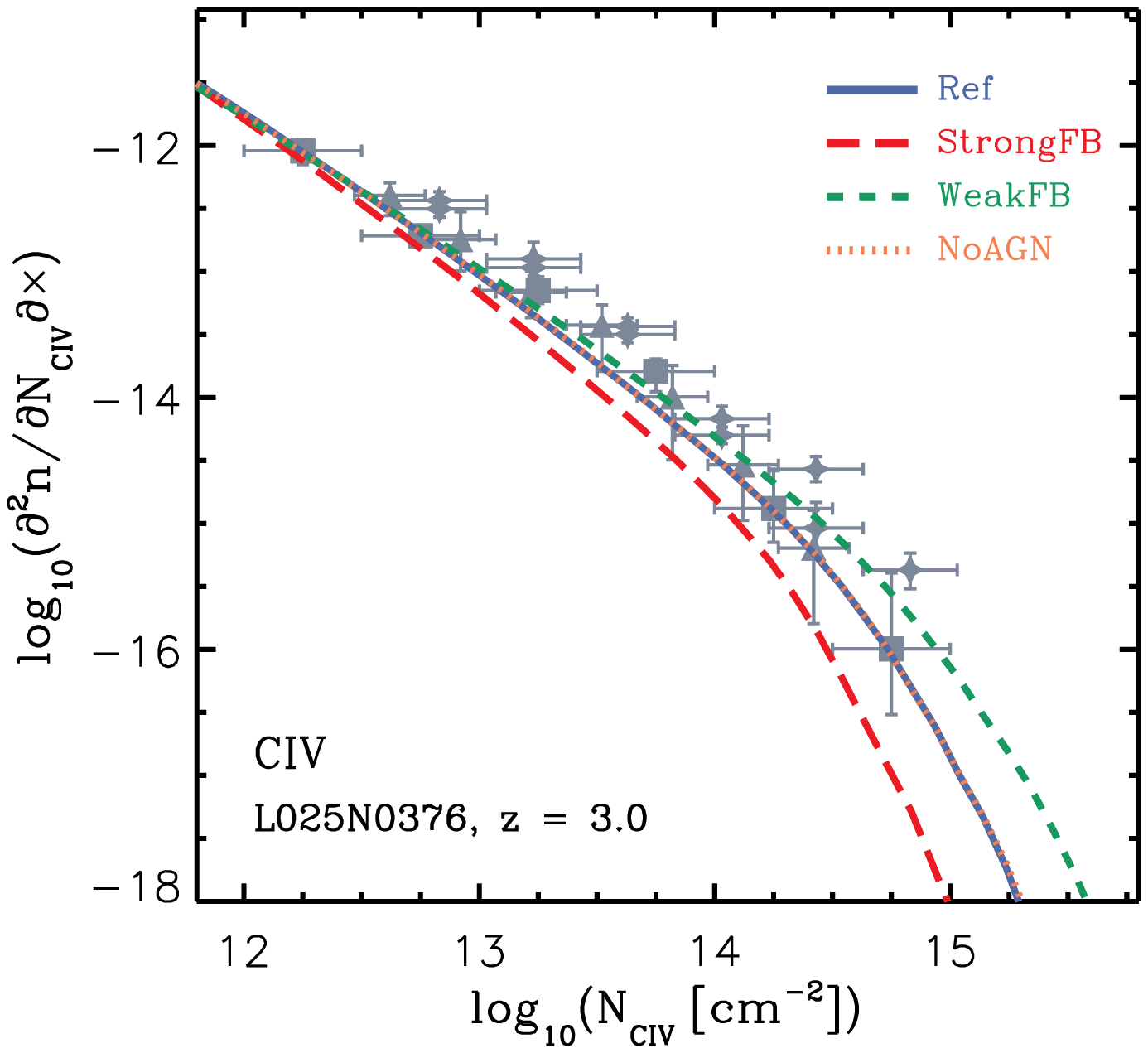}}}
             \hbox{{\includegraphics[width=0.45\textwidth]	
             {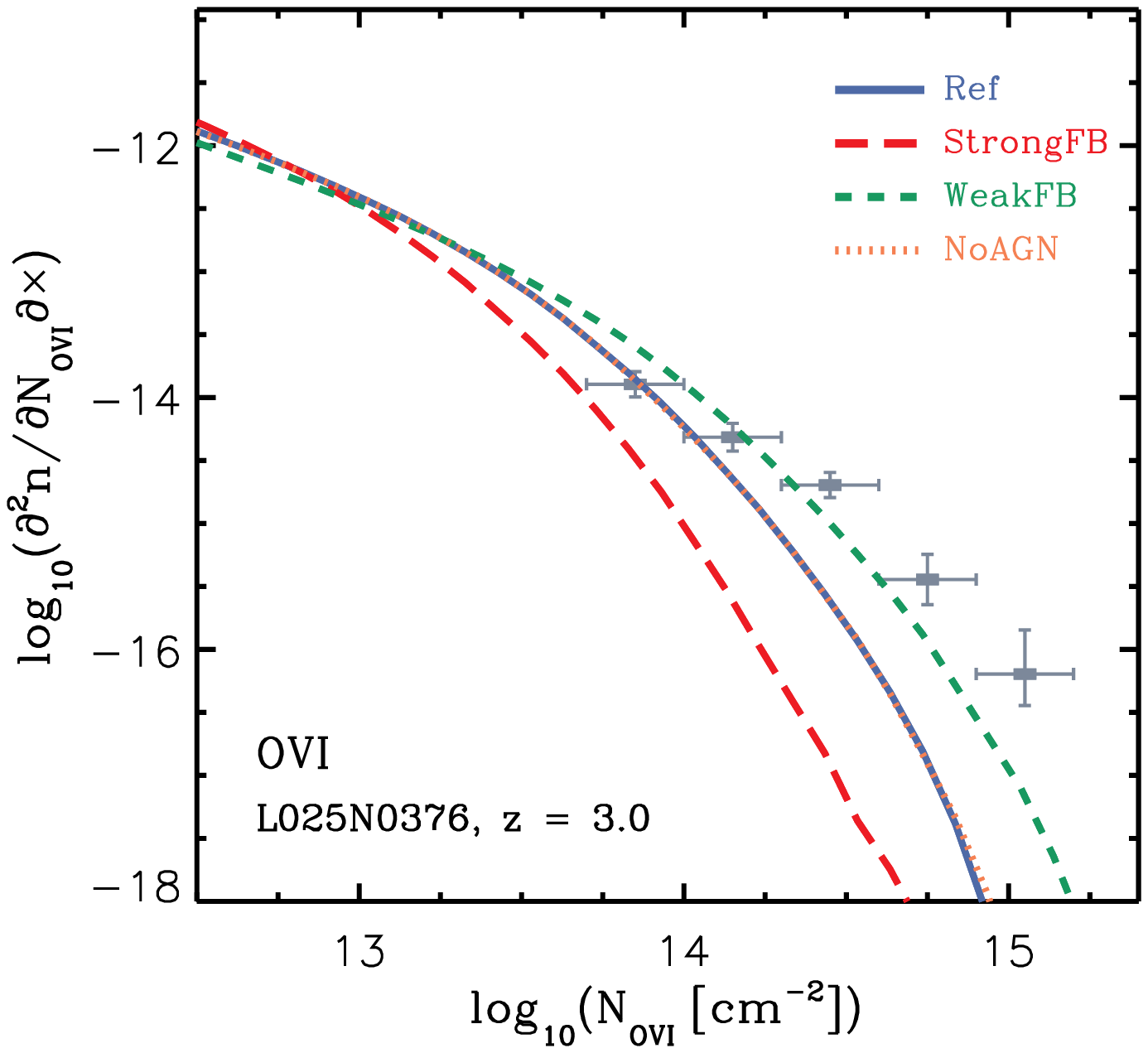}}}}
\centerline{\hbox{{\includegraphics[width=0.45\textwidth]
              {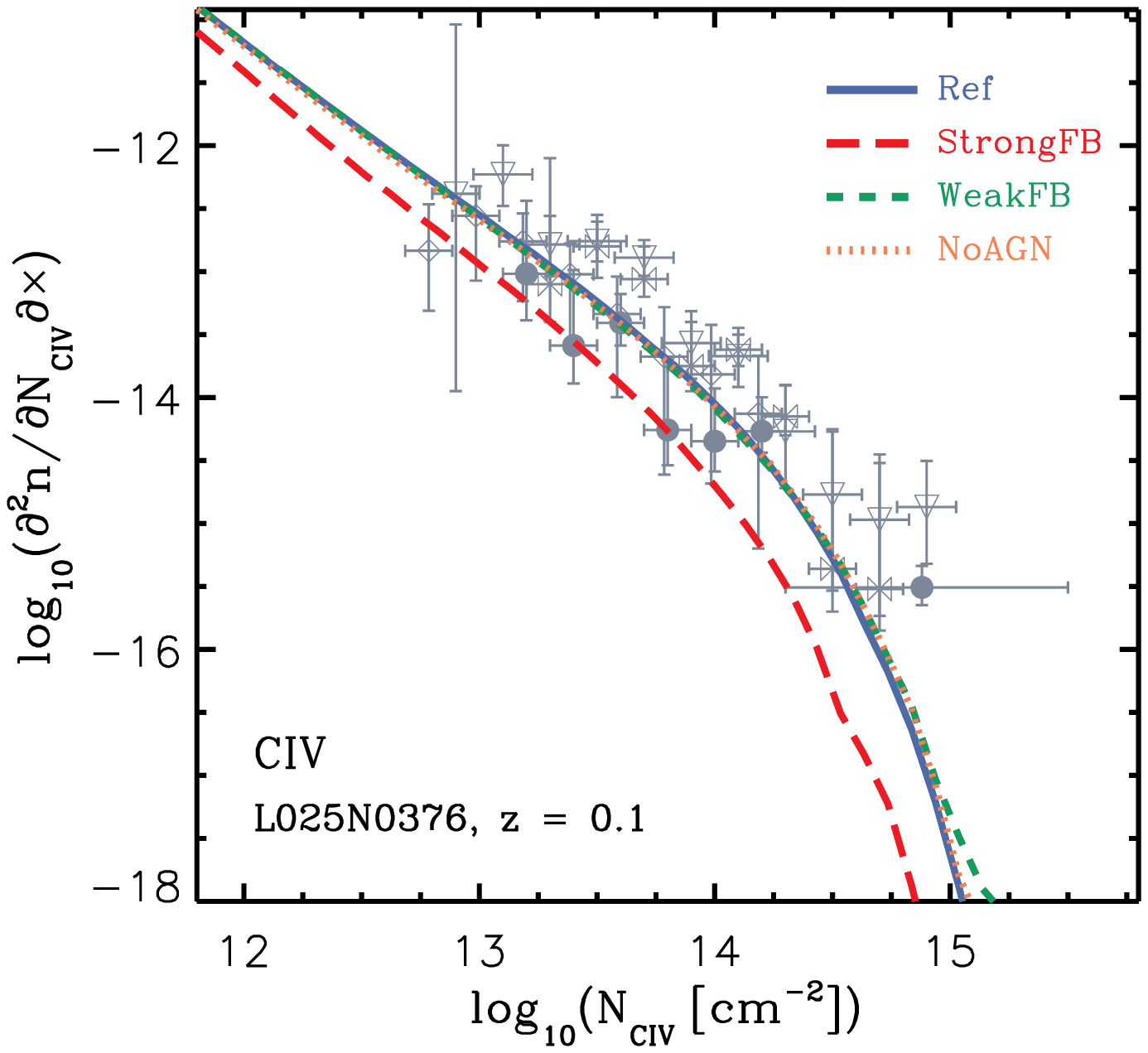}}}
             \hbox{{\includegraphics[width=0.45\textwidth]	
             {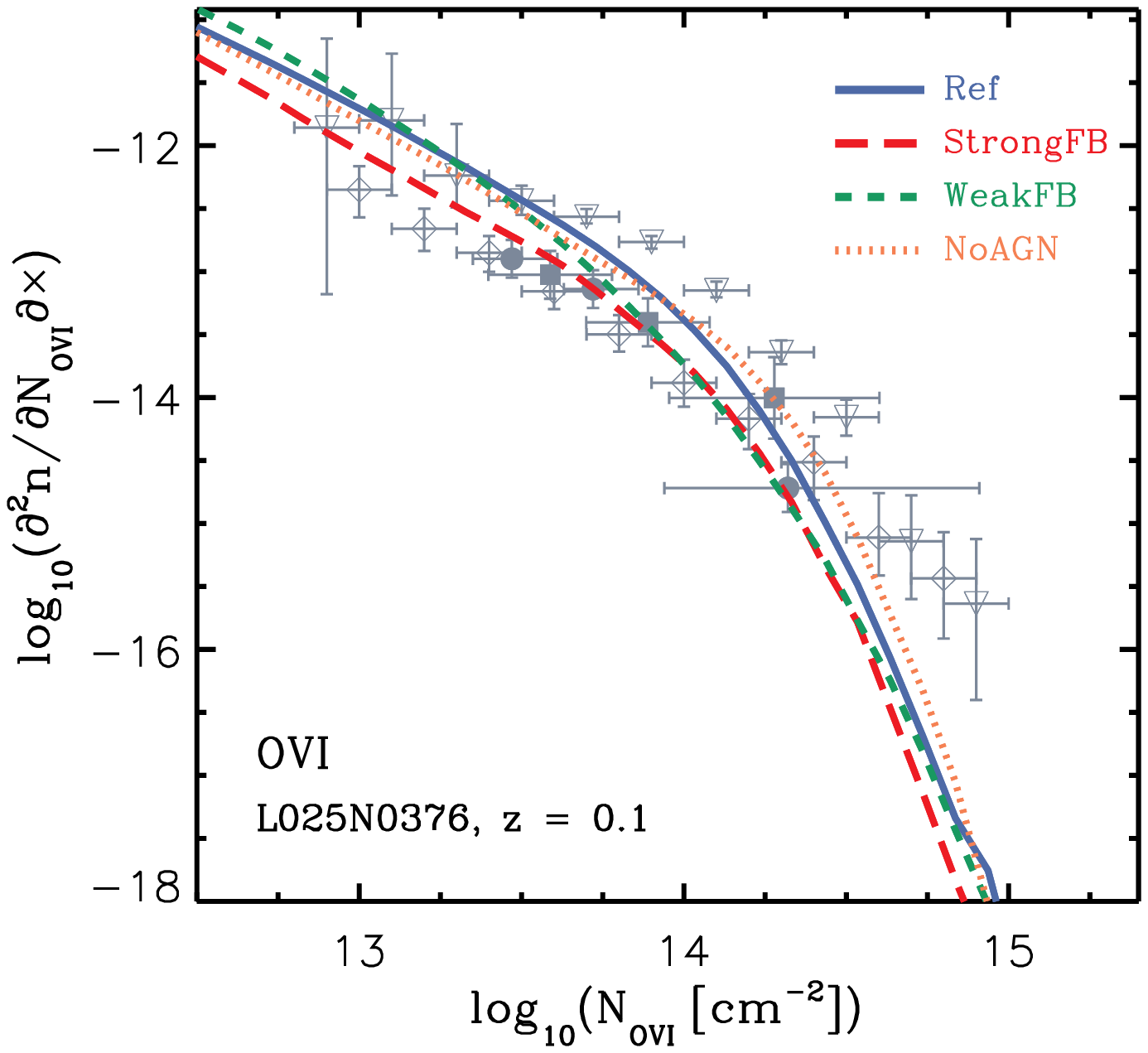}}}}
\caption{Column density distribution functions of $\CIV$ (left) and $\OVI$ (right) in the EAGLE simulations with different feedback strengths at $z = 3.0$ (top) and $z = 0.1$ (bottom). Blue (solid), red (long-dashed), green (dashed) and orange (dotted) curves show the results from the \emph{Ref}, \emph{StrongFB}, \emph{WeakFB} and \emph{NoAGN} simulations, respectively, all using a box size of 25 cMpc. The symbols show compilations of observational measurements, identical to those shown in Fig. \ref{fig:cddf-allz}. The differences between the CDDFs in different models are greater at higher column densities.}
\label{fig:CDDF-var}
\end{figure*}
\begin{figure*}
\centerline{\hbox{{\includegraphics[width=0.45\textwidth]
             {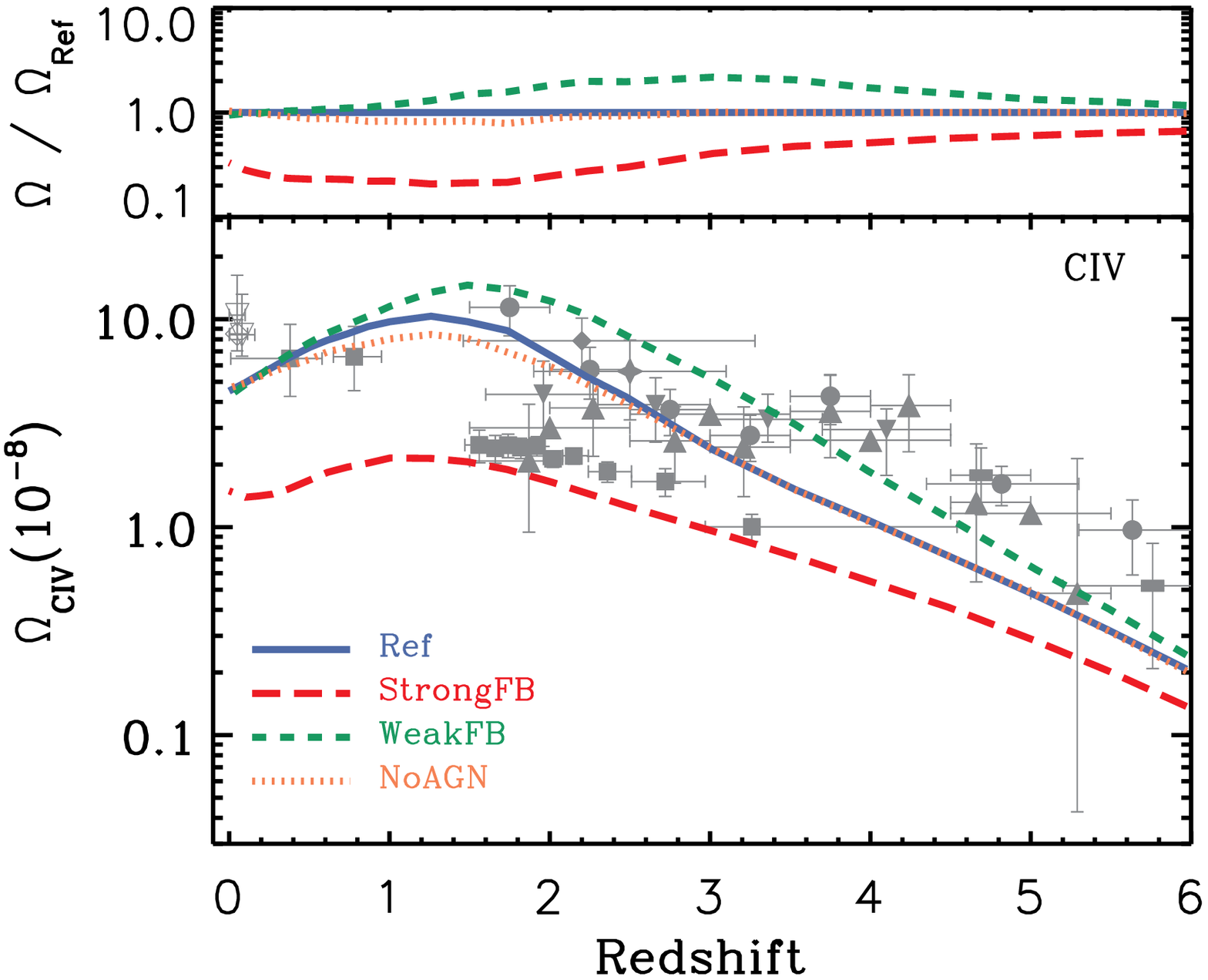}}}
             \hbox{{\includegraphics[width=0.45\textwidth]
             {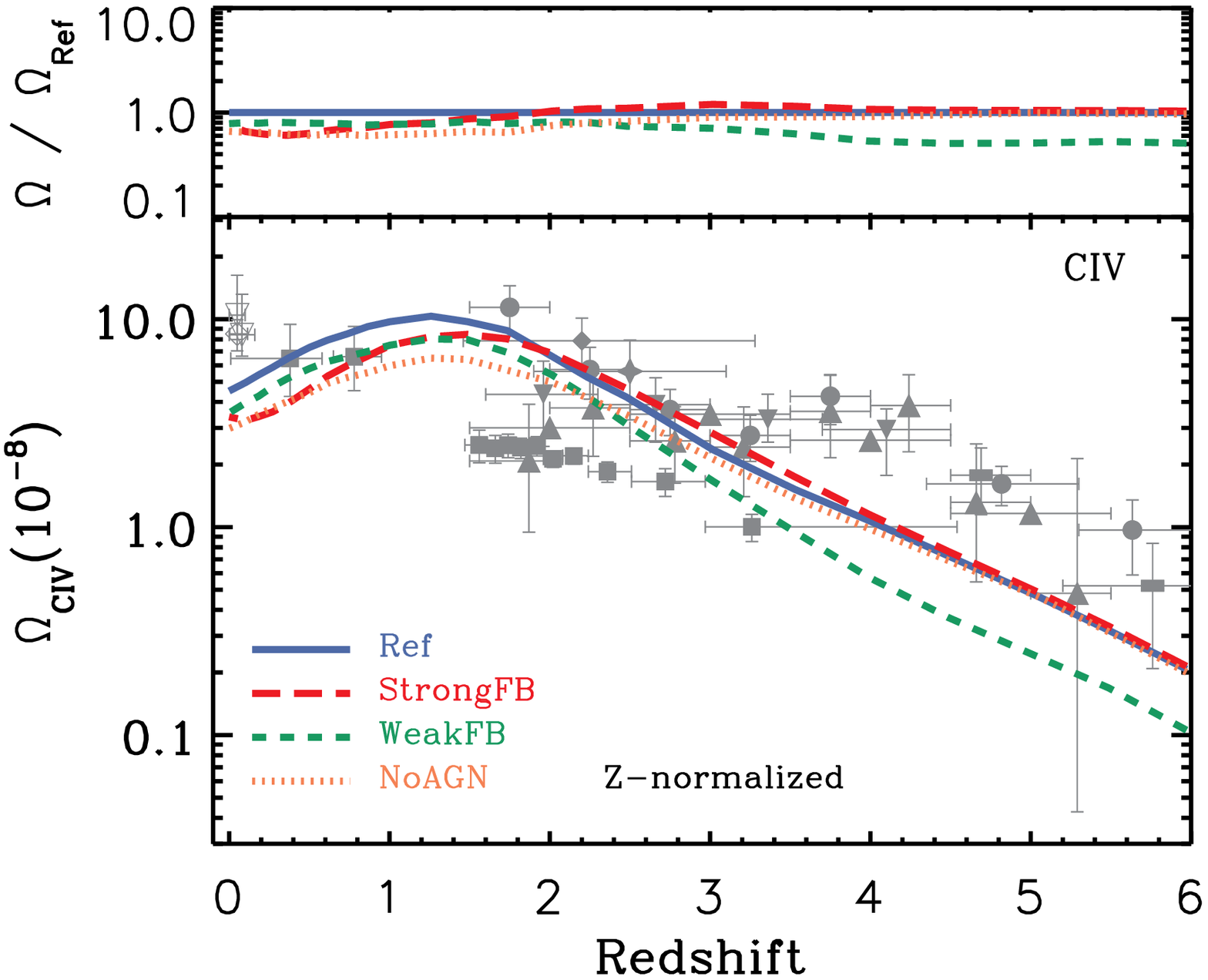}}}}
\caption{Evolution of the cosmic density of $\CIV$ in the EAGLE simulations for different feedback models. Blue (solid), red (long-dashed), green (dashed) and orange (dotted) curves show the results from the \emph{Ref}, \emph{StrongFB}, \emph{WeakFB} and \emph{NoAGN} simulations, respectively. The grey symbols show a compilation of observational measurements, identical to the one shown in Fig. \ref{fig:ion-dens-obs}. The top section of each panel shows the ratio between the ion density and that of the \emph{Ref} simulation. The cosmic density of ions is higher in models with weaker feedback which is mostly because of their higher metal production rates (i.e., star formation rates). To show this, the {\bf{right}} panel shows the ion density after normalizing them to match the cosmic elemental density of the \emph{Ref} simulation ($\Omega_{\rm{CIV,norm}} = \Omega_{\rm{CIV}} ~\frac{\Omega_{\rm{C,~Ref}}}{\Omega_{\rm{C}}}$). The agreement between the different simulations becomes much better after accounting for the differences between their metal abundances. This suggests that the main impact of varying the feedback on the cosmic abundance of metal ions in the IGM is caused by changes in the metal production rate.}
\label{fig:ion-dens-var}
\end{figure*}

The cumulative distribution of temperatures associated with absorbers with different column densities is shown in Fig. \ref{fig:cum-dist-T-z3} for $z = 3$ . Panels from top-left to bottom-right show $\SiIV$, $\CIV$, $\NV$, $\OVI$ ad $\NeVIII$, respectively and curves with different colors (thickness) show different column densities. The range of temperatures is relatively narrow compared to the density ranges (see Figs. \ref{fig:4dplot-all-ions} and \ref{fig:cum-dist-density-c4o6}), particularly for ions with lower ionization energies (see the top row). For $\SiIV$ and $\CIV$ absorbers, the temperature decreases slowly with increasing column density. This correlation becomes weaker for $\NV$ before becoming inverted for $\OVI$ and $\NeVIII$. Part of this trend can be understood by noting that lower ionization metals ($\SiIV$ and $\CIV$) are typically in density-temperature regimes where cooling is efficient and the temperature decreases with the density and hence with the column density (see Fig. \ref{fig:4dplot-all-ions}). Higher ionization metals ($\OVI$ and $\NeVIII$) on the other hand, are in density-temperature regimes where cooling is less efficient and shock heating dominates the relation between temperature and density (see Fig. \ref{fig:4dplot-all-ions}). As a result, the temperature of those absorbers increases with their (column) densities.

A dashed vertical line in each panel indicates the temperature at which the ion fraction peaks in CIE (see the left panel of Fig. \ref{fig:CIE}). Absorbers that have temperatures higher than or similar to the temperature indicated by this line are most likely primarily collisionally ionized. For metals with lower ionization energies ($\SiIV$ and $\CIV$), a negligible fraction of absorbers have temperatures high enough to be significantly affected by collisional ionization. However, this fraction increases as the ionization energy increases. This can be seen as steepening of the cumulative temperature distribution of $\NV$ absorbers with $10^{14}  < N_{\NV} \leq 10^{15} \cmsq$ and from the significant fraction of high column density $\OVI$ and $\NeVIII$ absorbers (with $N_{\rm{ion}} \geq 10^{13} \cmsq$) that have temperatures similar to the values indicated by the vertical dashed curves. 

Fig. \ref{fig:cum-dist-T-z0p1} shows the temperature distribution of metals at $z = 0.1$. Most of the trends discussed above remain qualitatively the same at $z = 0.1$. However, the fraction of absorbers that are affected by collisional ionization increases from $z = 3$ to $z = 0.1$, which is visible also in Fig. \ref{fig:4dplot-all-ions} as additional branches in the contours appearing at higher temperatures in the bottom panel. As the bottom left panel of Fig. \ref{fig:cum-dist-T-z0p1} suggests, $\lesssim 30\%$ of low-z $\OVI$ absorbers with $N_{\OVI} \geq 10^{13} \cmsq$ are in the temperature regimes where collisional ionization is the dominant source of ionization (see also \citealp{TG11}). This is in good agreement with recent low-z observations of $\OVI$ absorbers \citep{Savage14}. Note, however, that if we assume that all $\OVI$ absorbers with $T \gtrsim 10^5$ K (instead of $T \gtrsim 10^{5.5}$ K; see the vertical line) are collisionally ionized, then this fraction increases to more than $80\%$.

\section{The impact of feedback}
\label{sec:fdbk}

In this section, we investigate the impact of feedback on the results we showed in the previous section. For this purpose, we compare simulations that use different feedback implementations, a box size $L = 25$ cMpc, and our default resolution ($N = 2 \times 376^3$ particles for this box size; see Table \ref{tbl:sims}). In addition to our reference model, we use feedback models \emph{WeakFB} and \emph{StrongFB} with, respectively, half and twice the amount of stellar feedback compared to the \emph{Ref} simulation (see \citealp{Crain15} for more details). To investigate the impact of AGN feedback, we also use a model for which AGN feedback is turned off (\emph{NoAGN}).

\begin{figure*}
\centerline{\hbox{{\includegraphics[width=0.5\textwidth]
              {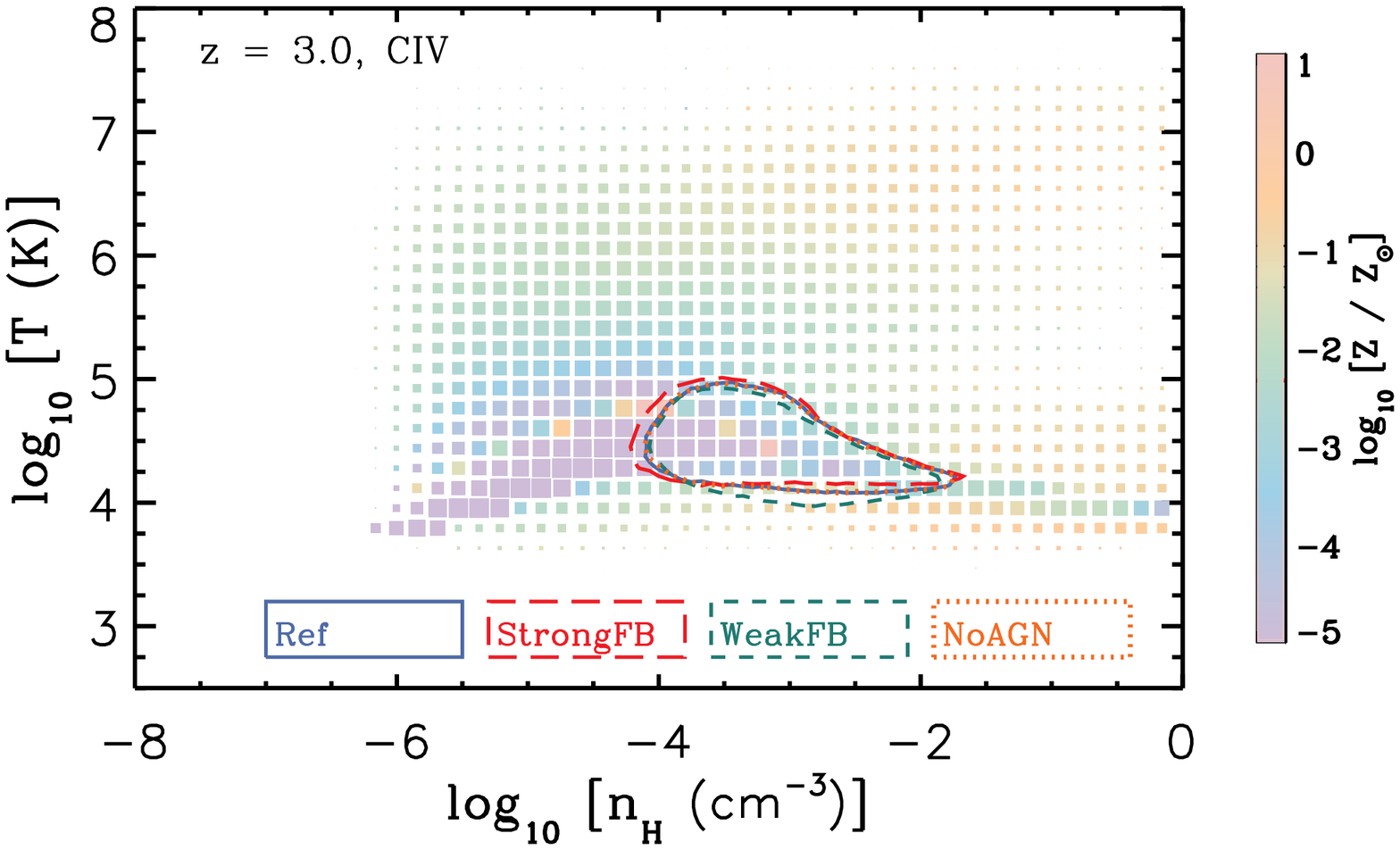}}}
             \hbox{{\includegraphics[width=0.5\textwidth]	
             {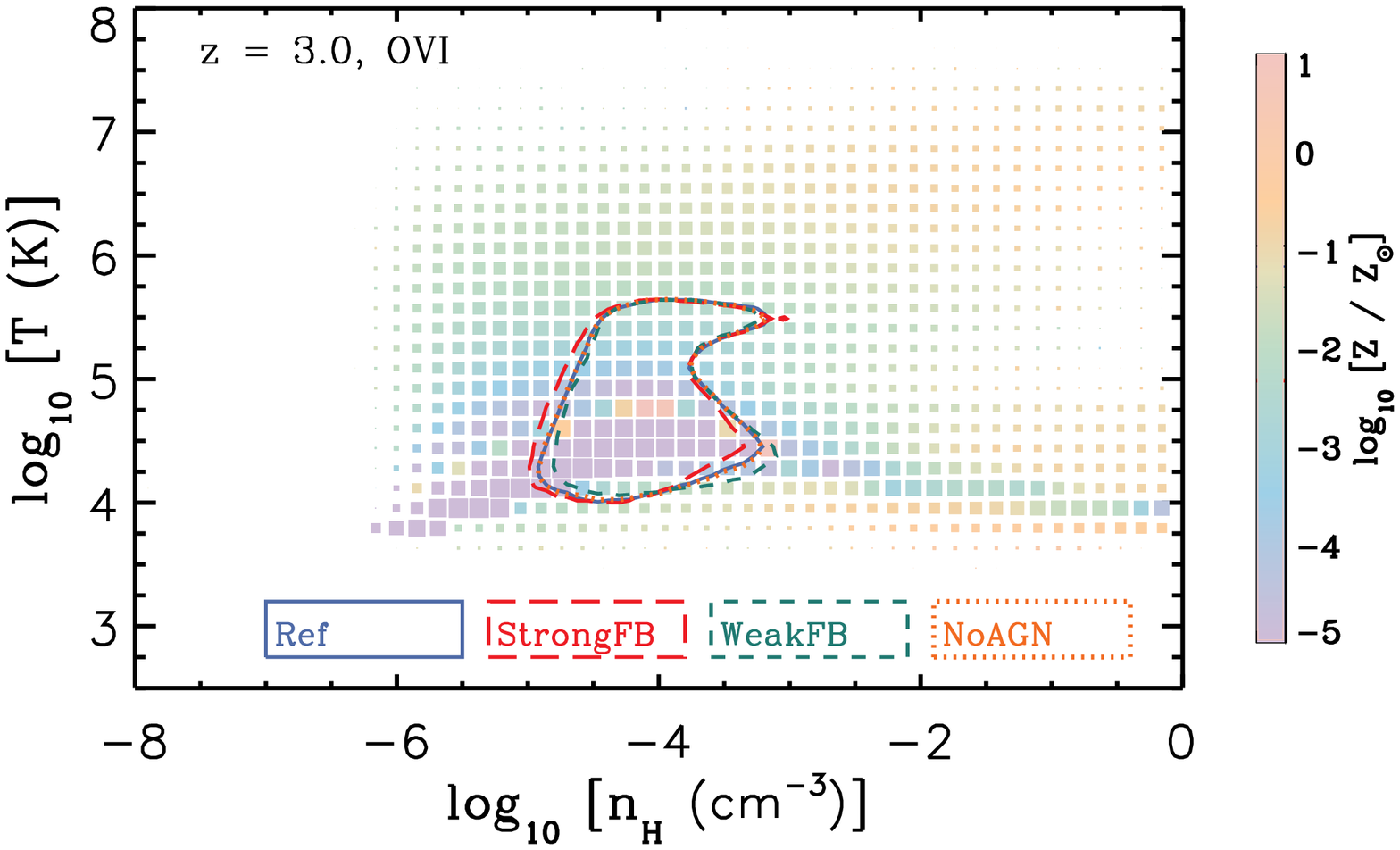}}}}
\centerline{\hbox{{\includegraphics[width=0.5\textwidth]
              {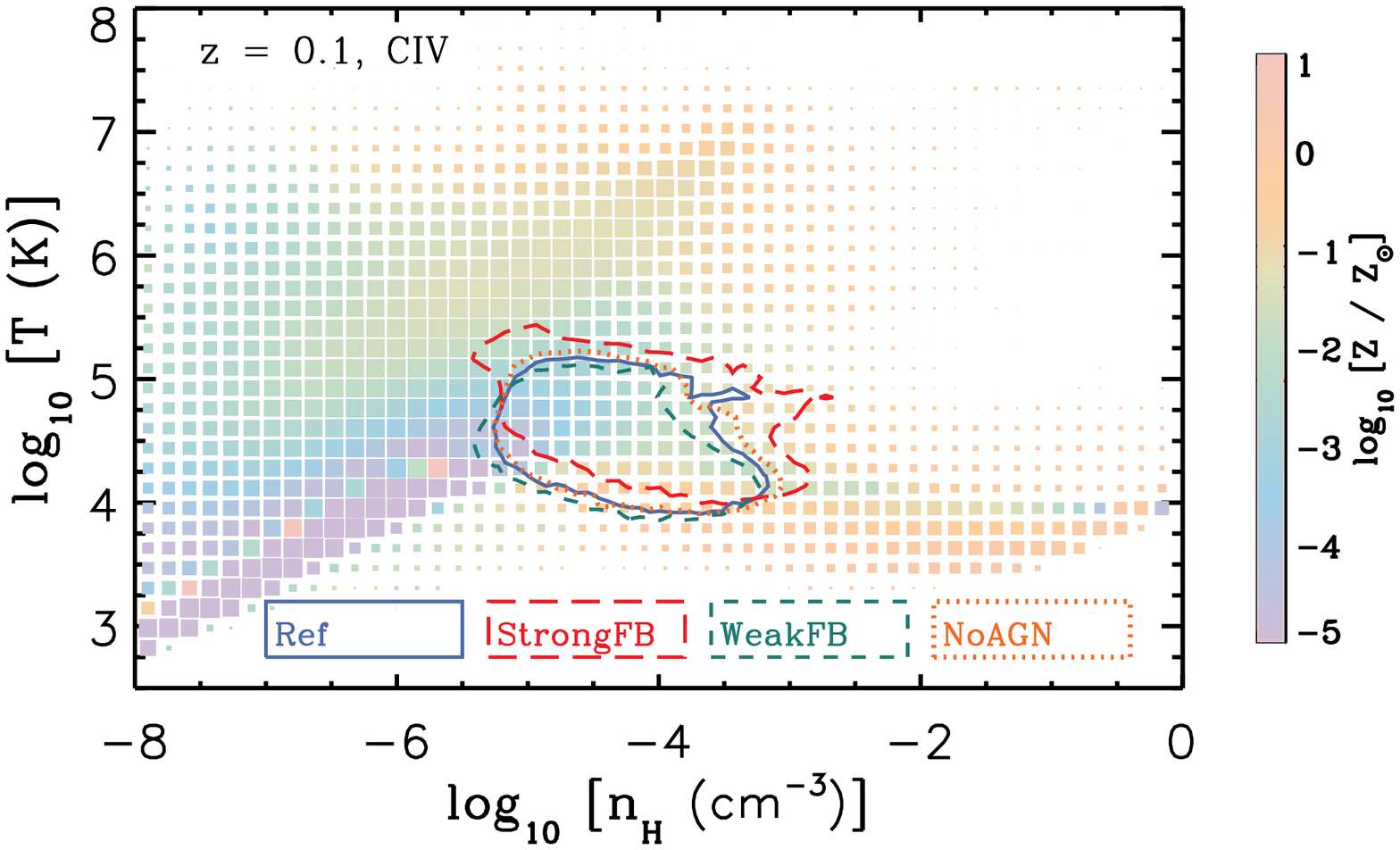}}}
             \hbox{{\includegraphics[width=0.5\textwidth]	
             {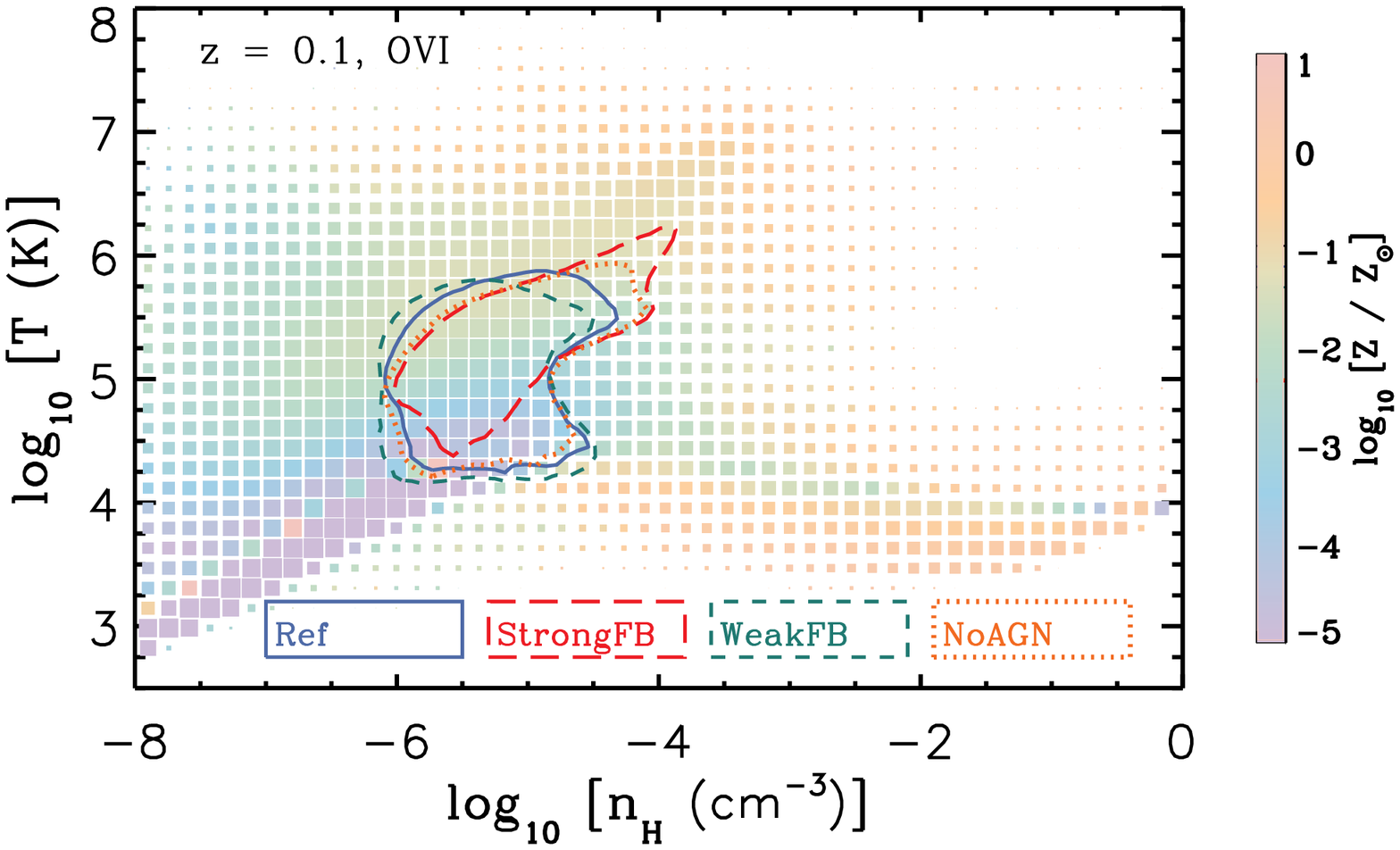}}}}
\caption{Temperature-density distribution of gas in the EAGLE \emph{Ref-L025N0376} simulation at $z = 3$ (top) and $z = 0.1$ (bottom). The size of each cell is proportional to the logarithm of the gas mass enclosed in it and its color shows its median metallicity. The areas enclosed by contours with different colors and line-styles show the range of temperature-densities that contain $80\%$ of $\CIV$ mass (left) and $\OVI$ mass (right) for different feedback models. Blue solid, red long-dashed, green dashed and orange dotted lines show \emph{Ref}, \emph{StrongFB}, \emph{WeakFB} and \emph{NoAGN} models, respectively. The temperature-density distribution of absorbers is very similar for different feedback models, particularly at $z = 3$. At low redshift, however, the temperature distribution of absorbers become more sensitive to variations of feedback.}
\label{fig:4dplot-c4-o6-var}
\end{figure*}
\begin{figure*}
\centerline{\hbox{{\includegraphics[width=0.45\textwidth]
              {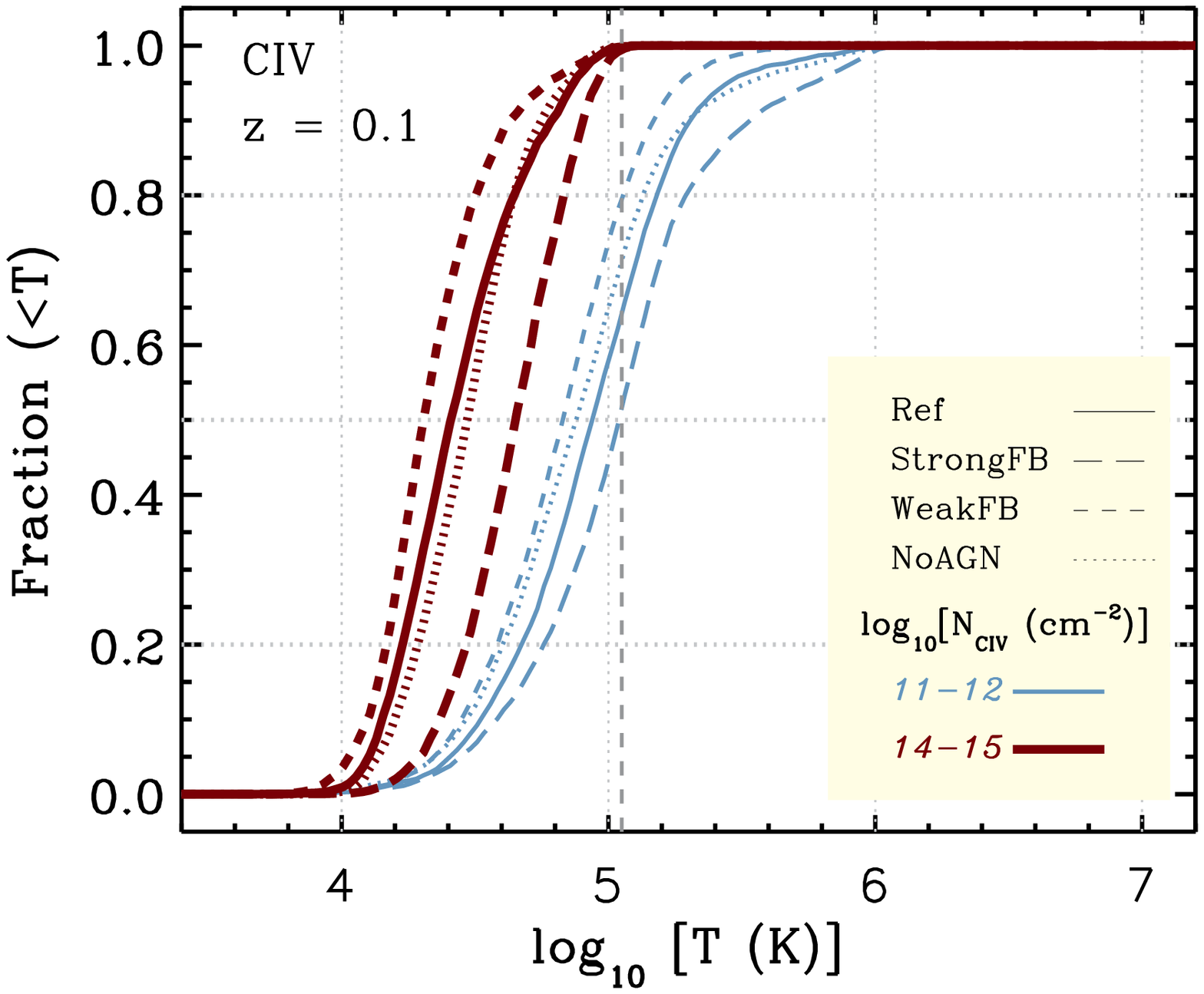}}}
             \hbox{{\includegraphics[width=0.45\textwidth]	
             {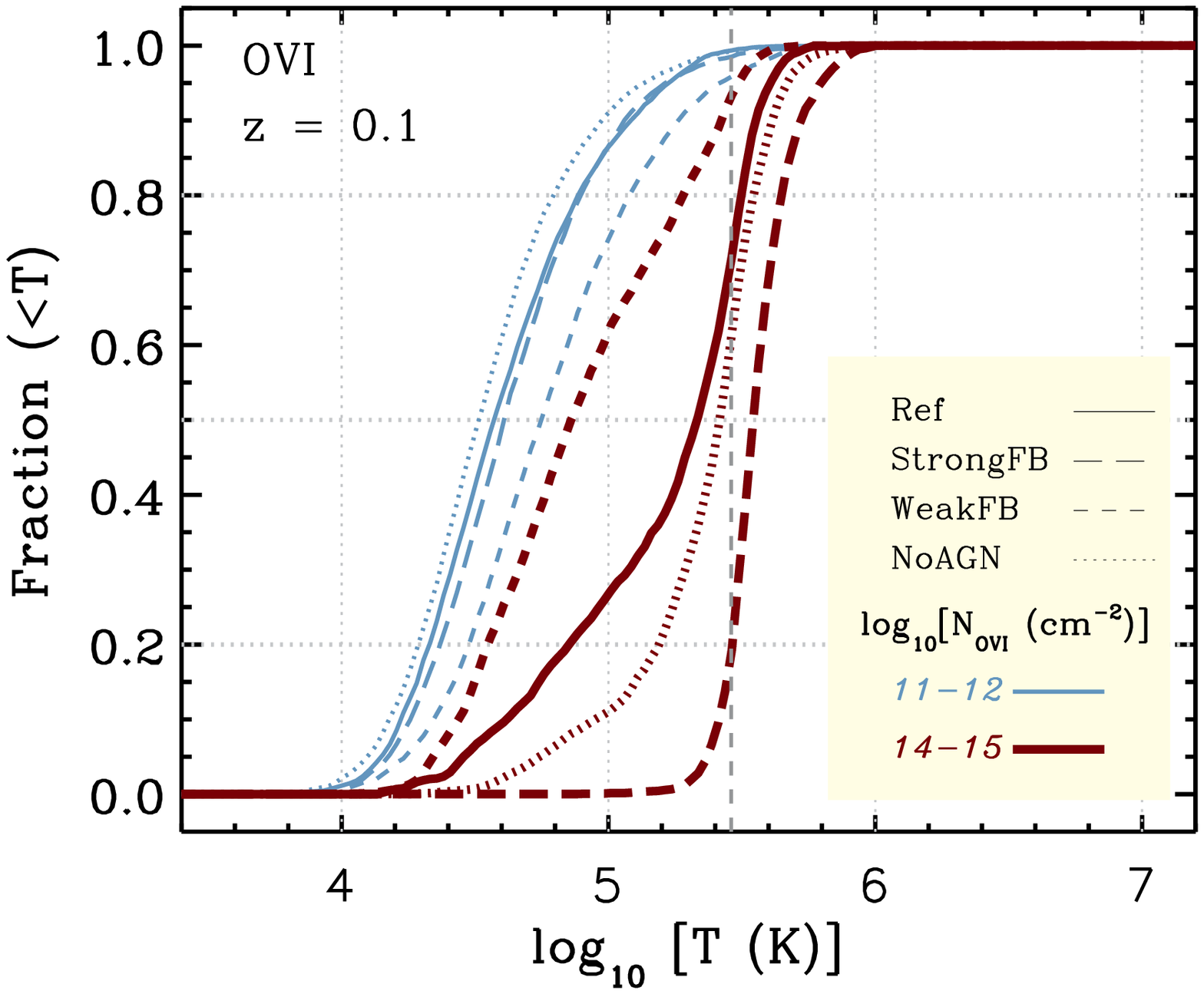}}}}
\caption{Cumulative fraction of $\CIV$ (left) and $\OVI$ (right) absorbers with ion-weighted temperatures below $T$ as a function of $T$ for  different column densities at $z = 0.1$. Thin blue and thick red curves show column density bins with $10^{11}  < N_{\rm{ion}}/ \cmsq \leq 10^{12}$ and $10^{14}  < N_{\rm{ion}}/ \cmsq \leq 10^{15}$, respectively. Solid, long dashed, dashed and dotted curves show the results for the \emph{Ref}, \emph{StrongFB}, \emph{WeakFB} and \emph{NoAGN} models, respectively. The vertical dashed curves show the temperature at which the ion fraction peaks in CIE.}
\label{fig:cum-dist-T-c4o6-var}
\end{figure*}

\subsection{CDDF of absorbers}
\label{sec:CDDF-var}

The impact of varying the feedback on the CDDFs of $\CIV$ and $\OVI$ absorbers is shown in the left and right panels of Fig. \ref{fig:CDDF-var}, respectively\footnote{Hereafter we only show various trends for $\CIV$ and $\OVI$ absorbers. The trends for $\CIV$ ($\OVI$) are very similar to those for $\SiIV$ ($\NeVIII$)}, for $z = 3$ (top panel) and $z = 0.1$ (bottom panel). At $z = 3$ the different feedback models result in very similar CDDFs at $N_{\rm{ion}} \lesssim 10^{13}\cmsq$. At higher column densities, however, the amplitude of the CDDFs seem to anti-correlate with the strength of stellar feedback. The increasing sensitivity of absorbers to feedback with increasing column density can be explained by noting that weaker absorbers are typically at lower densities and farther away from galaxies (see \citealp{Rahmati14}; Rahmati et al. in prep.). We note that sensitivity of CDDFs to feedback is more evident at $z = 3$, around the peak of the cosmic star formation rate when feedback is also at its peak action. At low redshift (e.g., z = 0.1), however, the differences between the CDDFs in the \emph{WeakFB} and \emph{Ref} models decreases. This is somewhat a coincidence which is related to the fact that both \emph{Ref} and \emph{WeakFB} models have very similar stellar/metal contents $z \lesssim 1$ despite having very different star formation histories at higher redshifts.

The differences between the CDDFs in different models can be reduced significantly by accounting for the differences in their metal contents. For instance, at $z = 3$, the total metal mass (in gas) in the \emph{StrongFB} (\emph{WeakFB}) simulation is $\approx 0.4$ dex lower (higher) than that of the \emph{Ref} model. Noting that most gaseous metals are associated with relatively high column density systems, one can verify the above statement by shifting the CDDF of the \emph{StrongFB} (\emph{WeakFB}) model towards the right (left). We note however that such a shift would increase the differences between the low ends of the CDDFs at $z = 3$.

The comparison between the \emph{Ref} and \emph{NoAGN} models shows that AGN feedback does not change the distribution of high ionization metals significantly, although there is some effect for $\OVI$ at $z = 0.1$. This suggests that the bulk of metal absorbers are produced by low-mass galaxies since those are less affected by AGN feedback \cite[e.g.,][]{Crain15}. The total metal mass in gas is nearly the same in both models (the gaseous metal mass in the \emph{AGN} model is only $\approx 10\%$ higher and lower than the \emph{Ref} model at $z = 3$ and $z = 0$, respectively). We note that using a larger simulation box (i.e., 50 cMpc instead of 25 cMpc), and thus a larger number density of massive galaxies that are affected significantly by AGN feedback, slightly enhances the difference made by the presence of AGN feedback but only at the highest column densities (at $N_{\rm{ion}} \gtrsim 10^{14.5} \cmsq$ for $\CIV$ and $\OVI$ absorbers). This difference between the effect of AGN in the two different box sizes is, however, comparable to the effect of box size for the reference model (see Fig. \ref{fig:cddf-box} in Appendix \ref{ap:res})

The impact of feedback on the evolution of the cosmic density of $\CIV$ is shown in the left panel of Fig. \ref{fig:ion-dens-var}. As expected from the sensitivity of the CDDFs to feedback, the cosmic density of $\CIV$ and other ions (not shown) decreases if the efficiency of stellar feedback increases. Inclusion of AGN feedback, on the other hand, does not change the cosmic density of ions significantly. The right panel of Fig. \ref{fig:ion-dens-var} shows the normalized $\CIV$ density evolution after matching the cosmic elemental densities of the different simulations (total density of metals in gas and stars) to that of the \emph{Ref} simulations (i.e., $\Omega_{\rm{CIV,norm}} = \Omega_{\rm{CIV}} ~\frac{\Omega_{\rm{C,~Ref}}}{\Omega_{\rm{C}}}$ at a given redshift). The agreement between the different simulations improves after accounting for the differences between their total metal abundances. This result, which also holds for ions other than $\CIV$, suggests that the main impact of varying the feedback on the cosmic abundance of metal ions is caused by changes in the metal production rate, as we argued earlier. Further support for this argument comes from the fact that other feedback variations among EAGLE simulations that have similar star formation histories and stellar contents, also result in very similar CDDF and cosmic densities for the ions we study here (not shown).

\subsection{Physical properties of absorbers}
\label{sec:phys-var}

Fig. \ref{fig:4dplot-c4-o6-var} shows physical conditions associated with $\CIV$ (left) and $\OVI$ (right) absorbers for different feedback models at $z = 3$ (top) and $z = 0.1$ (bottom). This figure is analogous to Fig. \ref{fig:4dplot-all-ions} where cell sizes indicate the gas mass contained in each cell and colors indicate the median metallicities. Contours with blue solid, red long-dashed, green dashed and orange dotted lines show the temperature-density regions that contain $80\%$ of the absorbers\footnote{We impose a uniform distribution of absorbers as a function of column density by resampling the actual distribution, before calculating the temperature-density region which contains $80\%$ of absorbers.} in the \emph{Ref}, \emph{StrongFB}, \emph{WeakFB} and \emph{NoAGN} models, respectively.

As the top panels of Fig. \ref{fig:4dplot-c4-o6-var} suggest, at $z = 3$ temperature and density distributions of $\CIV$ and (to a lesser degree) $\OVI$ absorbers are insensitive to factor of two changes in the efficiency of stellar feedback and/or the inclusion of AGN feedback. We note, however, that for the highest column densities ($10^{14} < N_{\rm{ion}} < 10^{15}\cmsq$), which translates into absorbers with higher densities, the sensitivity to feedback increases. 

As shown in the bottom panels of Fig. \ref{fig:4dplot-c4-o6-var}, in contrast to the CDDFs and/or cosmic densities, the physical conditions of highly ionized metals are relatively sensitive to feedback at lower redshifts. At $z = 0.1$ and at fixed column density, absorbers tend to have higher densities in the presence of stronger stellar feedback (not shown).

We note that the temperature of the absorbers does tend to be sensitive to variations in the feedback. As Fig. \ref{fig:cum-dist-T-c4o6-var} shows, the typical temperature of $\CIV$ and $\OVI$ absorbers at $z = 0.1$ increases with the strength of stellar feedback (the same is true for the other ions we study in this work). This leads to an increasing fraction of collisionally ionized absorbers for more efficient feedback at $z = 0.1$ (the vertical dashed lines around $T \approx 10^5$ and $T \approx 10^{5.5}$ K in the left and right panels of Fig. \ref{fig:cum-dist-T-c4o6-var} indicate the temperatures at which the ion fractions peak in the CIE). This is particularly evident for the absorbers with the highest column densities ($10^{14}  < N_{\rm{ion}}/ \cmsq \leq 10^{15}$), since they are systems with high densities that are located near star forming regions.

Model \emph{NoAGN}, which yields only smaller differences in the cosmic star formation rate, and hence the resulting metal production rate, compared to the \emph{Ref} model yields metal absorbers with physical conditions that are similar to those in the \emph{Ref} model (e.g., compare the dotted and solid curves in Fig. \ref{fig:cum-dist-T-c4o6-var}). 

The temperature distribution of high column density $\OVI$ absorbers ($N_{\rm{OVI}} \gtrsim 10^{13-14}\cmsq$) at low redshifts ($z \sim 0$) shows the strongest sensitivity to variations of feedback, even for models with similar metal production rates. This makes the low-redshift temperature distribution of strong $\OVI$ (and other high ionization species susceptible to collisional ionization, e.g., $\NeVIII$) a promising probe for differentiating between different feedback models that result in similar cosmic distribution of absorbers.

\section{Summary and conclusions}
\label{sec:conclusions}
In this work we studied the cosmic distribution of highly ionized metals in the EAGLE simulations (S15). The EAGLE reference simulation reproduces the galaxy stellar mass function and galaxy star formation rates over a wide range of redshifts \citep{Furlong14} which is critical for achieving reasonable metal production rates. The feedback implementation allows the galactic winds to develop naturally without disabling hydrodynamics or radiative cooling, meaning that the effect of winds on the circumgalactic gas, in particular its phase structure, is modelled self-consistently. Because the gas properties were not considered during the calibration of the simulations, they provide a good basis for testing the simulations.

Exploiting the capability of the simulations to follow individual elemental abundances, we calculated the ion fractions of $\SiIV$, $\CIV$, $\NV$, $\OVI$ and $\NeVIII$, assuming ionization equilibrium in an optically thin gas exposed to the \citet{HM01} UVB model. The same UVB model, and therefore ion fractions, were used to calculate radiative cooling/heating rates in the simulations. This makes our ion fractions self-consistent with the hydrodynamical evolution of the baryons in the simulation. Note that the relatively high ionization energy of these species, and the typical densities at which they appear, make them largely insensitive to self-shielding corrections that are necessary for simulating species with lower ionization potentials, such as $\HI$ \citep[e.g.,][]{Rahmati13a}. 

We used a projection technique to obtain the simulated true column densities of absorption systems using projection lengths of $\sim 200-400$ km/s. Observationally, however, column densities are measured by fitting Voigt profiles to the spectra and decomposing them into components with different widths and at different redshifts. While this may cause differences between the observationally derived column density distributions and what is predicted on the basis of measuring the true column densities of simulated systems, the differences are expected to be smaller than other important sources of uncertainty such as noise, continuum fitting errors, deviations from perfect Voigt profiles and contamination.

We showed that EAGLE's predictions for the evolution of the column density distribution functions (CDDFs) of highly ionized metals broadly agree with observations (see figures \ref{fig:cddf-allz} and \ref{fig:cddf-allz2}). Despite the overall agreement, we found some significant differences. For example, the predictions seem to underproduce the high-end of the observed $\OVI$ CDDF ($N_{\OVI} \approx 10^{15}\cmsq$) at $z \sim 0$. However, the significance of those differences remain uncertain as they are comparable to the differences between different observational datasets. Moreover, even for a fixed IMF, the existing factor of $\sim2$ uncertainty in the stellar nucleosynthetic yields \citep[e.g.,][]{Wiersma09b} translates into the same level of uncertainty in the simulated column densities.

We found that the evolution of the CDDFs for all the ions (except $\NeVIII$) becomes stronger with increasing column density, and that the high column density ends of the CDDFs evolve similarly to the cosmic star formation rate density. We also found that the evolution of the CDDFs, particularly at lower column densities ($N_{\rm{ion}} \lesssim 10^{13}\cmsq$) is stronger for ions with higher ionization energies: while the CDDF of $\SiIV$ at $N_{\SiIV} \lesssim 10^{13}\cmsq$ hardly evolves between $z = 5$ and 0, the CDDF of $\NeVIII$ increases by $\approx 2$ dex at similar column densities from $z = 5$ to the present time. 

We also found broad agreement between the normalization and evolution of the predicted and observed cosmic densities of $\SiIV$, $\CIV$, $\NV$, $\OVI$ and $\NeVIII$ (see Fig. \ref{fig:ion-dens-obs}). The agreement is best at intermediate redshifts, $0.5 \lesssim z < 4$, while the simulation underproduces the observed cosmic densities of $\SiIV$, $\CIV$ and $\OVI$ by up to $0.3$ dex at $z \approx 0$, and the cosmic densities of $\SiIV$ and $\CIV$ are underproduced by up to $0.5$ dex at $z \approx 4$. We note, however, that the differences between different observational surveys have similar magnitudes. Using simulations with higher resolution and/or using a weaker UVB model (e.g., HM12) improves the agreement between EAGLE and the observations.

The cosmic densities of all ions increase with time at high redshifts ($z \gtrsim 2$), but their evolutions change qualitatively at later times depending on the ionization energy of the ions. While the cosmic densities of $\NeVIII$ and $\OVI$, which have the highest ionization energies among the ions we study here, increase monotonically with time, the cosmic density of $\NV$ peaks at $z \approx 2$ and remains largely unchanged afterwards. The cosmic densities of $\SiIV$ and $\CIV$, which have lower ionization energies, also peak at $z \approx 2$ but drop to nearly half of their maximum values by $z = 0$. Analysing the evolution of the ion fractions, we showed that a decrease in the ion fractions dominates the evolution of the $\SiIV$ and $\CIV$ cosmic densities at $z \lesssim 2$, while for $\NV$ the decrease in the ion fraction is just enough to compensate the monotonically increasing metal content of the gas, resulting in a nearly constant $\NV$ cosmic density at low redshifts. The ion fractions of $\OVI$ and $\NeVIII$ are roughly constant and increasing with time, respectively. Noting that $\OVI$ and $\NeVIII$ are mostly collisionally ionized, this trend is due to the increasing fraction of shock heated gas (i.e., gas hot enough to be collisionally ionized; see Fig. \ref{fig:4dplot-all-ions}) with decreasing redshift, which is a consequence of structure formation.

We showed that different ions are found in different parts of the density-temperature diagram (see Fig. \ref{fig:4dplot-all-ions}). The typical physical gas density of all metal absorbers increases with their column density and redshift (see Fig. \ref{fig:cum-dist-density-c4o6}), while the typical metallicity increases with time (see Fig. \ref{fig:cum-dist-Z-c4o6}). The typical physical densities of absorbers decrease as their ionization energies increase. As a result, $\NeVIII$ and $\OVI$ absorbers have typical densities for which radiative cooling is inefficient and adiabatic expansion and shock heating mainly govern the hydrodynamics. The typical temperature of $\NeVIII$ and $\OVI$ absorbers, therefore, increases with their (column) density. On the other hand, $\SiIV$ and $\CIV$ absorbers, which have lower ionization energies, have typical densities for which radiative cooling is more efficient and the temperature of the gas therefore decreases with their (column) density. 

The fraction of collisionally ionized metal absorbers increases with time and ionization energy. At $z = 3$, a negligible fraction of low column density $\SiIV$ and $\CIV$ absorbers (i.e., $N_{\rm{ion}} \lesssim 10^{12} \cmsq$) are affected by collisional ionization while a significant fraction ($10-20\%$) of high column density $\NV$ and $\OVI$ absorbers (i.e., $N_{\rm{ion}} \gtrsim 10^{14} \cmsq$) are collisionally ionized, and almost all relatively high column density $\NeVIII$ absorbers (i.e., $N_{\rm{ion}} \gtrsim 10^{13} \cmsq$) are collisionally ionized (see Fig. \ref{fig:cum-dist-T-z3}). At low-redshift, on the other hand, the significance of collisional ionization increases for all the metals, especially for those with higher ionization energies, e.g., $\gtrsim 20\%$ of $\OVI$ absorbers at $z = 0.1$ have temperatures higher than the peak of collisional ionization fraction (see Fig. \ref{fig:cum-dist-T-z0p1}).

By comparing simulations with feedback models varied with respect to our reference model, we found that AGN feedback has little effect on the CDDF of high ionization metals and their cosmic densities (see Figures \ref{fig:CDDF-var} and \ref{fig:ion-dens-var}). Increasing (decreasing) the efficiency of stellar feedback by a factor of 2, on the other hand, significantly decreases (increases) the CDDFs and cosmic densities of high ionization metals, particularly around the peak of the cosmic star formation rate density, and at high column densities. We argued that the impact of feedback is mainly due to changes in the metal production rate, i.e., the star formation rate. We supported this argument by showing that normalizing the cosmic gas metallicities of simulations with different feedback efficiencies, removes most of the differences in their metal CDDFs and cosmic ion densities. Moreover, simulations with other feedback variations that yield similar star formation histories, also have similar metal CDDFs and cosmic ion densities. This may suggest that, at least among EAGLE simulations, the constraints on feedback based on the cosmic ion densities of metals do not add significantly to what is provided by the evolution of the galaxy stellar mass function and/or the cosmic star formation history.

After inspecting the physical properties of metal absorbers and their sensitivity to feedback, we found that the temperature distribution of highly ionized metals that are prone to significant collisional ionization is sensitive to feedback, particularly at low redshifts ($z \lesssim 1$). At $z = 0.1$ the temperature distribution of $\OVI$ absorbers with $N_{\OVI} \gtrsim 10^{14}\cmsq$ varies significantly with changes in the feedback (see Fig. \ref{fig:cum-dist-T-c4o6-var}). For example, the temperature distribution of those absorbers suggests that the fraction of collisionally ionized high-column density $\OVI$ absorbers increases from $\lesssim 10\%$ when AGN feedback is absent, to $\gtrsim 80\%$ when the efficiency of stellar feedback is twice as high as in the reference model. This provides a promising avenue for constraining feedback processes through analyzing the temperature distribution of high column density metal absorbers at relatively high column densities. 

In this work we investigated the CDDFs of metal ions, irrespective of the locations of galaxies. In future work, we will use EAGLE to study the relation between metal absorbers and galaxies.

\section*{Acknowledgments}
\addcontentsline{toc}{section}{Acknowledgments}
We thank the anonymous referee for useful suggestions. We also thank Simeon Bird, Richard Bower, Joseph Burchett, Romeel Dave, Stephan Frank, Piero Madau, Lucio Mayer, Evan Scannapieco and Benny Trakhtenbrot for useful discussions. We thank Simeon Bird and Charles Danforth for providing us with machine-readable versions of their published results. We thank all the members of the EAGLE collaboration. This work used the DiRAC Data Centric system at Durham University, operated by the Institute for Computational Cosmology on behalf of the STFC DiRAC HPC Facility (www.dirac.ac.uk). This equipment was funded by BIS National E-infrastructure capital grant ST/K00042X/1, STFC capital grant ST/H008519/1, and STFC DiRAC Operations grant ST/K003267/1 and Durham University. DiRAC is part of the National E-Infrastructure. We also gratefully acknowledge PRACE for awarding us access to the resource Curie based in France at Tr{\'e}s Grand Centre de Calcul. This work was sponsored in part with financial support from the Netherlands Organization for Scientific Research (NWO), from the European Research Council under the European Union's Seventh Framework Programme (FP7/2007-2013) / ERC Grant agreement 278594-GasAroundGalaxies, from the UK Science and Technology Facilities Council (grant numbers ST/F001166/1 and ST/I000976/1) and from the Interuniversity Attraction Poles Programme initiated by the Belgian Science Policy Office ([AP P7/08 CHARM]). RAC is a Royal Society University Research Fellow.

\appendix
\section{UVB and physical conditions of metal absobers}
\label{ap:UVB}

The column density distribution of high ionization metal absorbers that are primarily photoionized are expected to be sensitive to the properties of the UVB radiation field. As mentioned in $\S$\ref{sec:ingredients} and for consistency with our hydrodynamical simulations, which use the HM01 UVB model to calculate the heating/cooling rates, we opted to use the same UVB model for calculating the ion fractions. We note that our fiducial UVB model is consistent with constraints on the photoionization rate of the background radiation at different redshifts \citep[e.g.,][]{Becker13}, and has been used to successfully reproduce $\HI$ CDDF and its evolution \citep[e.g.,][]{Rahmati13a,Rahmati15}, the connection between strong $\HI$ absorbers and galaxies \citep{Rahmati14}, as well as relative strengths of different high ionization metal ions \citep[e.g.,][]{Aguirre08}. However, the constraints on properties of the UVB are also consistent with other UVB models such as the models proposed by HM12 and \citet{FG08} which predict different UVB radiation fields compared to HM01. 

\begin{figure*}
\centerline{\hbox{{\includegraphics[width=0.45\textwidth]
              {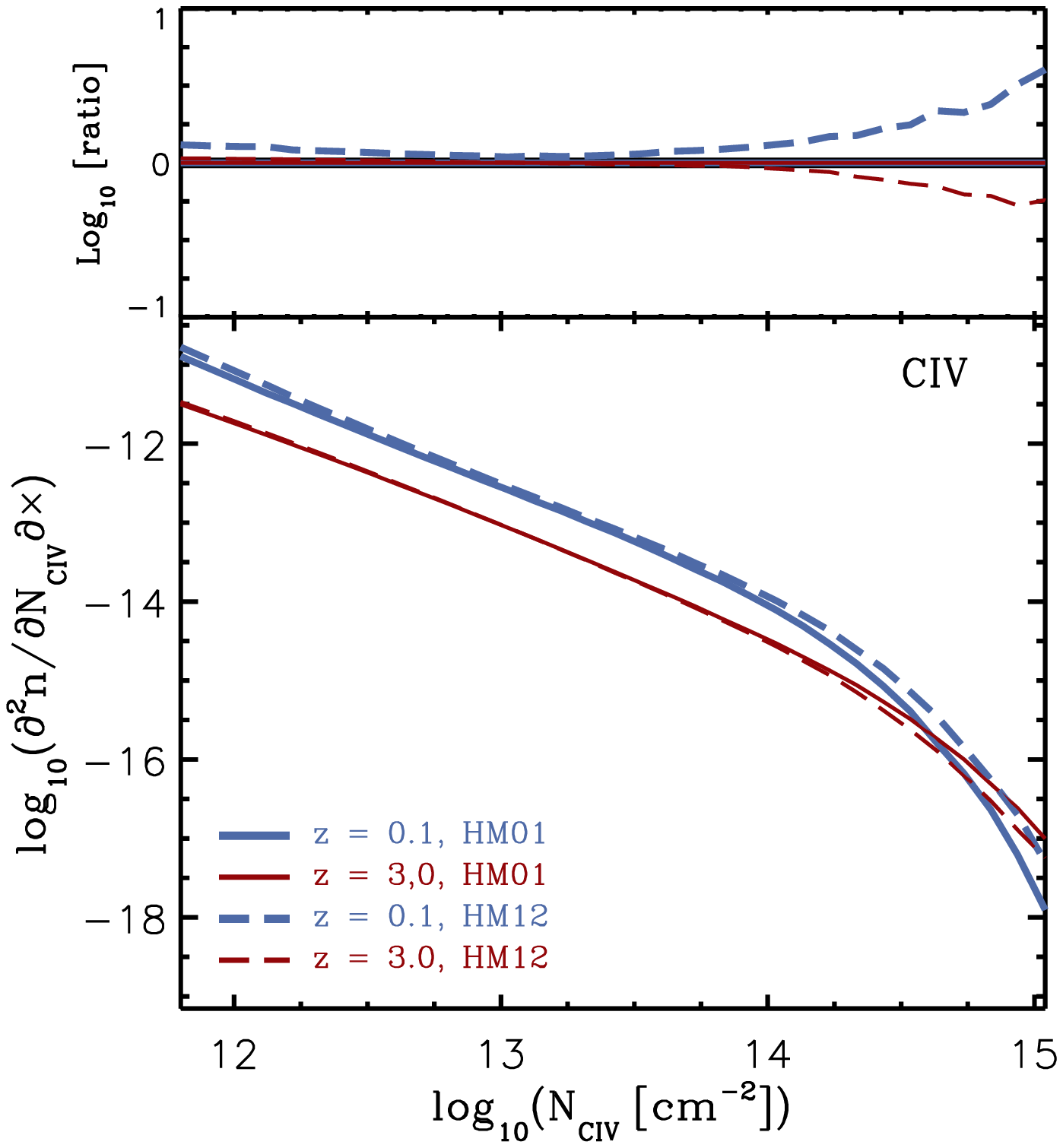}}}
             \hbox{{\includegraphics[width=0.45\textwidth]	
             {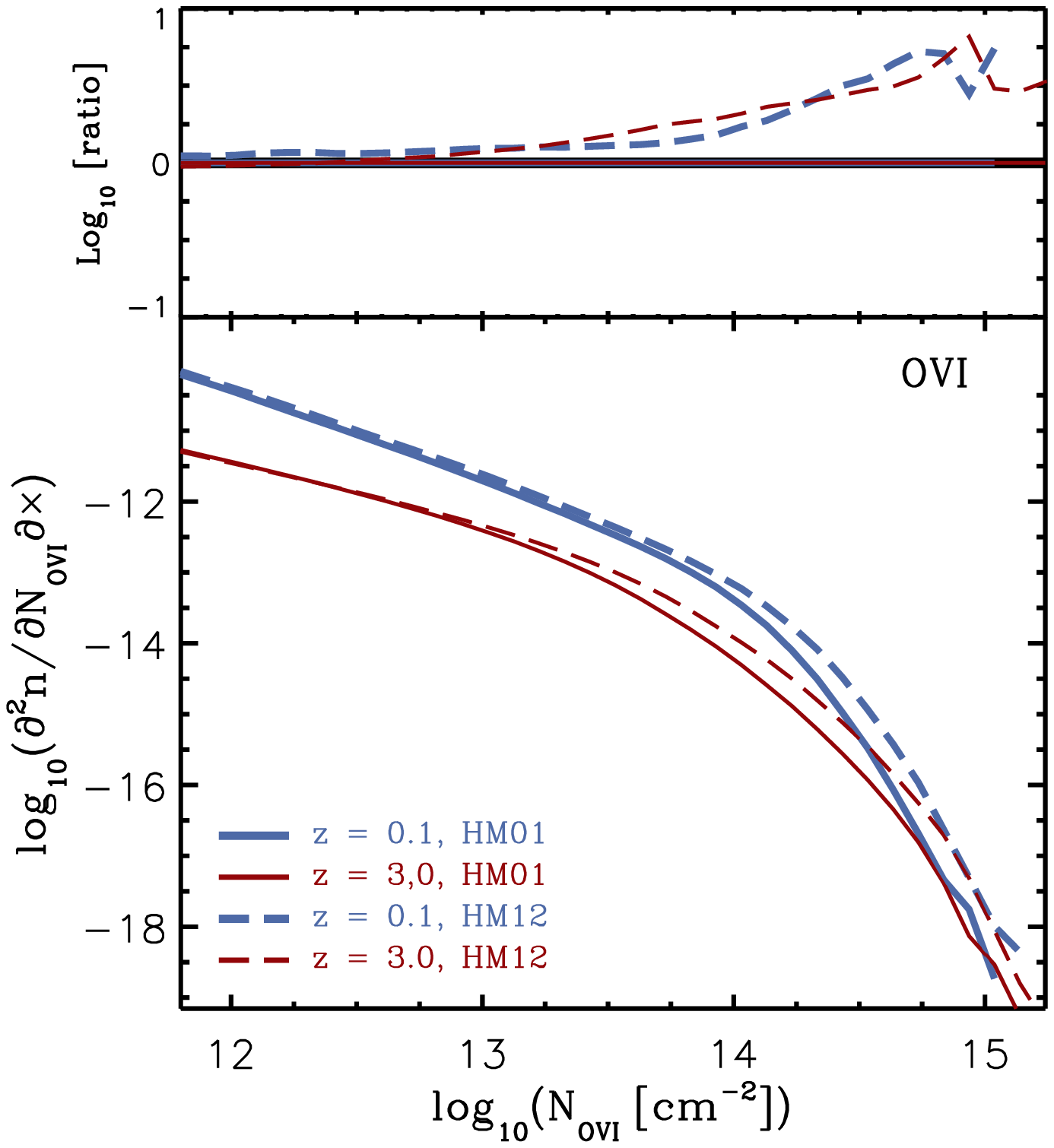}}}}
\caption{Column density distribution function of $\CIV$ (left panels) and $\OVI$ (right panels) in the EAGLE \emph{Ref-L025N0376} simulation using different UVB models. The solid curves in each plot show the result obtained using the \citet{HM01} UVB model (HM01) to calculate the ion fractions while the dashed curves show the results using the \citet{HM12} UVB model (HM12). Thin red and thick blue curves show the CDDFs at $z = 0.1$ and $z = 3$ respectively. The top section of each panel shows the ratio between the CDDF obtained using the HM12 model and that the fiducial UVB model, HM01. For both $\CIV$ and $\OVI$ absorbers, the highest column densities ($N_{\rm{ion}} \gtrsim 10^{14} \cmsq$) are more affected by changing the UVB model.}
\label{fig:CDDF-UVB}
\end{figure*}

To investigate the impact of using a different UVB radiation field on our results, we used the HM12 UVB model to calculate the ion fractions. The resulting CDDFs for $\CIV$ and $\OVI$ absorbers are compared with the results obtained by using our fiducial UVB model (e.g., HM01) in the left and right panels of Fig. \ref{fig:CDDF-UVB}, respectively. In each panel, the reference result using the HM01 UVB model is shown using solid curves while the long-dashed curves show the CDDF obtained using the HM12 model. Red thin curves indicate $z = 3$ and blue thick curves show $z = 0.1$. The top section of each panel shows the ratio between CDDFs obtained by using the HM12 and those obtained using HM01 for the same redshifts.

\begin{figure*}
\centerline{\hbox{{\includegraphics[width=0.5\textwidth]
              {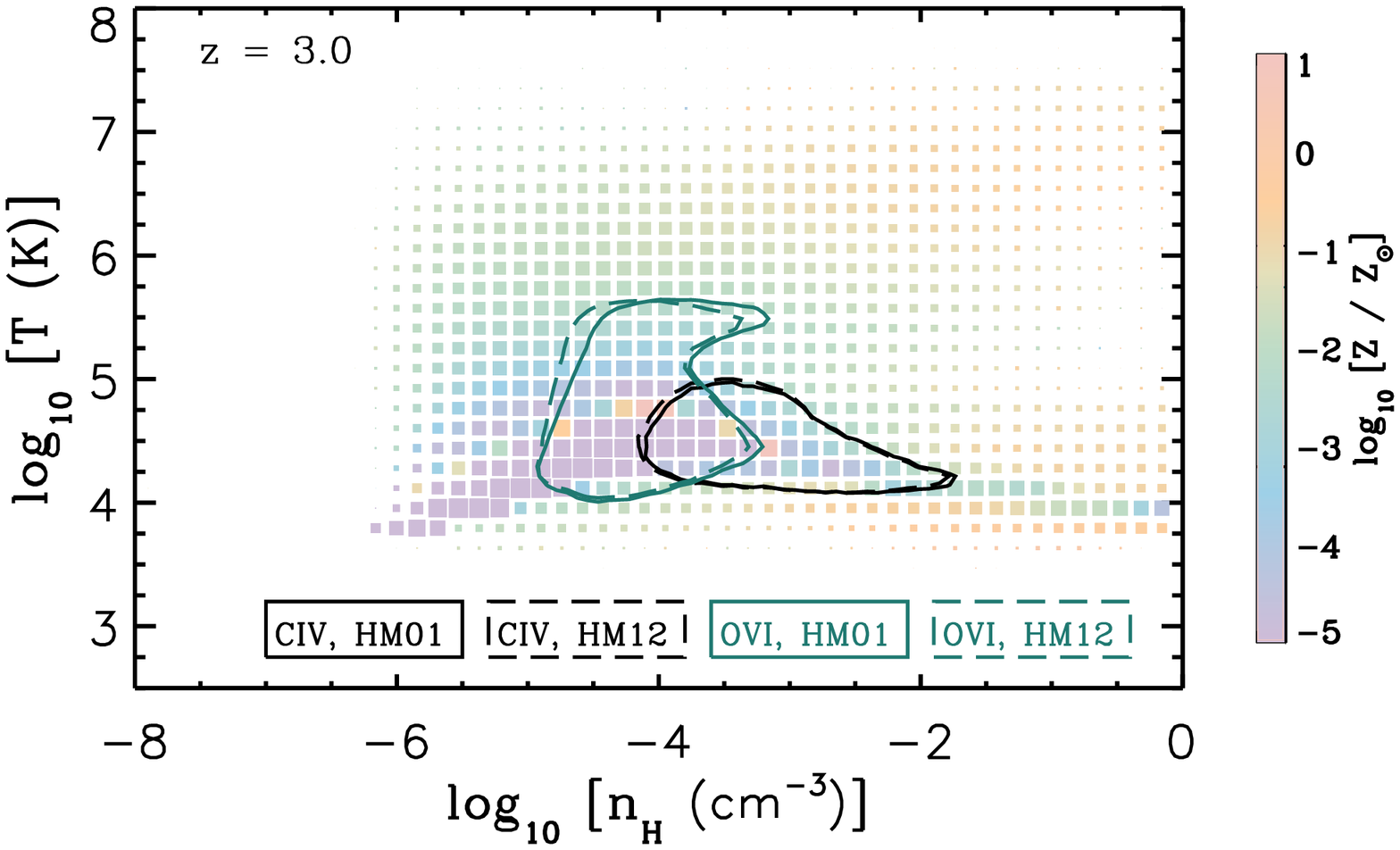}}}
             \hbox{{\includegraphics[width=0.5\textwidth]	
             {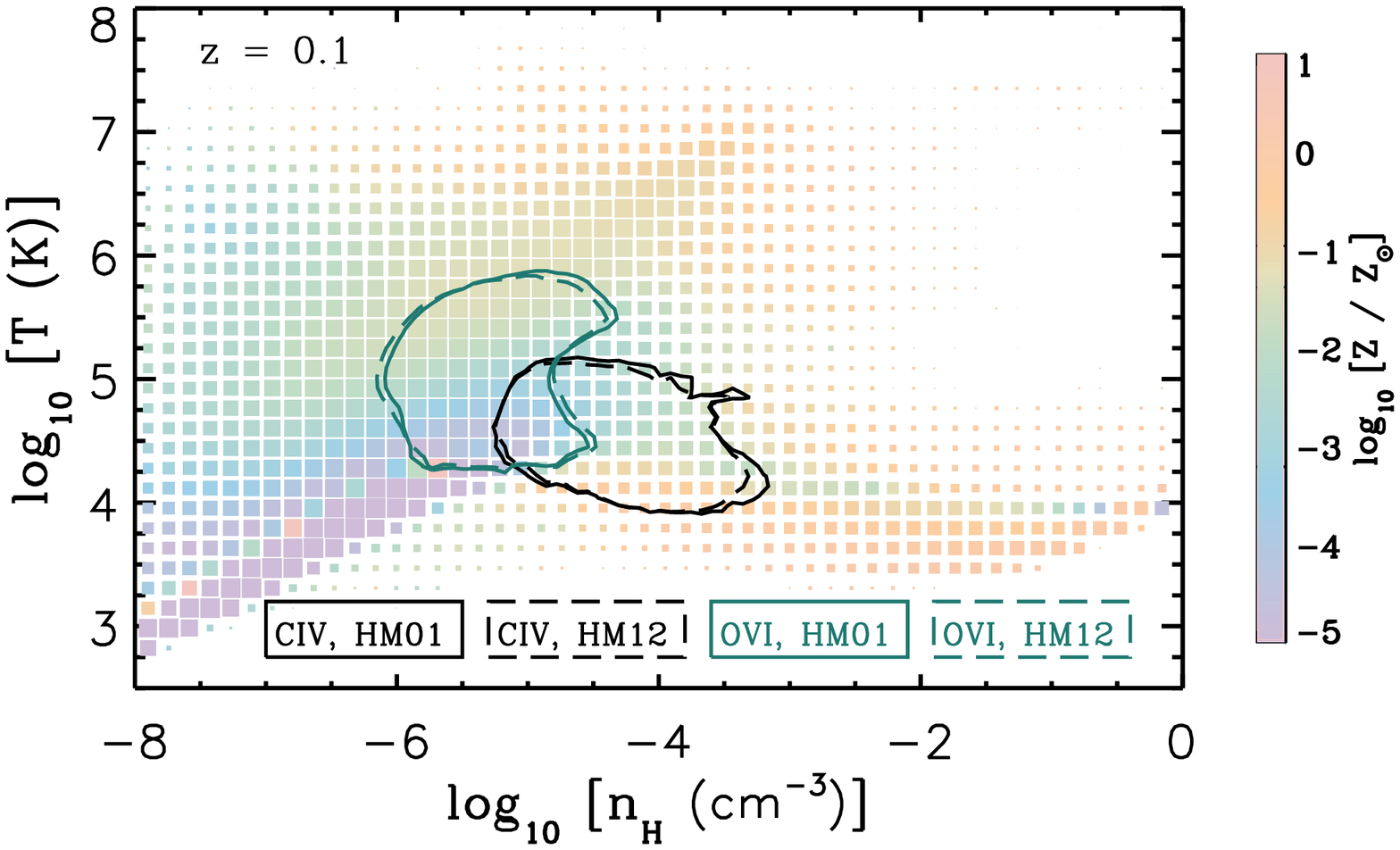}}}}
\caption{Temperature-density distribution of gas in the EAGLE \emph{Ref-L025N0376} simulation at $z = 3$ (left) and $z = 0.1$ (right) using different UVB models. The size of each cell is proportional to the logarithm of the gas mass enclosed in it and its color shows its median metallicity. The areas enclose by contours with different colors and line-styles show the range of temperature-densities that contain $80\%$ of $\CIV$ (black) and $\OVI$ (green) masses in the presence of different UVB models. The solid and dashed curves represent the HM01 and HM12 models, respectively. The physical properties of the absorbers change only weakly when the UVB is changed from HM01 to HM12.}
\label{fig:4dplot-c4-o6-UVB}
\end{figure*}
As Fig. \ref{fig:CDDF-UVB} shows, the impact of varying the UVB model is largest at high column densities, both for $\CIV$ (at $N_{\rm{CIV}} \gtrsim 10^{14}\cmsq$) and for $\OVI$ (at $N_{\rm{OVI}} \gtrsim 10^{13}\cmsq$). Using the HM12 model instead of our fiducial HM01 model results in a higher and lower $\CIV$ CDDF at $z= 0.1$ and $z = 3$, respectively. This is expected noting that the normalization of the HM12 UVB model at $\sim 4$ Ryd is higher (lower) than HM01 at $z = 0$ (3). For $\OVI$ using HM12 which at $\sim 8$ Ryd is stronger than HM01 at $z \lesssim 4$ increases the CDDF at both redshifts, and improves the agreement with the observations (see figures \ref{fig:cddf-allz} and \ref{fig:cddf-allz2}). The cosmic density of $\CIV$ is increased (decreased) by $\approx 40\%$ ($\approx 15\%$) at $z = 0.1$ ($z = 3$) by switching to HM12, while the cosmic density of $\OVI$ increases at $z = 0.1$ and $z= 3$ by $\approx 20\%$ and $\approx 70\%$, respectively. Both aforementioned changes would  improve the agreement with observations (see Fig. \ref{fig:ion-dens-obs}).

Hence, using the more recent HM12 model for the UVB improves the already reasonable agreement we found in $\S$\ref{sec:results} between our predictions (using the HM01 model) and observations. This is consistent with what we found to be the case for $\HI$ absorbers at $z \sim 2$ in \citep{Rahmati15}. The physical properties of the absorbers, on the other hand, are insensitive to whether HM01 or HM12 is used, as shown in Fig. \ref{fig:4dplot-c4-o6-UVB}.
\begin{figure*}
\centerline{\hbox{{\includegraphics[width=0.45\textwidth]
              {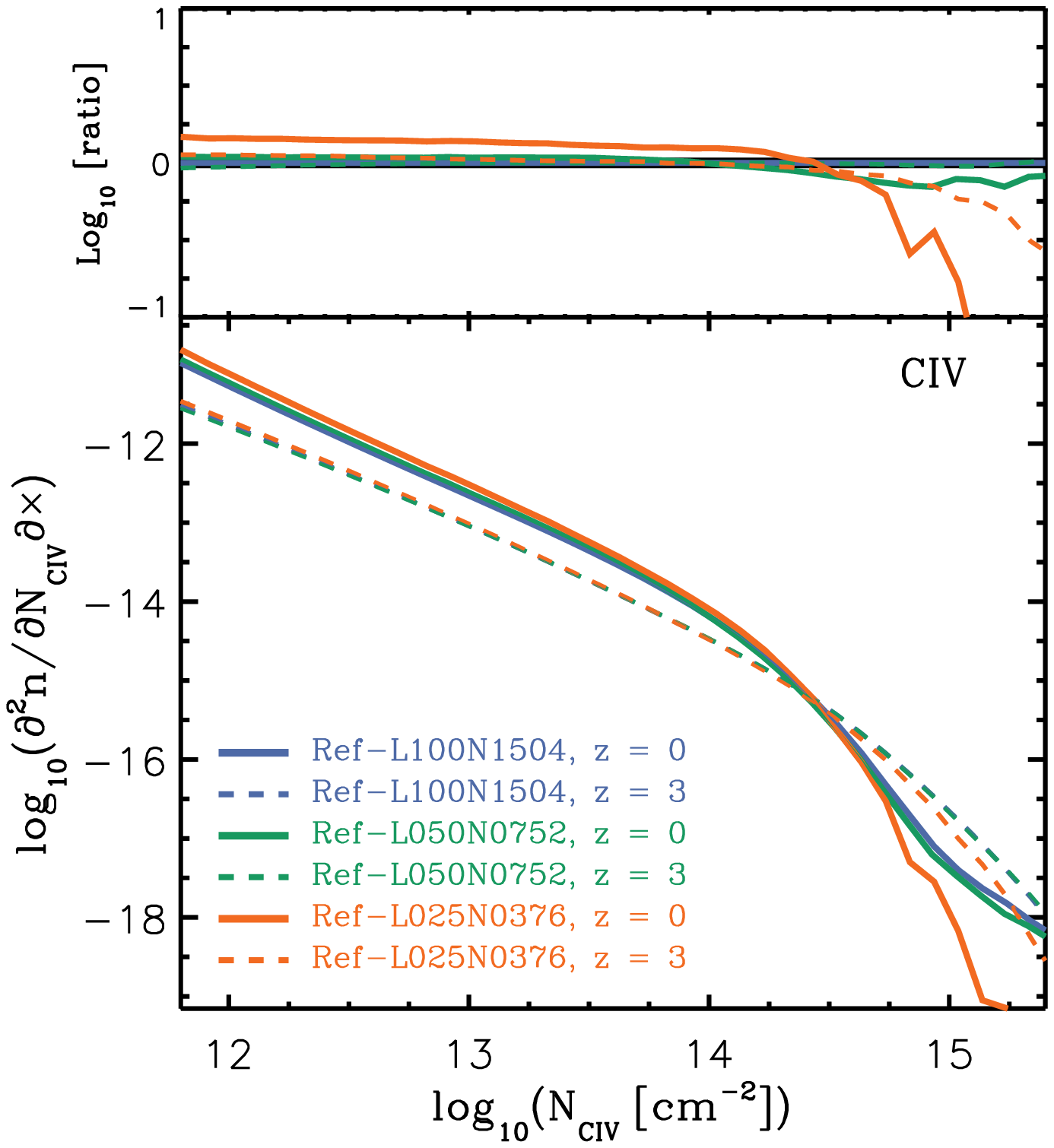}}}
             \hbox{{\includegraphics[width=0.45\textwidth]	
             {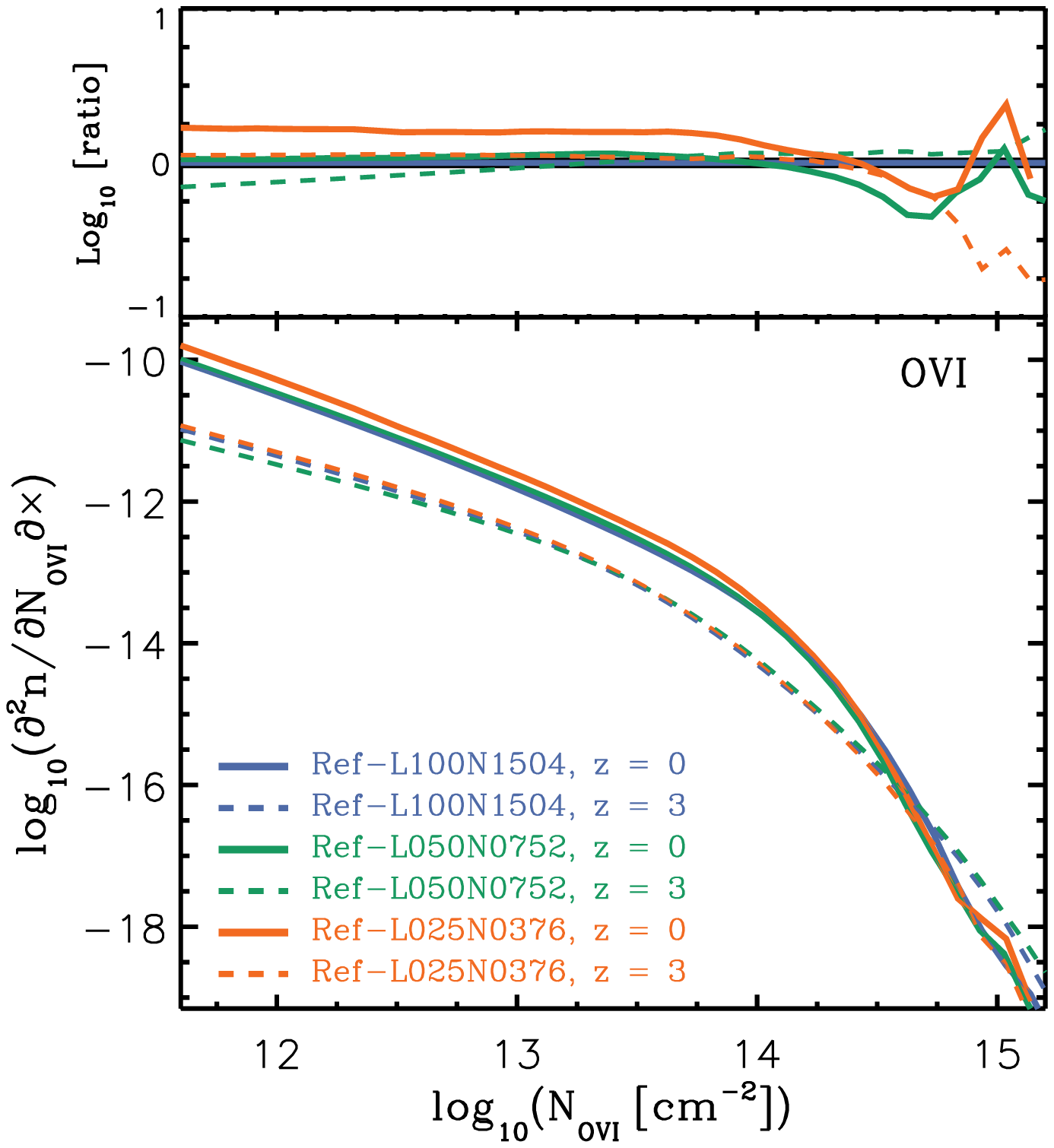}}}}
\caption{Column density distribution functions of $\CIV$ (left panel) and $\OVI$ (right panel) at $z = 3$ (dashed curves) and $z = 0$ (solid curves) in EAGLE simulations with different box sizes but a fixed resolution. Blue, green and red curves show the results from the \emph{Ref-L100N1504}, \emph{Ref-L050N0752} and \emph{Ref-L025N0376} simulations. The top section of the panels show the ratios of the CDDFs between different simulations and \emph{Ref-L100N1504} at a given redshift. The CDDFs of our fiducial simulations are converged with the simulation box size. The differences are largest at the highest column densities, suggesting that massive galaxies which are missing from smaller box simulations, are responsible for $\CIV$ and $\OVI$ absorbers with $\Nion \sim 10^{14} \cmsq$.}
\label{fig:cddf-box}
\end{figure*}
\begin{figure*}
\centerline{\hbox{{\includegraphics[width=0.45\textwidth]
              {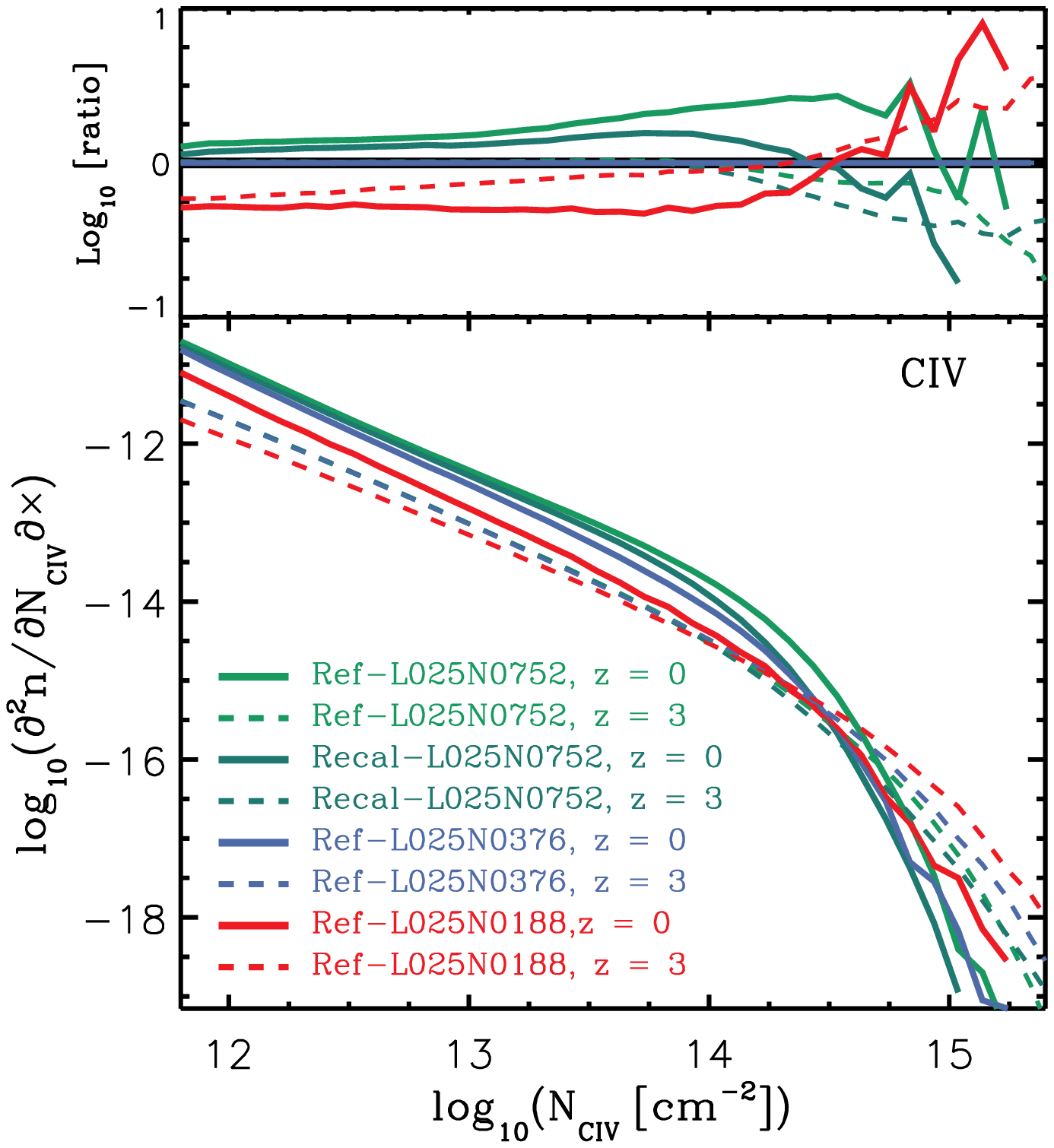}}}
             \hbox{{\includegraphics[width=0.45\textwidth]	
             {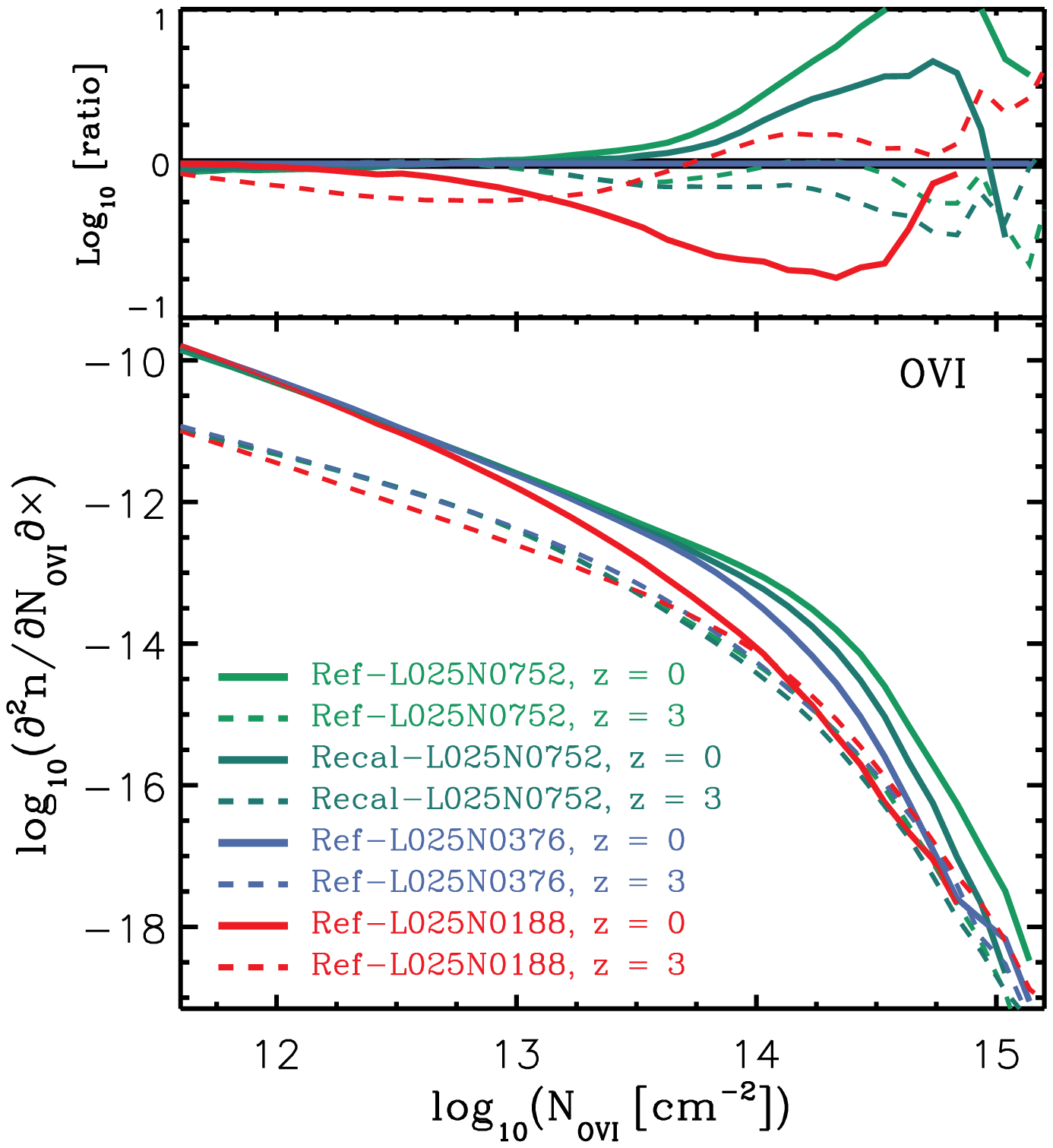}}}}
\caption{Column density distribution functions of $\CIV$ (left panel) and $\OVI$ (right panel) at $z = 3$ (dashed curves) and $z = 0$ (solid curves) in EAGLE simulations with different resolutions but the same box size. Light green, green, blue, and red curves show the results from the \emph{Ref-L025N0752}, \emph{Recal-L025N0752}, \emph{Ref-L025N0376} and \emph{Ref-L025N0188} simulations, respectively. The top section of the panels show the ratios between different CDDFs and that of the \emph{Ref-L025N0376} simulation at a given redshift. The convergence is best for $\Nion \lesssim 10^{14} \cmsq$ and at $z = 3$ (i.e., higher redshifts). Moreover, the weak convergence is better than the strong convergence (see the text).}
\label{fig:cddf-res}
\end{figure*}
\begin{figure*}
\centerline{\hbox{{\includegraphics[width=0.5\textwidth]
              {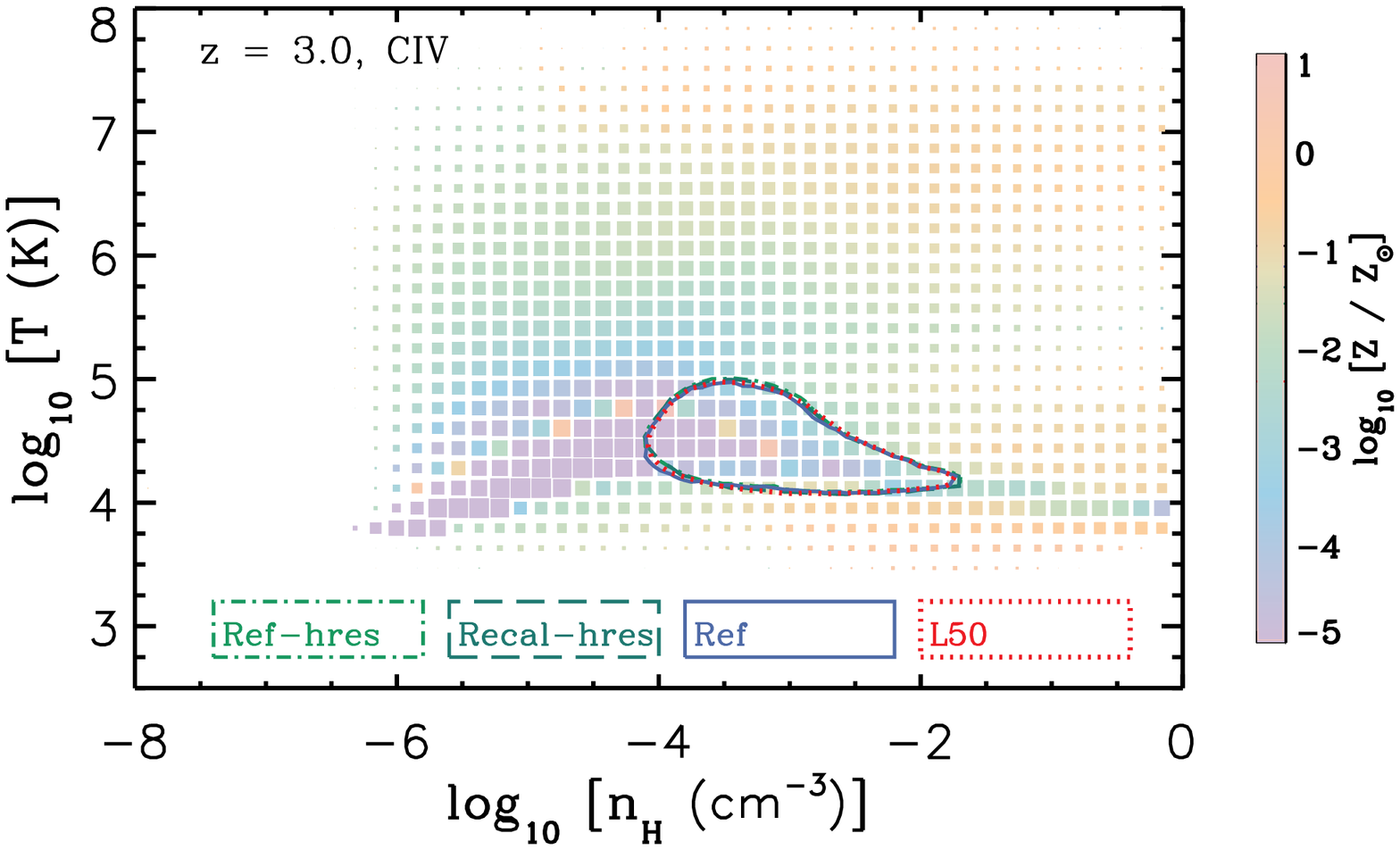}}}
             \hbox{{\includegraphics[width=0.5\textwidth]	
             {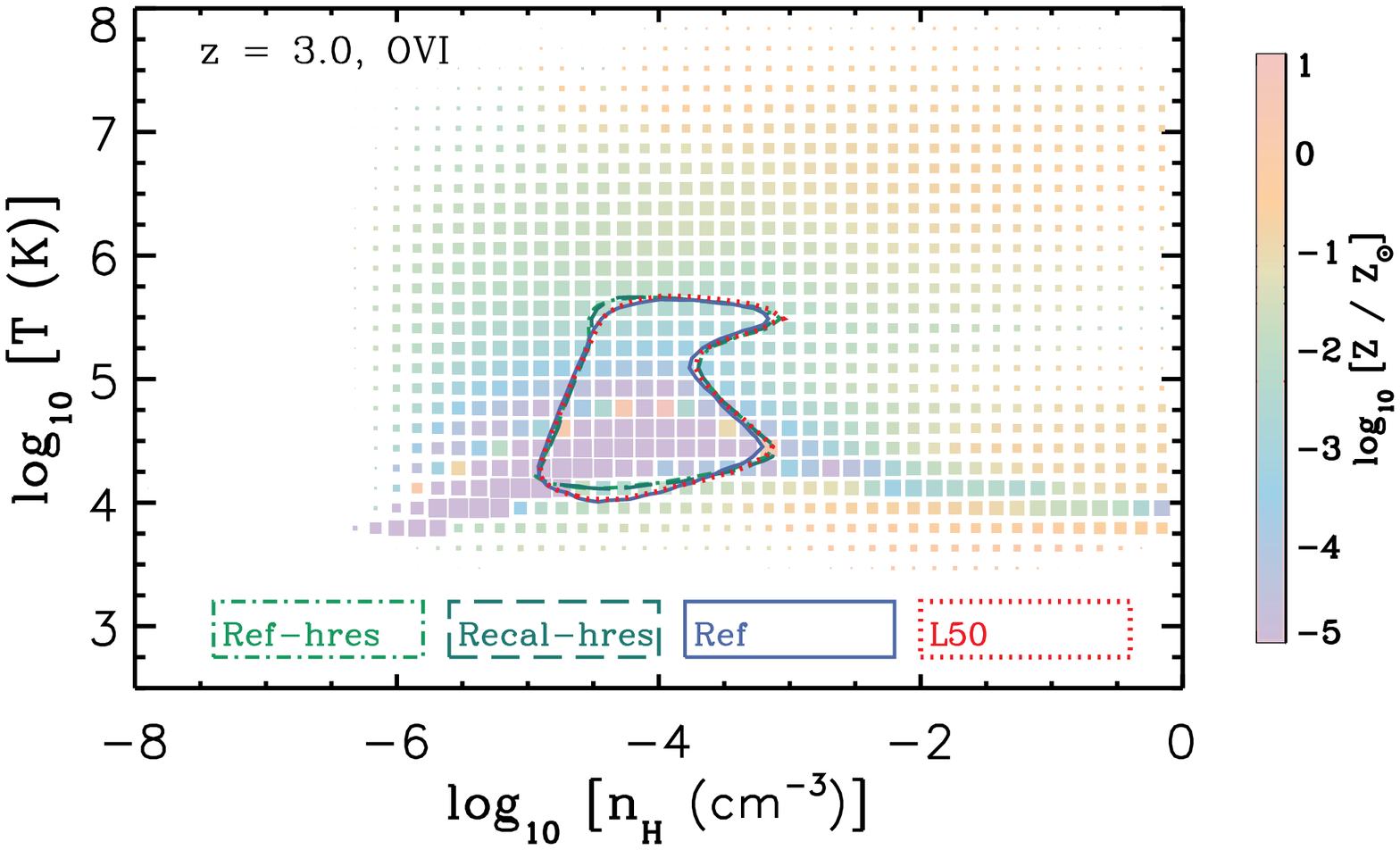}}}}
\centerline{\hbox{{\includegraphics[width=0.5\textwidth]
              {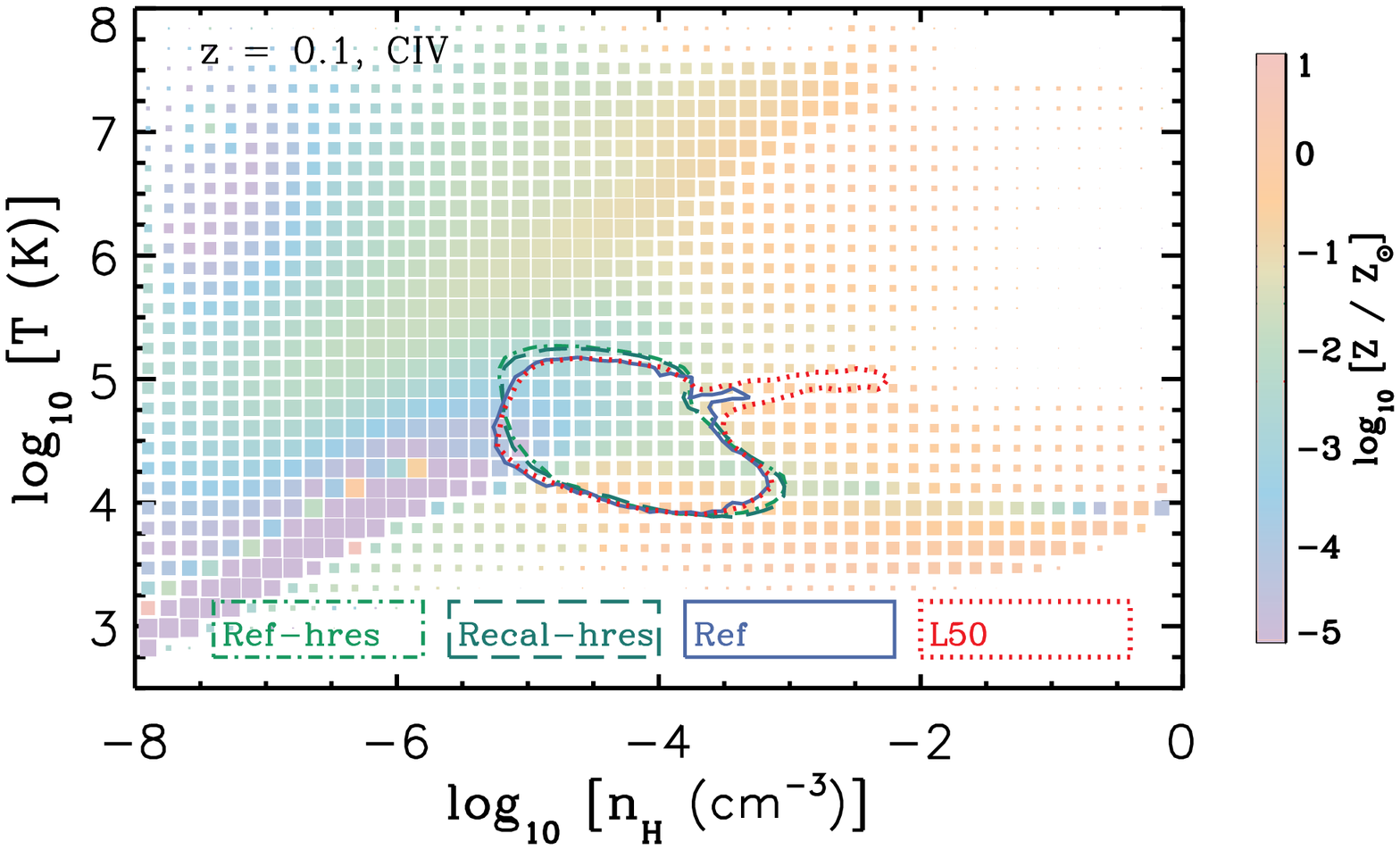}}}
             \hbox{{\includegraphics[width=0.5\textwidth]	
             {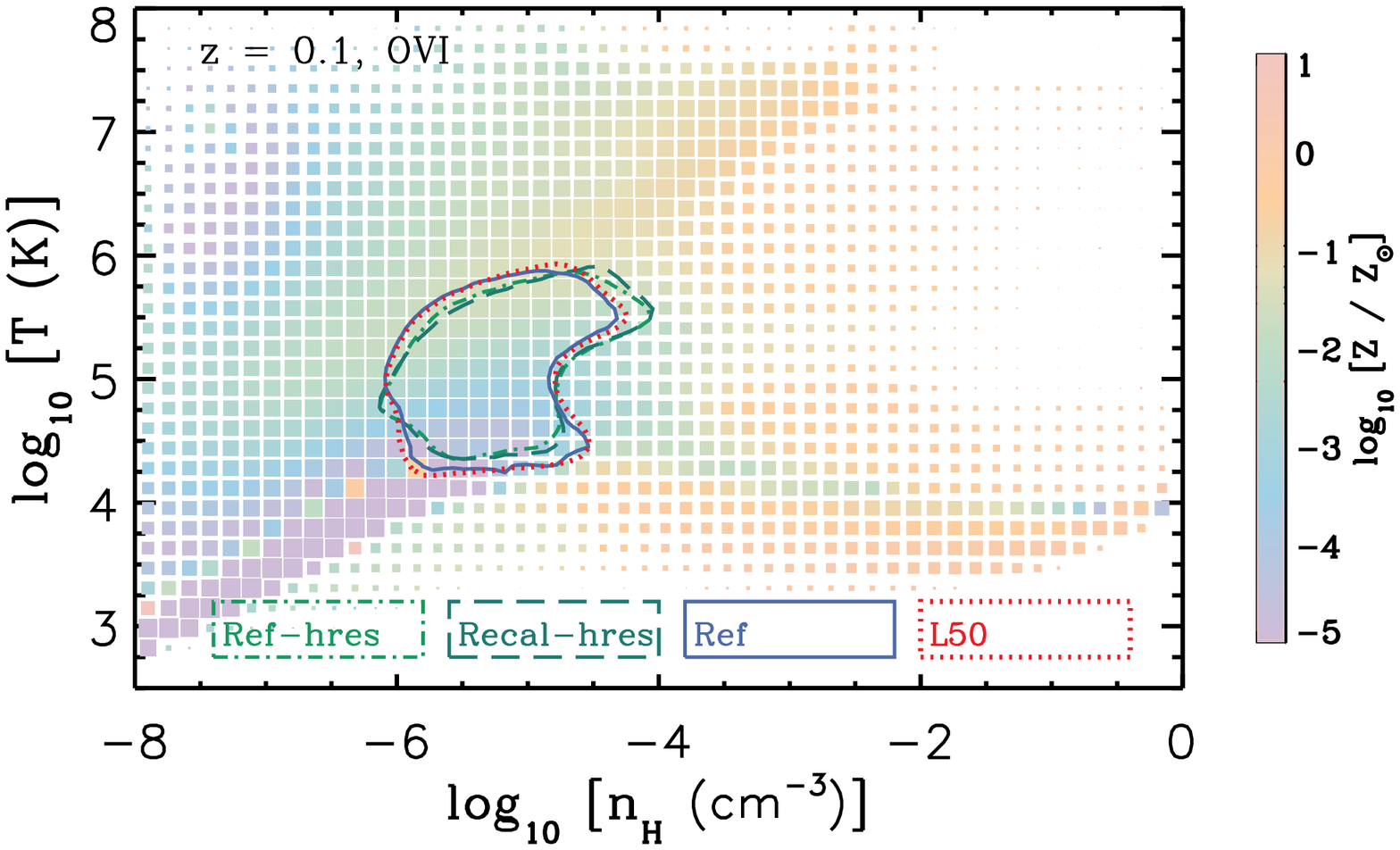}}}}
\caption{Temperature-density distribution of gas in the EAGLE \emph{Ref-L050N0752} simulation at $z = 3$ (top) and $z = 0.1$ (bottom) which has a numerical resolution identical to our fiducial simulation (\emph{Ref-L100N1504}). The size of each cell is proportional to the logarithm of the gas mass enclosed in it and its color shows its median metallicity. The areas enclose by contours with different colors and line-styles show the range of temperature-densities that contain $80\%$ of $\CIV$ (left) and $\OVI$ (right) masses in simulations with different resolutions and box-sizes. Blue solid, green long-dashed, green dot-dashed and red dotted lines show \emph{Ref-L025N0376} (REF), \emph{Recal-L025N0752} (Recal-hres), \emph{Ref-L025N0752} (REF-hres) and \emph{Ref-L050N0752} (L50) models, respectively. The temperature-density distribution of absorbers is insensitive to resolution and box size.}
\label{fig:4dplot-c4-o6-res}
\end{figure*}
\section{Box size and resolution tests}
\label{ap:res}

Fig. \ref{fig:cddf-box} shows the impact of the size of the simulation box on the CDDFs of $\CIV$ (left) and $\OVI$ (right) absorbers. In each panel, the solid and dashed curves show the CDDFs at $z = 0$ and $z = 3$, respectively. Blue, green and orange curves show the CDDFs in the \emph{Ref-L100N1503}, \emph{Ref-L050N0752} and \emph{Ref-L025N0376} simulations, respectively. The simulations use identical subgrid physical models and numerical resolution, they differ only in their box sizes. The differences between $L = 25$ and 50 cMpc are substantial, particularly for high column density but the CDDFs are converged for box sizes $L \ge 50$ cMpc at $z = 0$. The cosmic densities of the ions we study here also converge with box size for $L \ge 50$ cMpc (not shown).

Fig. \ref{fig:cddf-res} shows the impact of the numerical resolution on the CDDFs of $\CIV$ (left) and $\OVI$ (right) absorbers. In each panel, the solid and dashed curves show the CDDFs at $z = 0$ and $z = 3$, respectively. Green, dark green, blue and red curves show the CDDFs in the \emph{Ref-L025N0752}, \emph{Recal-L025N0752}, \emph{Ref-L025N0376} and \emph{Ref-L025N0188} simulations, respectively. The simulations use identical box sizes but different resolutions. The three \emph{Ref-L025NXXXX} simulations use the reference model, but for the \emph{Recal-L025N0752} simulation the subgrid model for feedback was recalibrated to better match the $z \sim 0$ stellar mass function (to account for resolution dependence of feedback and star-formation). We use the simulations that have identical models and different resolutions to perform \emph{strong} convergence test. In addition, to perform \emph{weak} convergence test, we compare simulations with different resolutions and different models that account for the resolution dependencies in the sub-grid model and produce similar $z = 0$ stellar mass functions (see S15). The CDDFs are not fully converged with resolution at high column densities, especially at $z = 0$, with higher-resolution simulations resulting in more high column density metal absorbers. Note that the numerical convergence looks better if we consider the \emph{weak} convergence instead of \emph{strong} convergence. However, we note that much of the difference is most likely caused by differences in metal production rates in simulations with different resolutions. Indeed, the cosmic densities of ions, which increase with increasing resolution at late times shows a very weak resolution dependence after normalizing the different simulations to the same elemental abundances (see $\S$\ref{sec:CDDF-var}).
\end{document}